\documentclass[a4paper,eqsecnum,nofootinbib]{revtex4}
\usepackage{slashed,bbold,amsmath,amssymb,mathtools,wrapfig} 
\usepackage{color,epsfig}
\usepackage{hyperref}

\newcommand{\nn}{\nonumber}
\newcommand{\beq} {\begin{equation}}
\newcommand{\eeq} {\end{equation}}
\newcommand{\beqa} {\begin{eqnarray}}
\newcommand{\eeqa} {\end{eqnarray}}
\newcommand{\bal} {\begin{align}}
\newcommand{\eal}{\end{align}}

\newcommand{\cf}{{\it cf.}}
\newcommand{\ie}{{\it i.e.}}

\newcommand{\eg}{{\it e.g.}}

\newcommand{\as}{{\alpha_s}}
\newcommand{\lqcd}{\Lambda_{QCD}}

\newcommand{\la}{\Lambda}

\newcommand{\vphi}{\varphi}
\newcommand{\veps}{\varepsilon}
\newcommand{\order}[1]{${\cal O}\left(#1 \right)$}
\newcommand{\morder}[1]{{\cal O}\left(#1 \right)}
\newcommand{\eq}[1]{(\ref{#1})}
\newcommand{\fig}[1]{Fig.~\ref{#1}}
\newcommand{\lsim}{\lesssim}
\newcommand{\gsim}{\gtrsim}

\newcommand{\inv}[1]{\frac{1}{#1}}
\newcommand{\halft}{{\textstyle \frac{1}{2}}}

\newcommand{\quart}{{\textstyle \frac{1}{4}}}
\newcommand{\sfrac}[2]{{\textstyle\frac{#1}{#2}}}
\newcommand{\ket}[1]{\left\vert{#1}\right\rangle}
\newcommand{\bra}[1]{\langle{#1}\vert}

\newcommand{\com}[2]{\left[{#1},{#2}\right]}
\newcommand{\comb}[2]{\big[{#1},{#2}\big]}
\newcommand{\acom}[2]{\left\{{#1},{#2}\right\}}
\newcommand{\acomb}[2]{\big\{{#1},{#2}\big\}}
\newcommand{\tr}{\mathrm{Tr}\,}

\newcommand{\bs}[1]{\boldsymbol{#1}}

\newcommand{\psl}{{\slashed{p}}}
\newcommand{\Psl}{{\slashed{P}}}
\newcommand{\Asl}{{\slashed{A}}}
\newcommand{\Pisl}{{\slashed{\Pi}}}

\newcommand{\mA}{\mathcal{A}}
\newcommand{\mB}{\mathcal{B}}

\newcommand{\bC}{\mathbb{C}}

\newcommand{\mE}{\mathcal{E}}
\newcommand{\mH}{\mathcal{H}}
\newcommand{\mJ}{\mathcal{J}}
\newcommand{\mK}{\mathcal{K}}

\newcommand{\mM}{\mathcal{M}}
\newcommand{\mP}{\mathcal{P}}
\newcommand{\bP}{\mathbb{P}}

\newcommand{\xv}{{\bs{x}}} 
\newcommand{\yv}{{\bs{y}}}

\newcommand{\pv}{{\bs{p}}}
\newcommand{\kv}{{\bs{k}}}
\newcommand{\qv}{{\bs{q}}}

\newcommand{\Pv}{{\bs{P}}}
\newcommand{\Av}{{\bs{A}}}

\newcommand{\gv}{\bs{\gamma}}

\newcommand{\gz}{\gamma^0}
\newcommand{\go}{\gamma^1}

\newcommand{\gff}{\gamma_5}

\newcommand{\nv}{\bs{\nabla}}
\newcommand{\sv}{\bs{\sigma}}

\newcommand{\moe}{{m}}

\newcommand{\kum}{{\,{_1}F_1}}

\newcommand{\wtB}{{\widetilde{B}}}
\newcommand{\wtD}{{\widetilde{D}}}

\newcommand{\rar}{\rightarrow}
\newcommand{\lar}{\leftarrow}

\newcommand{\rder}{{\buildrel\rar\over{\partial}}}
\newcommand{\lder}{{\buildrel\lar\over{\partial}}}

\newcommand{\rnab}{{\buildrel\rar\over{\nv}}}
\newcommand{\lnab}{{\buildrel\lar\over{\nv}}}

\newcommand{\dsi}{\sigma}
\newcommand{\tm}{{\tt{m}}}
\newcommand{\jz}{\lambda}

\newcommand{\xtr}{\bs{x}^\perp}
\newcommand{\ntr}{\bs{\nabla}^\perp}
\newcommand{\atr}{{\bs{\alpha}^\perp}}
\newcommand{\alv}{{\bs{\alpha}}}
\newcommand{\apa}{{\alpha^\parallel}}

\newcommand{\wfxi}{\Phi^{(\xi)}}
\newcommand{\wfr}{{\Phi^{(0)}}}

\begin{document}

\title{Lectures on Bound states\footnote{Based on lectures presented during 2014-15 at NIKHEF, Amsterdam; IPhT Saclay; CP$^3$, Odense; GSI, Darmstadt; Bloomington, IN and Subatech, Nantes.}}

\author{Paul Hoyer}
\affiliation{Department of Physics, POB 64, FIN-00014 University of Helsinki, Finland}

\begin{abstract} 

Even a first approximation of bound states requires contributions of all powers in the coupling. This means that the concept of ``lowest order bound state'' needs to be defined. In these lectures I discuss the ``Born'' (no loop, lowest order in $\hbar$) approximation. Born level states are bound by gauge fields which satisfy the classical field equations.

As a check of the method, Positronium states of any momentum are determined as eigenstates of the QED Hamiltonian, quantized at equal time. Analogously, states bound by a strong external field $A^\mu(\xv)$ are found as eigenstates of the Dirac Hamiltonian. Their Fock states have dynamically created $e^+e^-$ pairs, whose distribution is determined by the Dirac wave function. The linear potential of $D=1+1$ dimensions confines electrons but repels positrons. As a result, the mass spectrum is continuous and the wave functions have features of both bound states and plane waves.

The classical solutions of Gauss' law are explored for hadrons in QCD. A non-vanishing boundary condition at spatial infinity generates a constant \order{\alpha_s^0} color electric field between quarks of specific colors. Poincar\'e invariance limits the spectrum to color singlet $q\bar q$ and $qqq$ states, which do not generate an external color field. This restricts the \order{\alpha_s^0} interactions between hadrons to string breaking dynamics as in dual diagrams. Light mesons lie on linear Regge and parallel daughter trajectories. There are massless states which may be significant for chiral symmetry breaking. Since the bound states are defined at equal time in all frames they have a non-trivial Lorentz covariance.

\end{abstract}


\maketitle

\vspace{-.5cm}

\tableofcontents

\parindent 0cm
\vspace{-.2cm}

\section{Introduction and Summary \label{intro}}

Perturbation theory is a central tool in studies of the Standard Model of particle physics. Basic principles, \eg, analyticity and the factorization of hard QCD subprocesses, are supported by generic properties of Feynman diagrams. Scattering amplitudes are generally well approximated by contributions of low order in the coupling. 

Bound states are different: No Feynman diagram has a bound state pole. Bound states are generated by the {\em divergence} of the perturbative expansion.

\subsubsection{Basics of bound state perturbation theory \label{basics}}

Introductory Quantum Mechanics describes the Hydrogen atom by postulating the Schr\"odinger equation with the classical potential $V(r)$,
\beq\label{classpot}
V(r)=-\alpha/r \hspace{2cm} \Phi_0(r) = N\,\exp(-\alpha\, m\, r)
\eeq
The wave function $\Phi_0(r)$ is exponential in $\alpha$, whereas Feynman diagrams are of fixed order in $\alpha$. It is important to understand how the QM description of bound states emerges from the underlying Quantum Field Theory (QED) \cite{Lepage:1978hz}. 

Why does the perturbative expansion diverge for bound states -- no matter how small is $\alpha$? The typical momentum transfer between the atomic constituents is of the order of the inverse atomic radius, $p \sim \alpha m$. This Bohr momentum scale brings powers of $\alpha$ into the denominators of electron and photon propagators. They can balance the powers of $\alpha$ from the vertices.

\begin{figure}[h]
\includegraphics[width=1.\columnwidth]{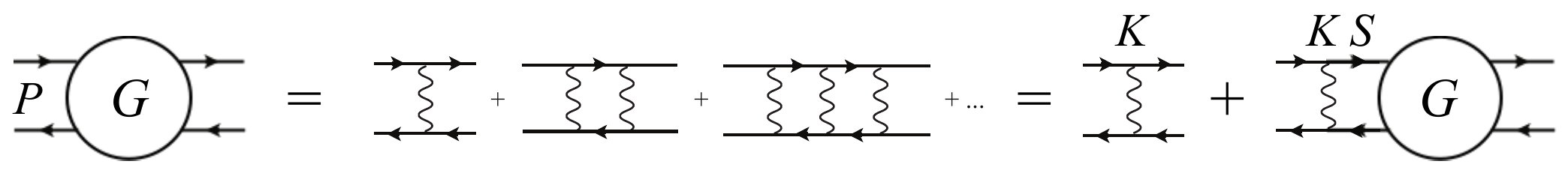}
\caption{For momentum transfers $p=\morder{\alpha m}$ the leading contribution to the $e^+e^- \to e^+e^-$ scattering amplitude (left) comes from ladder diagrams (middle). Their sum can be expressed as a convolution between single photon exchange and the scattering amplitude itself (right). This gives the ``Dyson-Schwinger equation'' \eq{DSeq} for the $e^+e^- \to e^+e^-$ amplitude.
{\label{sladder4}}}
\end{figure}

It is straightforward to verify\footnote{See, for example, section II A of my previous lecture notes \cite{Hoyer:2014gna}. Those notes give more details also on other issues mentioned here.} that in the $e^+e^- \to e^+e^-$ amplitude the powers of $\alpha$ balance for the ``ladder'' diagrams in \fig{sladder4} (and only for them). As indicated, the sum can be expressed as a convolution between single photon exchange and the amplitude itself. This is called a ``Dyson-Schwinger equation''.

\begin{figure}[h]
\includegraphics[width=1.\columnwidth]{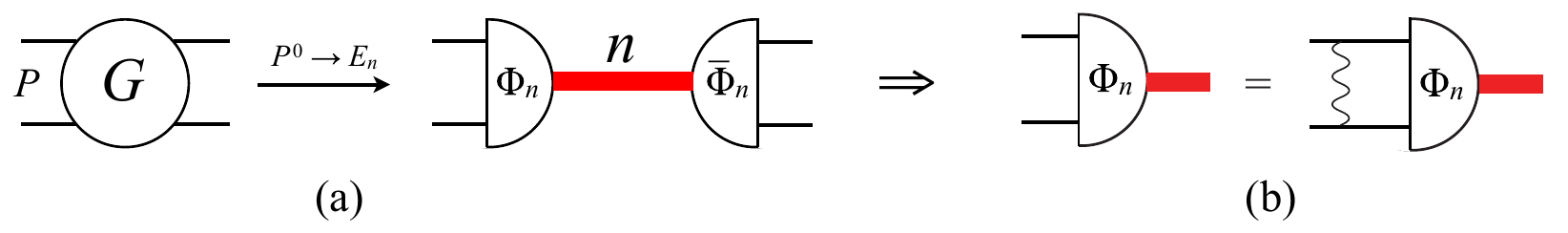}
\caption{(a) The residue of a bound state pole in a scattering amplitude $G$ factorizes into a product of wave functions for the initial and final state. (b) The Dyson-Schwinger relation in \fig{sladder4} implies a ``Bethe-Salpeter equation'' for the wave function.
{\label{BS-rel}}}
\end{figure}

The loop integrals in the ladder diagrams range over all momenta $p$. Only loop momenta $p \sim \alpha m$ contribute at leading order. The intermediate $e^+e^-$ propagators may then be treated as being on-shell. This reduces the ladder sum to a geometric sum of single photon exchange amplitudes, analogous to the familiar
\beq\label{geomsum}
G(x) = x + x^2 + x^3 + \ldots = x+ xG(x) \hspace{1cm} \Longrightarrow \hspace{1cm} G(x) = \frac{x}{1-x}
\eeq
The pole of $G(x)$ at $x=1$ is analogous to a bound state pole. More precisely, let $G(P)$ be the Green function of total momentum $P$ with external propagators removed, $K$ the similarly truncated single photon exchange amplitude and $S$ the two-particle ($e^+e^-$) propagator. Then the Dyson-Schwinger equation of \fig{sladder4} is
\beq\label{DSeq}
G(P) = K+KS\,K+KS\,KSK+\ldots = K+K\,S\,G(P)
\eeq
Each product denotes a convolution integral over the relative momentum between the $e^+$ and the $e^-$. Let $G(P)$ have a bound state pole at $P^0=E_n$, where $E_n = \sqrt{M_n^2+\Pv^2}$ is the energy of the bound state. On general grounds the residue of the pole is the product of the bound state wave functions $\Phi_n$ for the initial and final states. Because neither $K$ nor $S$ have bound state poles the Dyson-Schwinger relation \eq{DSeq} implies a ``Bethe-Salpeter'' equation for the wave function shown in \fig{BS-rel}(b),
\beq\label{BSeq}
G(P)=\frac{\Phi_n\Phi_n^\dag}{P^0-E_n}+\ldots \hspace{1cm}  {\color{red}  \Longrightarrow} \hspace{1cm}  \Phi_n = K\,S\,\Phi_n
\eeq

The Bethe-Salpeter equation is based on Feynman diagrams and thus explicitly Lorentz covariant. In the rest frame it reduces to the Schr\"odinger equation in the limit $\alpha\to 0$. Loop corrections to the single photon exchange kernel $K$ and to the two-particle propagator $S$ can be added perturbatively. In this way the Bethe-Salpeter equation provides an in principle exact, Lorentz-covariant framework for field theory calculations of bound states. Many topical studies of QCD and hadron physics are based on the Dyson-Schwinger equations \cite{Roberts:2015lja}. 

Explicit Lorentz covariance is very helpful in evaluating scattering amplitudes, but has turned out to complicate bound state calculations. In the bound state rest frame the momentum transfers are $|\pv| \sim \alpha\,m$ and $p^0 \sim \alpha^2 m$. The photon propagator $\sim 1/p^2 \simeq -1/\pv^2$ is thus independent of $p^0$ at lowest order: The interaction is instantaneous in time. This simplification is specific to the rest frame, in a general frame $p^0 \sim |\pv| \propto \alpha\,m$. In a Lorentz covariant formulation we must therefore keep the $p^0$ dependence of the propagator. This makes the equation hard to solve -- in fact no analytic solution of the Bethe-Salpeter equation is known, even for a single photon exchange kernel. The calculation of higher order corrections is progressively more difficult \cite{Karmanov:2013rga}.

For reasons such as this the Bethe-Salpeter framework is impractical in precision calculations of atomic structure. The preferred method is Non-Relativistic QED (NRQED), which is an effective field theory based on expanding the QED action in powers of $|\pv|/m$ \cite{Kinoshita}. At lowest order this gives the Schr\"odinger equation of Introductory QM. NRQED is applicable only in the rest frame of the bound state, but this is sufficient to determine energy levels. Positronium hyperfine structure at \order{\alpha^7\log\alpha} provides one of the most stringent tests of the Standard Model \cite{Baker:2014sua}.

\subsubsection{Bound state Born term \label{born}}

We have seen that already the first approximation of a bound state wave function \eq{classpot} requires all powers of $\alpha$. The higher order corrections are also expressed as powers of $\alpha$. This makes the choice of the first approximation ambiguous. We can apparently rearrange the infinite series by moving some of the ``higher order corrections'' into the ``lowest order'', or {\it vice versa}. We already saw a hint of this in the handling of the $p^0$ dependence of the single photon exchange kernel.

It turns out that the Dyson-Schwinger equation can be formulated in many equivalent ways \cite{Lepage:1978hz}. In its exact version the equation contains power series in $\alpha$ for both the kernel $K$ and the two-particle propagator $S$. Either of the expansions can be freely chosen, then the other one is fixed by the known expansion of the Green function $G$.

In considering this ambiguity it is helpful to recall a more familiar situation where the perturbative expansion diverges even more dramatically than for Positronium: Classical E\,\&\.M. Phenomena well described by classical electromagnetic fields involve very low momentum transfers, $|\pv| \ll \alpha\,m$. In this regime the relevant degrees of freedom are not individual photons but their collective fields, which obey Maxwell's equations. The classical dynamics is unambiguous.

Classical field theory emerges in the $\hbar \to 0$ limit of the quantum theory. The Green function of a scalar field is in the functional integral formulation given by
\beq\label{green}
G(x_1,\ldots,x_n) = \int [d\vphi]\,\exp\big(iS[\vphi]/\hbar\big)\,\vphi(x_1)\ldots\vphi(x_n)
\eeq
The main contribution to the functional integral in the $\hbar \to 0$ limit is from classical field configurations for which the action is stationary, $\delta S[\vphi]/\delta\vphi = 0$. Fluctuations of the field $\vphi$ around the classical configurations are suppressed by powers of $\hbar$. They correspond to loop corrections in the perturbative expansion.

The $\hbar$ expansion seems equivalent to the expansion in $\alpha$: Each loop correction brings one power of $\hbar$ and one of $\alpha$ \cite{Brodsky:2010zk}. However, the $\hbar$ and $\alpha$ powers do not match in the lowest order (Born) term. Tree-level Feynman diagrams have no loops, whereas their power of $\alpha$ depends on the number of external legs. A bound state is built from repeated scattering and thus has an unlimited number of external legs. The geometric sum of single photon exchange amplitudes in \fig{sladder4} is equivalent to a sum of \order{\hbar^0} tree diagrams, it has no loop contributions but all powers of $\alpha$. A loop correction to the ladder diagrams does bring a factor $\hbar\alpha$.

The functional integral \eq{green} provides a physically motivated definition of ``lowest order'' bound states: They are of lowest order in $\hbar$, not in $\alpha$. Born-level bound states are characterized by interactions mediated by a classical gauge field. This is regarded as evident in Introductory QM, where the classical potential \eq{classpot} of Hydrogen is adopted. 

Perturbation theory expands around free fields. The infinite sum of ladder diagrams builds the classical  potential. Consider the expression for the $S$-matrix,
\beq\label{smatrix}
S_{fi}={}_{out}\bra{f,\,t\to \infty}\left\{ {\rm T}\exp\Big[-i\int_{-\infty}^\infty dt\,H_I(t)\Big]\right\}\ket{i,\,t\to -\infty}_{in}
\eeq
The interaction Hamiltonian is proportional to the coupling constant, in QED 
\beq\label{Hint}
H_I = e\int d\xv\, \bar\psi\,e\Asl\,\psi
\eeq
Feynman diagrams of \order{e^n} are given by an $n$th order expansion of the time-ordered exponential. The $\ket{\ \ }_{in}$ and ${}_{out}\bra{\ \ }$ states are \order{e^0} {\it free} states at asymptotic times. Thus the charged particles in $\ket{e^+e^-}_{in} = b^\dag\,d^\dag\ket{0}$ have no associated photon field. This violates the classical field equations. If the electron is at $\xv = \xv_1$ and the positron at $\xv_2$ Gauss' law for $A^0$ requires
\beq\label{gausslaw0}
-\nv^2 A^0(\xv) = e\big[\delta(\xv-\xv_1)-\delta(\xv-\xv_2)\big]
\eeq
Nevertheless, the expression \eq{smatrix} for the $S$-matrix is formally exact since the free $\ket{e^+e^-}_{in}$ state at $t \to -\infty$ has an unlimited time to relax to the physical $\ket{e^+e^-}$ state, via repeated interactions specified by $H_I$. Building the classical gauge field requires an infinite number of interactions, hence the need for the infinite sum of ladder diagrams in \fig{sladder4}. 

The classical gauge field provides the binding potential in bound states. In ``hard'' QED interactions, $|\pv| \gg \alpha m$, the classical field is of secondary importance. Nevertheless, its absence in the $in$ and $out$ states manifests itself in the appearance of infrared singularities in the perturbative expansion. The soft photons decouple from neutral atoms at momenta corresponding to the atomic radius, $|\pv| \sim \alpha m$. This gives rise to $\log\alpha$ contributions in bound state perturbation theory.

In conclusion, describing Positronium in terms of Feynman diagrams requires a divergent sum to generate the classical $-\alpha/r$ potential. The $\hbar$ expansion provides a physically motivated first approximation. The
gauge field operator in the Hamiltonian takes the value of the classical field,
given by the equations of motion \eq{gausslaw0}. Loop corrections to the Born level bound states may be evaluated by using them as $in$ and $out$ states in the expression \eq{smatrix} of the $S$-matrix.

\subsubsection{Positronium bound by its classical gauge field \label{classpos}}

In section \ref{posisec} I determine the Positronium state and binding energy at Born level according to the above principles. The state \eq{bstate} is expressed in terms of the electron and positron field operators taken at equal time and with a spatial distribution described by a c-numbered ($4\times 4$) wave function $\Phi$. The requirement \eq{opseq} that this state be stationary in time, \ie, be an eigenstate of the QED Hamiltonian, determines the bound state equation for $\Phi$. Each Fock state $\ket{e^-(\xv_1)e^+(\xv_2)}$ generates a specific classical field. The Positronium state reduces to \eq{nrwf2} in the limit of weak binding, $\alpha \to 0$.

The use of the classical gauge field $A^\mu$ in the interaction Hamiltonian \eq{Hint} merits some remarks. In the Positronium rest frame ($\Pv$=0) the $A^0$ field dominates and is given by Gauss' law \eq{gausslaw0},
\beq\label{qedfield}
A^0(\xv)= \frac{e}{4\pi}\left(\inv{|\xv-\xv_1|}-\inv{|\xv-\xv_2|}\right)
\eeq
Note that:
\begin{enumerate}

\item The $e^\mp$ positions $\xv_1,\xv_2$ determine $A^0(\xv)$ for all $\xv$ at each instant of time.

\item Each position $\xv_1,\xv_2$ is associated with a distinct $A^0(\xv)$ field.

\item The potential energy of the electron is 
\beq\label{eleE}
eA^0(\xv_1) = -\frac{\alpha}{|\xv_1-\xv_2|}
\eeq 
plus an infinite ``self-energy'' $\alpha/0$. The latter is independent of $\xv_1,\xv_2$ and can be subtracted. 

\item The potential energy of the positron equals that of the electron,
\beq\label{posE}
-eA^0(\xv_2) = -\frac{\alpha}{|\xv_1-\xv_2|}
\eeq 

\item The field energy calculated from \eq{qedfield} is
\beq\label{fieldE}
\inv{4} \int d^3\xv F_{\mu\nu}(\xv)F^{\mu\nu}(\xv) = \frac{\alpha}{|\xv_1-\xv_2|}
\eeq
The sum of \eq{eleE}\,--\,\eq{fieldE} gives the Positronium potential $V(r)$ in \eq{classpot}.

\end{enumerate}

Each order of the $\hbar$ expansion is Poincar\'e invariant. In section \ref{scheq} I determine how the Positronium wave function and energy depend on its momentum $\Pv$ by Lorentz transforming the classical gauge field as in \eq{Aboost}. The outcome confirms expectations: $E=\sqrt{\Pv^2+M^2}$ and the wave function Lorentz contracts as in classical relativity. The same result was found using the Bethe-Salpeter equation \cite{Jarvinen:2004pi}.

\subsubsection{Dirac states \label{dirac}}

The Born level Positronium state \eq{nrwf2} consists of a single $e^+e^-$ pair and its associated classical gauge field. The situation is different for a relativistic electron bound in an external field, described by the Dirac equation. Relativistic dynamics involves virtual pairs created by the field (\cf\ \fig{diracdiag}b), as first demonstrated by the Klein Paradox \cite{Klein:1929zz}. This obscures the relation between the ostensibly single-particle Dirac wave function and the multiparticle state. I consider the structure of Dirac states in section \ref{diracsec}.

A Dirac state \eq{statedef} can be expressed in terms of the electron field $\psi(t,\xv)$ operating on a non-trivial ground state $\ket{\Omega}$. The Dirac wave function specifies the spatial distribution of the $e^\pm$ creation and annihilation operators. The ground state \eq{vac2} is a superposition of $e^+e^-$ pairs, determined by the complete set of Dirac wave functions.

In section \ref{diracsubsec2} I apply the general expressions to the case of Dirac states in $D=1+1$ dimensions, where the classical external field $eA^0(x) = V'|x|$ is linear and $A^1=0$. The analytic solution \eq{fgsol} has features of both bound and scattering states: The energy spectrum is continuous and the norm of the wave function approaches a constant value as $|x| \to \infty$. These features are generic for power potentials $V(r) \propto r^{\pm n}$ in all dimensions -- with the exception of the $1/r$ potential in $D=3+1$ \cite{plesset}. 

The linear potential $eA^0(x) = V'|x|$ confines electrons but repels positrons. Thus the electron and positron in an $e^+e^-$ pair are distributed asymmetrically, with the $e^-$ at small and the $e^+$ at large $|x|$. In particular, the constant norm of the wave function as $|x| \to \infty$ is due only to the positrons. The kinetic energy of a positron grows with $|x|$, due to  its increasingly negative potential. 

For a sizeable electron mass $m$ (in units of $\sqrt{V'}$) the Dirac spectrum is similar to the quantized Schr\"odinger spectrum for nearly all values of the parameter $\beta$ in the solution \eq{fgsol}. The continuous part of the Dirac spectrum is limited to states with wave functions that are suppressed at small $|x|$. The full set of Dirac states is complete and orthonormal, with $\delta$-function normalization as for plane waves.

\subsubsection{Application to hadrons in QCD \label{hadrons}}

Hadrons are very different from atoms -- which makes their similarities surprising. The fact that most hadrons can be classified as $q\bar q$ and $qqq$ states is indicative of a perturbative coupling. The likeness of the quarkonium and atomic spectra makes this rather compelling. I review these and other phenomenological aspects of hadron physics in section \ref{secmotiv}. 

In the perturbative approach to QED bound states described here the emphasis is on a proper definition of the lowest order contribution. For bound states this must be understood as lowest order in $\hbar$, not in $\alpha$. At Born level the binding is due to a classical gauge field. In the Positronium rest frame the instantaneity of $A^0$ avoids issues of retardation. The state can be boosted by Lorentz transforming the classical gauge field. 

This approach may apply also to hadrons in QCD. The classical gauge field then needs to cause confinement and chiral symmetry breaking. The solution of Gauss' law is unique up to homogeneous solutions, specified by the boundary condition at spatial infinity. A non-vanishing gauge field for $\xv \to \infty$ generally breaks Poincar\'e invariance. In section \ref{confsubsubsec} I list the arguments which point to a single acceptable solution. It allows bound states only in the form of color singlet mesons \eq{mfield} and baryons \eq{bfield}. It is characterized by one dimensionful parameter $\la$ related to the strength of the classical field. The homogeneous solution is of \order{\alpha_s^0}, \ie, independent of the QCD coupling.

In analogy to the QED case \eq{qedfield} there is a specific color field for each quark configuration of the hadron. Since the overall state is a singlet under (global) gauge transformations the gluon field vanishes when summed over the quark colors in mesons \eq{mfieldsum} as well as in baryons \eq{bfieldsum3}. Hence  the classical field is invisible to external observers (such as other hadrons). The only \order{\alpha_s^0} interactions between hadrons occur through quark overlap: the creation/annihilation of $q\bar q$ pairs. The dynamics resembles that described by dual diagrams \cite{Zweig:2015gpa,Harari:1981nn}, such as \fig{dualdiag}.

The classical field results in a linear potential \eq{mpot} for mesons. The baryon potential \eq{bpot} agrees with the meson one when two of the quarks are at the same position. The two color components $a=3,8$ confine along $\xv_1-\xv_2$ and $\xv_1+\xv_2-2\xv_3$ respectively, as seen from \eq{fenb2}. There is no analogous solution for states with more than three quarks.

The classical field is non-perturbative in the sense that a comparable solution cannot be found for single quarks or gluons -- only color singlet states are physical, in analogy to QED$_2$ \cite{Coleman:1976uz}. Perturbative corrections are expected to be calculable using the standard $S$-matrix expression \eq{smatrix}, with the \order{\alpha_s^0} mesons and baryons as $in$ and $out$ states. The strong coupling $\as(Q^2)$ will freeze as $Q^2 \to 0$ if the $\hbar$ expansion is relevant (loop corrections are suppressed). The indications that $\as$ does freeze are briefly discussed in section \ref{asfreeze}.

Since $\hbar$ is a fundamental parameter the full symmetries of the exact result, as well as unitarity, must be realized at each order of the expansion. Poincar\'e invariance is addressed in sections \ref{symsubsubsec} and \ref{framesec}, and a non-trivial example of gauge invariance given in section \ref{bsboost11subsec}. Unitarity is an infinite set of non-linear relations between bound state scattering amplitudes that can only be satisfied order by order in an expansion. The solution appears to be analogous to the familiar one based on Feynman diagrams. The tree approximation (discussed further below) implies calculable hadron loop corrections, corresponding to squaring the dual diagram in \fig{dualdiag}. The hadron loop expansion is equivalent to an expansion in $1/N_c$. Unitarity may be satisfied at each order of $1/N_c$.

\subsubsection{Hadrons at tree level \label{tree}}

Two distinct levels of approximation for hadrons emerge. The first one ignores perturbative effects of \order{\alpha_s}. In the \order{\alpha_s^0} sector there is what might be called a tree level, where string breaking caused by contributions such as \fig{dualdiag} (and the associated hadron loops) are neglected. Tree level results are straightforward to calculate. The first task is then to study whether they provide a reasonable first approximation to hadron physics, motivating further studies.

In section \ref{mes11subsec} I discuss the spectrum and wave functions of mesons in $D=1+1$ dimensions. The solutions are similar to those of the Dirac equation in section \ref{diracsubsec2}, but there is an important difference. Some components of the $2\times 2$ meson wave functions are generally singular at $M=V(x)$, where $M$ is the bound state mass and $V(x)$ the linear potential (see \eq{phi23}). Taking the quark mass $m \to \infty$ shows that requiring the relativistic wave function to be regular at $M=V$ is analogous to requiring normalizability of the non-relativistic Schr\"odinger wave function. At finite $m$ the wave functions are regular only for specific values of the bound state mass $M$ (section \ref{discretesubsubsec}). The meson spectrum is thus discrete, in contrast to the continuous Dirac spectrum.

The completeness and orthonormality of the Dirac states relies on a $\delta$-function normalization, as appropriate for a continuous spectrum. The discrete meson states are not complete and orthogonal, since two meson states have a non-vanishing overlap with a single one through string-breaking as in \fig{dualdiag}. Unitarity and completeness can emerge only when hadron loop effects are included.

The tree level meson states can approximate physical phenomena via duality between bound and scattering states. As previously mentioned, the Dirac wave functions have aspects of both bound and plane wave states: The linear potential confines electrons but repels positrons. In section \ref{dualitysubsubsec} I discuss how meson states reduce to plane wave quark states in certain limits, as in the parton model. Specifically, the wave functions accomodate the duality illustrated in \fig{dual}.

Meson states in $D=3+1$ dimensions can be classified according to their spin, parity and charge conjugation (section \ref{mes31subsec}, \eq{31traj}). For vanishing quark mass $(m=0)$ the spectrum resembles that of dual models \cite{Veneziano:1968yb,Lovelace:1969se}, with straight Regge and daughter trajectories.

Aspects of chiral symmetry are discussed in section \ref{cs31subsec}. For $m=0$ the action has exact chiral invariance. Since the ground state is chirally symmetric each meson should have a partner with the same mass but opposite parity. This is also seen in the solutions. However, in some cases the condition that the wave function be regular is incompatible with chiral invariance. Then there is no parity degeneracy.

There are states with zero mass $M$ that have a regular wave function at $M-V(r)= V'r=0$. String breaking effects may be particularly important for them, since the whole wave function is in the classically forbidden domain $V(r) > M$. Nevertheless, these solutions may be relevant for chiral symmetry breaking. In particular, the $M=0,\ J^{PC}= 0^{++}$ meson can mix with the chirally symmetric ground state without breaking Poincar\'e invariance. Then also the massless $J^{PC}= 0^{-+}$ pion couples to the axial current. The existence of massless bound states is intriguing, and their relevance for chiral symmetry breaking in QCD merits further study.

\section{Positronium at Born level \label{posisec}}

\setcounter{subsubsection}{0}

\subsubsection{State definition \label{posistate}}

The first step in our quest is the derivation of a Positronium state $n$ at Born level directly from the QED action. Bound states are stationary in time and thus by definition eigenstates of the Hamiltonian,
\beq\label{opseq}
H\ket{n,\Pv}=E\ket{n,\Pv}
\eeq
The operator fields of the QED Hamiltonian $H$ are taken at the same instant of time $t$ as the Positronium state $\ket{n,\Pv}$.
The Poincar\'e invariance of the QED action ensures that the energy eigenvalue has the proper dependence on the momentum $\Pv$ of the state,
\beq\label{epdep}
E=\sqrt{M^2+\Pv^2}
\eeq
This is not explicit in \eq{opseq} since the concept of equal time is frame dependent. The Hamiltonian transforms under boosts, and so do the states. Since we may consider $\hbar$ to be a free parameter the correct energy dependence must hold at each order of $\hbar$, and so also for the Born contribution. Eq. \eq{epdep} provides a non-trivial check that the Born state is correctly evaluated.

Due to the non-relativistic internal dynamics Fock states with more than one $e^+e^-$ pair do not contribute, so the state should have the structure
\beq\label{bstate}
\ket{n,\Pv}=\int d\xv_1 d\xv_2 \, e^{i\Pv\cdot(\xv_1 +\xv_2)/2}\bar\psi_\alpha(\xv_1)\Phi_{\alpha\beta}(\xv_1 -\xv_2)\psi_\beta(\xv_2)\ket{0}
\eeq
Remarks:
\begin{itemize}
\item Each state $\ket{e^-(\xv_1)e^+(\xv_2)}$ is associated with a classical gauge field $A^\mu(\xv;\xv_1,\xv_2)$. The gauge field operator in the Hamiltonian takes the value of the classical field when imposing \eq{opseq} at Born level (section \ref{classpos}).

\item The operator valued electron field satisfies the canonical anticommutation relation
\beq\label{canon}
\acom{\psi^\dag_\alpha(t,\xv)}{\psi_\beta(t,\yv)}=\delta_{\alpha,\beta}\,\delta^3(\xv-\yv)
\eeq
and may be expanded in the free state basis,
\beqa\label{eop}
\psi_\alpha(t=0,\xv)&=&\int \frac{d\kv}{(2\pi)^32E_k}\sum_\lambda\Big[u_\alpha(\kv,\lambda)e^{i\kv\cdot\xv}b_{\kv,\lambda}+v_\alpha(\kv,\lambda)e^{-i\kv\cdot\xv}d^\dag_{\kv,\lambda}\Big]\\[2mm]
\acom{b_{\pv,\lambda}}{b^\dag_{\qv,\lambda'}} &=& \acom{d_{\pv,\lambda}}{d^\dag_{\qv,\lambda'}} = 2E_p\,(2\pi)^3 \delta^3(\pv-\qv) \delta_{\lambda,\lambda'} \label{acom2}
\eeqa

\item The time $t$ is implicit in \eq{bstate} and may be take to be $t=0$.
Time evolution gives the state a phase $\exp(-iEt)$ and shifts the electron fields to $t$.

\item The $c$-numbered wave function $\Phi_{\alpha\beta}(\xv_1 -\xv_2)$ depends implicitly on $\Pv$ and on the quantum numbers $n$ of the Positronium state. Its $4\times 4$ Dirac matrix structure simplifies for non-relativistic internal dynamics. The momentum space ansatz
\beq\label{nrwf}
\Phi_{\alpha\beta}(\xv) = \int \frac{d\kv}{(2\pi)^3}\,e^{i\kv\cdot\xv}\,\prescript{}{\alpha}{\Big{[}}\gamma^0 u(\halft\Pv+\kv,\lambda)\Big]\, \Big[\bar v(\halft \Pv-\kv,\lambda')\gamma^0\Big]_\beta\, \phi_P(\kv)
\eeq
where the scalar wave function $\phi_P(\kv)$ is helicity independent, gives in \eq{bstate}
\beq\label{nrwf2}
\ket{n,\Pv}=\int \frac{d\kv}{(2\pi)^3}\, \phi_P(\kv)\,b^\dag(\halft\Pv+\kv,\lambda)\,d^\dag(\halft\Pv-\kv,\lambda')\ket{0}
\eeq
We shall see that this state solves \eq{opseq} for Positronium.

\item In a space translation $\xv_i \to \xv_i+\bs{a}\ (i=1,2)$ the integrand of \eq{bstate} picks up the phase $\exp(i\Pv\cdot\bs{a})$, implying that the state has overall momentum $\Pv$. More directly, \eq{bstate} is an eigenstate of the momentum operator (see section \ref{symsubsubsec}),
\beq\label{momop}
\bs{\mP}= -i\int d\xv\, \psi^\dag(\xv)\nv\psi(\xv)
\eeq
with eigenvalue $\Pv$. Similarly, the symmetry of $H$ under rotations ensures that the wave function $\Phi(\xv_1 -\xv_2)$ transforms according to the angular momentum representation of $n$ in the rest frame, $\Pv=0$.

\end{itemize}

\subsubsection{Classical Gauge Field \label{classA}}

Maxwell's equations specify the classical electromagnetic field in the presence of charges,
\beq\label{maxeq}
\partial_\mu F^{\mu\nu}(x) = j^\nu(x)
\eeq
In the rest frame of Positronium ($\Pv=0$) its electron and positron constituents move non-relativistically, $v \simeq \alpha \ll 1$. Hence the vector field $\Av$ is of \order{\alpha} compared to the Coulomb field $A^0$ (up to gauge transformations). Once we know the field in the rest frame we can boost it to a frame where $\Pv\neq 0$.

For $\Av=0$ \eq{maxeq} gives Gauss' law \eq{gausslaw0}, which determines the expression \eq{qedfield} for $A^0$.
I recall some features of this solution, which were already mentioned in section \ref{classpos} and are due to the instantaneous nature of $A^0$:
\begin{itemize}
\item[({\it i})] Each component $\ket{e^-(\xv_1,\lambda)e^+(\xv_2,\lambda')}$ of the Positronium state carries a specific classical field.
\item[({\it ii})] The value of $A^0(\xv)$ is determined instantaneously for all $\xv$.
\end{itemize}

The gauge field may be boosted in the usual way. For a boost $\xi$ in the $z$-direction 
\beqa
P &=& (M\cosh\xi,0,0,M\sinh\xi) \\[2mm] A^0(\xv) &=& \cosh\xi\, A^0_R(\xv_R;\xv_{1R},\xv_{2R}) \label{Pframe}\nn\\ 
A^1(\xv) &=& A^2(\xv) =0 \label{Aboost}\\
A^3(\xv) &=& \sinh\xi\, A^0_R(\xv_R;\xv_{1R},\xv_{2R}) \nn
\eeqa
where the rest frame quantities are marked by $R$. Since the distances are measured at equal time they are Lorentz contracted in the direction of the boost. Conversely, the rest frame coordinates in terms of the boosted ones are
\beq\label{restcoord}
\xv_R = (x,y,z)_R = (x,y,z\,\cosh\xi) \equiv (\xv_\perp,x_\parallel\,\cosh \xi)
\eeq
The boosted field, expressed in the coordinates of the moving frame, is determined by $A^0$ \eq{qedfield} in the rest frame,
\beq\label{boostpot}
\left(\begin{array}{c} A^0(\xv) \\ A^3(\xv) \end{array} \right)=
\left(\begin{array}{c} \cosh\xi \\ \sinh\xi \end{array} \right)
\times \frac{e}{4\pi}\left[\inv{\sqrt{(\xv-\xv_1)_\perp^2+(x-x_1)_\parallel^2 \cosh^2\xi}}-\inv{\sqrt{(\xv-\xv_2)_\perp^2+(x-x_2)_\parallel^2 \cosh^2\xi}}\right]
\eeq
In terms of the rest frame coordinates \eq{restcoord} this may be expressed as
\beq\label{boostpot2}
\slashed{A}=\frac{\slashed{P}}{M}\,\frac{e}{4\pi}\left[\inv{|\xv-\xv_1|_R}-\inv{|\xv-\xv_2|_R}\right]
\eeq

\subsubsection{The Hamiltonian \label{ham}}

It is convenient to divide the Hamiltonian into a kinetic and interaction part,
\beq
H=H_0+H_I
\eeq
In the absence of physical, transverse photons only the fermions contribute to the kinetic energy,
\beq\label{Hfree}
H_0 = \int d\xv\, \bar\psi(\xv)\big(i\slashed{\partial}-m\big)\psi(\xv) = \int \frac{d\pv}{(2\pi)^3\,2E_p}\,E_p \sum_\lambda\big[b_{\pv,\lambda}^\dag b_{\pv,\lambda}+d_{\pv,\lambda}^\dag d_{\pv,\lambda}\big]
\eeq
The classical field contributes to the energy through the terms $\quart F_{\mu\nu}F^{\mu\nu}$ and $\bar\psi\,e\slashed{A}\psi$ of the Hamiltonian density (their signs are opposite compared to the Lagrangian). In the rest frame, using \eq{gausslaw0},
\beq\label{efield}
\quart \int d\xv\,F_{\mu\nu}F^{\mu\nu} =  -\halft \int d\xv\, \nv A^0\cdot \nv A^0 = \halft \int d\xv\, A^0\cdot \nv^2 A^0 = -\halft e\, \big[A^0(\xv_1)-A^0(\xv_1)\big] = \frac{e^2}{4\pi|\xv_1-\xv_2|}
\eeq
In the last equality I discarded the infinite ``self-energies'' $\propto \alpha/0$, which are independent of $\xv_1,\xv_2$ and thus irrelevant. The result has the same form as the Coulomb potential $-\alpha/r$ but is of opposite sign. It removes the double counting otherwise arising from the electron feeling the positron field and {\it vice versa}, \cf\ \eq{eleE}\,--\,\eq{fieldE}.

The contribution \eq{efield} is conveniently taken into account using the equation of motion for the gauge field operators,
\beq
\partial_\mu F^{\mu\nu}(x)=e\bar\psi(x)\gamma^\nu\psi(x)
\eeq 
In the absence of time dependent (propagating) gauge fields we can partially integrate in the expression for the field energy (now regarded as an operator expression),
\beq\label{efield2}
\quart \int d\xv\,F_{\mu\nu}F^{\mu\nu} = -\halft \int d\xv\, A_\nu\partial_\mu F^{\mu\nu} = -\halft\int d\xv\, \bar\psi\,e\slashed{A}\psi
\eeq
This removes half of the fermion interaction term, so that the total interaction Hamiltonian becomes
\beq\label{Hint1}
H_I = \halft\int d\xv\, \bar\psi\,e\slashed{A}\psi
\eeq

\subsubsection{Schr\"odinger Equation for Positronium in motion \label{scheq}}

I determine the bound state equation by imposing \eq{opseq} in the general frame \eq{Pframe}. Thus we find the bound state equation for Positronium in motion. The present Hamiltonian formulation with a classical gauge field is equivalent to the approach in \cite{Jarvinen:2004pi}, which used a time-ordered Bethe-Salpeter equation. 

The kinetic energy is given by applying $H_0$ \eq{Hfree} to the Positronium state \eq{nrwf2},
\beq\label{H0act}
H_0\ket{n,\Pv} = \int \frac{d\kv}{(2\pi)^3}\, \phi_P(\kv)(E_+ +E_-)\,b^\dag(\halft\Pv+\kv,\lambda)\,d^\dag(\halft\Pv-\kv,\lambda')\ket{0} \hspace{1cm} E_\pm \equiv\sqrt{(\halft\Pv\pm\kv)^2+m^2}
\eeq
The binding energy $E_b$ is defined as the difference between the bound state mass $M$ and twice the electron mass $m$,
\beq\label{ebind}
M=2m+E_b
\eeq
Expanding the energy eigenvalue $E$ in powers of the \order{\alpha^2} $E_b$ gives
\beq\label{Eexpand}
E=\sqrt{\Pv^2+(2m+E_b)^2} \simeq \mE + \frac{2m E_b}{\mE} \hspace{2cm} \mE \equiv \sqrt{\Pv^2+4m^2}
\eeq
Expressing $E(\Pv)$ in this form implies that the bound state equation should give a $\Pv$-independent result for $E_b$. Expanding the fermion energies $E_\pm$ in powers of the \order{\alpha} relative momentum $\kv$ gives at \order{\alpha^2},
\beq\label{Epmexpand}
E_\pm \simeq \halft\mE+ \frac{\pm \Pv\cdot\kv+\kv^2}{\mE}-\frac{(\Pv\cdot\kv)^2}{\mE^3} 
\eeq
I denote the component of $\kv$ along $\Pv$ by $k_\parallel$ and the orthogonal components by $\kv_\perp$. The difference between the total energy $E$ and the $e^\pm$ kinetic energies is of \order{\alpha^2} as it should be 
\beq\label{ediff}
E-E_+-E_- \simeq \inv{\cosh\xi}\Big[E_b-\inv{m}\Big(\kv_\perp^2+\frac{k_\parallel^2}{\cosh^2\xi}\Big)\Big] = \inv{\cosh\xi}\Big(E_b-\frac{\kv_R^2}{m}\Big)
\eeq
where the rest frame momentum is conjugate to $\xv_R$ in \eq{restcoord},
\beq\label{restmom}
\kv_R = \Big(\kv_\perp,\frac{k_\parallel}{\cosh\xi}\Big) \hspace{2cm} \kv_R\cdot \xv_R = \kv\cdot \xv
\eeq
I brought the boost $\xi$ into \eq{ediff} through the approximation
\beq\label{boostappr}
\cosh\xi = \frac{\sqrt{\Pv^2+(2m+E_b)^2}}{2m+E_b} \simeq \frac{\mE}{2m}
\eeq
which is allowed since \eq{ediff} is already of \order{\alpha^2}, the accuracy of this calculation. Thus
\beq\label{H0Ediff}
(H_0-E)\ket{n,\Pv} = \inv{\cosh\xi} \int \frac{d\kv}{(2\pi)^3}\, \Big[\inv{m}\Big(\kv_\perp^2+\frac{k_\parallel^2}{\cosh^2\xi}\Big)-E_b\Big]\phi_P(\kv)\,b^\dag(\halft\Pv+\kv,\lambda)\,d^\dag(\halft\Pv-\kv,\lambda')\ket{0}
\eeq

As $H_I$ \eq{Hint1} acts on the bound state \eq{bstate} we get contributions only for its commutation with the fermion operators since pair production effects may be neglected,
\beq
\com{H_I}{\bar\psi(\xv_1)} =\halft e\, \bar\psi(\xv_1)\Asl(\xv_1)\gamma^0 \hspace{2cm}
\com{H_I}{\psi(\xv_2)} =-\halft e\, \gamma^0 \Asl(\xv_2)\bar\psi(\xv_1)
\eeq
Up to the $\xv_1,\xv_2$-independent self-energies $\propto \alpha/0$ \eq{boostpot2} gives
\beq\label{potcont}
e\Asl(\xv_1)=-e\Asl(\xv_2)= \frac{\Psl}{M}\,\frac{-\alpha}{|\xv_1-\xv_2|_R} \equiv \frac{\Psl}{M}\,V(\xv_{1R}-\xv_{2R})
\eeq
Hence
\beq\label{Hintact}
H_I\ket{n,\Pv} =\inv{2M}\int d\xv_1 d\xv_2\,e^{i\Pv\cdot(\xv_1+\xv_2)/2}\,\bar\psi(\xv_1)\big[\Psl\gamma^0\Phi(\xv_1-\xv_2)+\Phi(\xv_1-\xv_2)\gamma^0\Psl\big] V(\xv_{1R}-\xv_{2R})\psi(\xv_2)\ket{0}
\eeq
Since this contribution is of \order{\alpha^2} we may neglect the \order{\alpha} $\kv$-dependence of the $u$- and $\bar v$-spinors in the expression \eq{nrwf} of the wave function,
\beq\label{phiapprox}
\Phi_{\alpha\beta}(\xv_1-\xv_2) = \prescript{}{\alpha}{\Big{[}}\gamma^0 u(\halft\Pv,\lambda)\Big]\, \Big[\bar v(\halft \Pv,\lambda')\gamma^0\Big]_\beta\, \phi_P(\xv_1-\xv_2)\big[1+\morder{\alpha}\big]
\eeq
where $\phi_P(\xv)$ is the Fourier transform of $\phi_P(\kv)$. The Dirac structure in \eq{Hintact} is then explicit,
\beqa
\Psl\gamma^0\Phi(\xv) &=& 2m\,u(\halft \Pv,\lambda)\,\bar v(\halft \Pv,\lambda')\,\phi_P(\xv)\nn\\[2mm]
\Phi(\xv)\gamma^0\Psl &=& -2m\,\gamma^0 u(\halft \Pv,\lambda)\,\bar v(\halft \Pv,\lambda')\,\phi_P(\xv)
\eeqa

To combine $H_I\ket{n,\Pv}$ with $(H_0-E)\ket{n,\Pv}$ I express \eq{H0Ediff} in coordinate space. Expanding the fermion fields as in \eq{eop} it is readily seen that
\beq\label{H0Ediff2}
(H_0-E)\ket{n,\Pv} = \inv{2}\int d\xv_1 d\xv_2\,e^{i\Pv\cdot(\xv_1+\xv_2)/2}\,\bar\psi(\xv_1)\Big[-\inv{m}\Big(\bs{\partial}_{1\perp}^2+\frac{\partial_{1\parallel}^2}{\cosh^2\xi}\Big)-E_b\Big]\big[\gamma^0\Phi(\xv_1-\xv_2)-\Phi(\xv_1-\xv_2)\gamma^0\big]\psi(\xv_2)\ket{0}
\eeq
is equivalent to \eq{H0Ediff} (the $\gamma^0$'s cause the factor $1/\cosh\xi$). Together with \eq{Hintact}, using also \eq{phiapprox}, the bound station condition becomes
\beqa\label{HEdiff}
(H_0-E+H_I)\ket{n,\Pv}&=& \inv{2}\int d\xv_1 d\xv_2\,e^{i\Pv\cdot(\xv_1+\xv_2)/2}\,\bar\psi(\xv_1)\Big[-\inv{m}\Big(\bs{\partial}_{1\perp}^2+\frac{\partial_{1\parallel}^2}{\cosh^2\xi}\Big)+V(\xv_{1R}-\xv_{2R})-E_b\Big]\phi_P(\xv_1-\xv_2) \nn\\[2mm]
&\times& \big[u(\halft\Pv,\lambda)\, \bar v(\halft \Pv,\lambda')\gamma^0 -\gamma^0u(\halft\Pv,\lambda)\, \bar v(\halft \Pv,\lambda')\big]\psi(\xv_2)\ket{0} =0
\eeqa
The bound state equation is thus
\beq
\Big(-\frac{\nv_R^2}{m}-\frac{\alpha}{|\xv_R|}\Big)\phi_P(\xv)=E_b\,\phi_P(\xv)
\eeq
which implies that the scalar wave function $\phi_P(\xv)$ defined by \eq{nrwf2} for a bound state with momentum $\Pv$ is given by the Lorentz contracted rest frame Schr\"odinger wave function $\phi$,
\beq
\phi_P(\xv)=\phi(\xv_R)
\eeq
The relation between $\xv$ and $\xv_R$ is given in \eq{restcoord}. The fact that $E_b$ is independent of $\Pv$ ensures that the bound state energy \eq{Eexpand} is Lorentz covariant. This is a dynamical (not explicit) feature of Hamiltonian formulations, and provides a non-trivial check that the evaluation is complete at leading order.

\section{Dirac states \label{diracsec}}

\subsection{General analysis \label{diracsubsec1}}


In the previous section I discussed the frame dependence of a non-relativistic bound state (Positronium) governed by the Schr\"odinger equation. Strongly coupled states cannot be addressed within perturbative QED. For the binding energy $-\quart m \alpha^2$ to be commensurate with the electron mass $m$ the coupling $\alpha$ must be of \order{1} -- but then a perturbative treatment is no longer meaningful. Our understanding of relativistic bound states is consequently model dependent.

The Dirac equation provides a familiar model for relativistic binding. It describes the bound states of an electron in a strong external gauge field $A^\mu(\xv)$, which I take to be independent of time. The essential simplifications compared to full QED are: 
\begin{itemize}
\item[({\it i})] $A^\mu(\xv)$ is a fixed classical field, unaffected by the electron.
\item[({\it ii})]  The electron interacts only with the external field. Loop corrections are neglected.
\end{itemize}
The $\xv$-dependence of $A^\mu$ means that space translation invariance is lost, so there is no conserved momentum nor boost covariance. Binding energies are well defined since a static field preserves time translation invariance. 

\subsubsection{Dirac equation and wave function \label{dwf}}

It is useful to consider separately the Dirac (4-spinor) wave functions $\Phi_n$ with positive eigenvalues $E_n > 0$, and wave functions $\bar\Phi_n$ with negative eigenvalues $-\bar E_n <0 $. The corresponding Dirac equations are
\beqa
\big(-i\nv\cdot\gv+m+e\slashed{A}\big)\Phi_n(\xv) &=& E_n\gz\Phi_n(\xv) \label{dir1} \\[2mm]
\big(-i\nv\cdot\gv+m+e\slashed{A}\big)\bar\Phi_n(\xv) &=& -\bar E_n\gz\bar\Phi_n(\xv) \label{dir2}
\eeqa

\begin{figure}[h]
\includegraphics[width=.6\columnwidth]{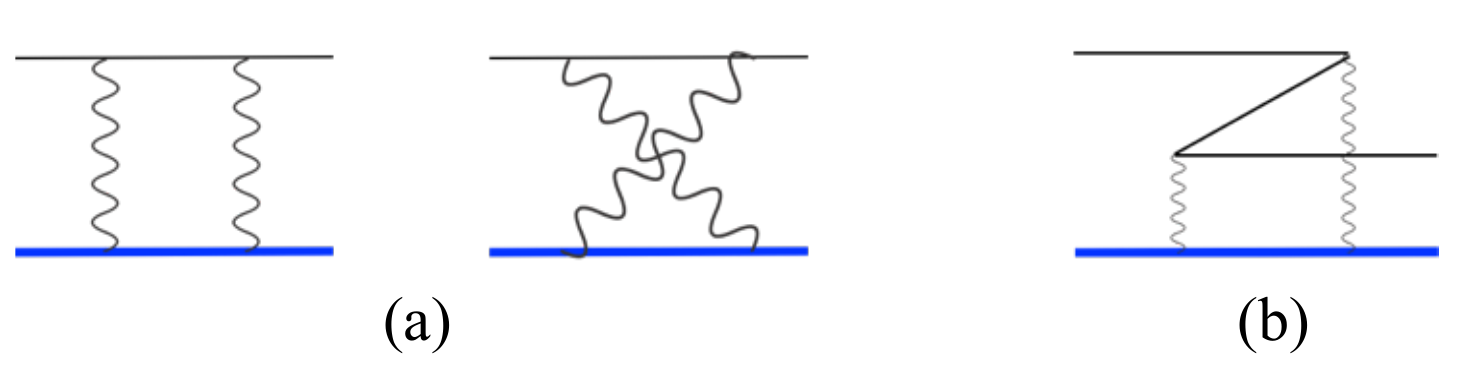}
\caption{(a) Feynman diagrams contributing to the Dirac states in the limit where the mass of the lower particle (thick blue line) tends to infinity. In the rest frame of the heavy particle the wavy lines represent instantaneous Coulomb photons. (b) When the crossed-photon diagram in (a) is time-ordered it reveals the creation and subsequent annihilation of an $e^+e^-$ pair.
{\label{diracdiag}}}
\end{figure}

The Positronium states in $e^+e^-$ scattering are at lowest order described by the sum of ladder diagrams in \fig{sladder4}. Dirac states with a Coulomb ($A^0$) potential can be analogously obtained from the scattering of an electron on a heavy particle at rest \cite{diracref}. As indicated in \fig{diracdiag}(a) also crossed photon diagrams must then be included. In the limit of infinite target mass only instantaneous Coulomb photons are exchanged. When time-ordered, the crossed photon diagram in (a) takes the appearance of \fig{diracdiag}(b): at an intermediate time the electron is accompanied by an $e^+e^-$ pair. For $N$ photon exchanges $N!$ diagrams contribute, with up to $N-1$ simultaneous $e^+e^-$ pairs.

Soon after Dirac proposed his equation it was realized that the wave function cannot be a single particle probability amplitude in the same sense as the Schr\"odinger wave function. In electron scattering on potentials $V \gsim m$ the Dirac wave function does not conserve probability density, a phenomenon that is known as ``Klein's paradox'' \cite{Klein:1929zz}. The paradox is related to the $e^+e^-$ pairs in the Dirac state. A proper description requires the methods of quantum field theory as explained, \eg, in \cite{Hansen:1980nc}.

Dirac bound states provide a model for the dynamics of relativistic binding which is relevant also for hadrons. Like hadrons, Dirac states have Fock components with any number of constituents, yet their quantum numbers reflect only the ``valence'' electron. How does the ostensibly single particle wave function $\Phi(\xv)$ in \eq{dir1} specify the higher Fock states? The Hamiltonian formalism I used for Positronium in Section \ref{posisec} is helpful in answering this question.

\subsubsection{Dirac states {\rm \cite{blaizot}} \label{dstatesec}}

The Dirac Hamiltonian has the same appearance as the fermion Hamiltonian of QED,
\beq\label{ham1}
H_D=\int d\xv\,\bar\psi(\xv)\big[-i\nv\cdot\gv+m+e\slashed{A}\big]\psi(\xv)
\eeq
However, now the field $A^\mu$ is a fixed classical background field. There are no physical (propagating) photons. Since the Hamiltonian is quadratic in the fermion operator fields it can be diagonalized \cite{Blaizot-Ripka}.

The positive \eq{dir1} and negative \eq{dir2} energy Dirac wave functions determine $e^-$ and $e^+$ bound states defined as
\beqa\label{statedef}
\ket{n} &=& \int d\xv\,\psi^\dag_\alpha(\xv)\Phi_{n\,\alpha}(\xv)\ket{\Omega}
\equiv c_n^\dag\ket{\Omega} \nn \\[2mm]
\ket{\bar n} &=& \int d\xv\,{\bar\Phi}_{n\,\alpha}^\dag(\xv)\psi_\alpha(\xv)\ket{\Omega} \equiv \bar c_n^\dag\ket{\Omega}
\eeqa
where $\psi(\xv)$ is the electron field operator \eq{eop} and there is a sum over the Dirac index $\alpha$. The vacuum state $\ket{\Omega}$ is an eigenstate of the Hamiltonian with eigenvalue taken to be zero,
\beq\label{vac1}
H_D\ket{\Omega} =0
\eeq
Two equivalent expressions for $\ket{\Omega}$ are given in Eqs. \eq{vac2} and \eq{vac4} below. Using
\beq\label{hdcom}
\com{H_D}{\psi^\dag(\xv)} = \psi^\dag(\xv)\gamma^0(i\lnab\cdot\gv+m+e\Asl) \hspace{2cm}
\com{H_D}{\psi(\xv)} = -\gamma^0(-i\rnab\cdot\gv+m+e\Asl)\psi(\xv)
\eeq
we see that the states \eq{statedef} are eigenstates of the Dirac Hamiltonian with {\it positive} eigenvalues,
\beqa\label{hdeigen}
H_D\ket{n} &=& E_n\ket{n} \hspace{1cm} E_n>0 \nn \\[2mm]
H_D\ket{\bar n} &=& \bar E_n\ket{\bar n} \hspace{1cm} \bar E_n>0
\eeqa

In terms of the wave functions in momentum space,
\beq\label{fourierdef}
\Phi_n(\xv) = \int \frac{d\pv}{(2\pi)^3}\, \Phi_n(\pv) e^{i\pv\cdot\xv}
 \hspace{2cm} \bar\Phi_n(\xv) = \int \frac{d\pv}{(2\pi)^3}\, \bar\Phi_n(\pv) e^{i\pv\cdot\xv}
\eeq
the eigenstate operators defined in \eq{statedef} can be expressed as
\beqa
c_n &=& \sum_\pv \Phi_n^\dag(\pv)\big[ u(\pv,\lambda)b_{\pv,\lambda}+ v(-\pv,\lambda)d_{-\pv,\lambda}^\dag\big] \equiv B_{np}b_p+D_{np}d^\dag_p  \label{cndef1}\\[-3mm]
 &&\hspace{10cm} \sum_\pv \equiv \int\frac{d\pv}{(2\pi)^3 2E_p}\sum_\lambda \nn \\[-3mm]
\bar c_n &=& \sum_\pv \big[b_{\pv,\lambda}^\dag u^\dag(\pv,\lambda)+ d_{-\pv,\lambda} v^\dag(-\pv,\lambda)\big]\bar\Phi_n(\pv)
\equiv \bar B_{np}b_p^\dag+\bar D_{np}d_p \label{cndef2}
\eeqa
In the second expressions on the rhs. a sum over the repeated index $p \equiv (\pv,\lambda)$ is implied. In the weak binding limit ($|\pv| \ll m$) the positive energy wave function $\Phi_n$ has only upper Dirac components, whereas $\bar\Phi_n$ has only lower Dirac components. Then $\ket{n}$ is a single electron state, whereas $\ket{\bar n}$ is a single positron state.

The operators $c_n$ and $\bar c_n$ are related to $b,d$ via a Bogoliubov transformation. Using the commutation relations \eq{acom2} and the orthonormality of the Dirac wave functions we see that they obey standard anticommutation relations,
\beqa\label{cmnortho}
\acom{c_m}{c_n^\dag} &=& \sum_\pv \Phi_{m,\alpha}^\dag(\pv) \big[u_\alpha(\pv,\lambda)u_\beta^\dag(\pv,\lambda) + v_\alpha(-\pv,\lambda)v_\beta^\dag(-\pv,\lambda)\big]\Phi_{n,\beta}(\pv)
= \int\frac{d\pv}{(2\pi)^3}\Phi_{m,\alpha}^\dag(\pv)\Phi_{n,\alpha}(\pv) =\delta_{mn}\nn \\[2mm]
\acom{\bar c_m}{c_n^\dag} &=&0 \nn \\[2mm]
\acom{\bar c_m}{\bar c_n^\dag} &=& \int\frac{d\pv}{(2\pi)^3}\bar \Phi_{m,\alpha}^\dag(\pv)\bar\Phi_{n,\alpha}(\pv) =\delta_{mn}
\eeqa

Inserting the completeness condition for the Dirac wave functions,
\beq\label{comp2a}
\sum_n \big[\Phi_{n,\alpha}(\xv)\Phi_{n,\beta}^\dag(\yv) +
\bar\Phi_{n,\alpha}(\xv)\bar\Phi_{n,\beta}^\dag(\yv)\big]=\delta_{\alpha\beta}\delta^3(\xv-\yv)
\eeq
into the Dirac Hamiltonian \eq{ham1} we find, recalling that the wave functions satisfy \eq{dir1} and \eq{dir2},
\beqa
H_D &=& \int d\xv\,d\yv\,\bar\psi_{\alpha'}(\xv)\big[-i\nv_\xv\cdot\gv+m+e\slashed{A}(\xv)\big]_{\alpha'\alpha}
\sum_n \big[\Phi_{n,\alpha}(\xv)\Phi_{n,\beta}^\dag(\yv) +
\bar\Phi_{n,\alpha}(\xv)\bar\Phi_{n,\beta}^\dag(\yv)\big]\psi_\beta(\yv)
\nn\\[2mm]
&=& \sum_n\int d\xv\,d\yv\,\psi_\alpha^\dag(\xv)\big[E_n \Phi_{n,\alpha}(\xv)\Phi_{n,\beta}^\dag(\yv)-\bar E_n \bar\Phi_{n,\alpha}(\xv)\bar\Phi_{n,\beta}^\dag(\yv)\big]\psi_\beta(\yv)\nn\\[2mm]
&=& \sum_n \big[E_n c_n^\dag c_n - \bar E_n \bar c_n \bar c_n^\dag\big] \to
\sum_n \big[E_n c_n^\dag c_n + \bar E_n \bar c_n^\dag \bar c_n\big] \label{hdiag}
\eeqa
In the last step I normal-ordered the operators, neglecting the zero-point energies according to \eq{vac1}.

\subsubsection{The vacuum state \label{vacsubsubsec}}

The expression for the vacuum state may be found using the methods in \cite{Blaizot-Ripka}. In terms of the coefficients defined in \eq{cndef1},
\beq\label{vac2}
\ket{\Omega} = N_0\exp\Big[-b_q^\dag \big(B^{-1}\big)_{qm}D_{mr}d_r^\dag\Big]\ket{0}
\eeq 
Sums over the repeated indices $q,m,r$ are implied in the exponent, and the normalization factor $N_0$ ensures that $\bra{\Omega}\Omega\rangle = 1$. The vacuum state may alternatively be expressed in terms of $\bar B, \bar D$ using the relation
\beq\label{BbarBid}
B_{mp}\bar B_{np}+D_{mp}\bar D_{np}=\sum_\pv \Phi_{m,\alpha}^\dag(\pv)\big[u_\alpha(\pv,\lambda)u_\beta^\dag(\pv,\lambda) + v_\alpha(-\pv,\lambda)v_\beta^\dag(-\pv,\lambda)\big]\bar\Phi_{n,\beta}(\pv)
= \int\frac{d\pv}{(2\pi)^3}\Phi_{m,\alpha}^\dag(\pv)\bar\Phi_{n,\alpha}(\pv) =0
\eeq
Multiplying by $(B^{-1})_{qm}({\bar D}^{-1})_{rn}$ and summing over $m,n$ gives
\beq
-(B^{-1})_{qm}D_{mr} = ({\bar D}^{-1})_{rn}{\bar B}_{nq}
\eeq
Using also $b^\dag_q d^\dag_r = -d^\dag_r b^\dag_q$ we find the form of the vacuum equivalent to \eq{vac2},
\beq\label{vac4}
\ket{\Omega} = N_0\exp\Big[-d_r^\dag \big({\bar D}^{-1})_{rn}{\bar B}_{nq} b_q^\dag\Big]\ket{0}
\eeq 

In order to verify that $c_n\ket{\Omega}=0$ we note that since $b_p$ essentially differentiates the exponent in \eq{vac2},
\beq
B_{nq}b_q\ket{\Omega}=-B_{nq}\big(B^{-1}\big)_{qm}D_{mr}d_r^\dag\ket{\Omega} = -D_{nr}d_r^\dag\ket{\Omega}
\eeq
This cancels the contribution of the second term in the definition \eq{cndef1} of $c_n$. The demonstration that $\bar c_n$ annihilates the vacuum is simular. Thus
\beq
c_n\ket{\Omega}=\bar c_n\ket{\Omega}=H\ket{\Omega}=0
\eeq

According to \eq{cndef1} the $e^-$ bound state takes the form,
\beq\label{nstate}
c_n^\dag\ket{\Omega} \equiv \big[b_r^\dag B_{rn}+d_r D_{rn}\big]\ket{\Omega}
\eeq
which also serves to define the coefficients $B_{rn}$ and $D_{rn}$. The ``sea'' contribution $\propto d_r$ is
\beq\label{seadist}
d_r D_{rn}N_0\exp\Big[-b_q^\dag \big(B^{-1}\big)_{qm}D_{mr}d_r^\dag\Big]\ket{0}
= b_q^\dag \big(B^{-1}\big)_{qm}D_{mr}D_{rn}\ket{\Omega}
\eeq
Similarly as in \eq{BbarBid} we have
\beq\label{BBid}
\sum_\pv (B_{mp}\,B_{pn}+D_{mp}\,D_{pn})= \int\frac{d\pv}{(2\pi)^3}\Phi_{m}^\dag(\pv)\,\Phi_{n}(\pv) =\delta_{mn}
\eeq
Using this in \eq{seadist} gives
\beq
d_r D_{rn}\ket{\Omega}=\big[b_q^\dag(B^{-1})_{qn}-b_r^\dag B_{rn}\big]\ket{\Omega}
\eeq
Hence the fermion state \eq{nstate} can be alternatively expressed as
\beq\label{nstate2}
c_n^\dag\ket{\Omega}= b_p^\dag(B^{-1})_{pn}\ket{\Omega}
\eeq
The vacuum state $\ket{\Omega}$ \eq{vac2} describes the distribution of the $e^+e^-$ pairs that appear in the bound state through perturbative diagrams such as \fig{diracdiag}(b). It is a formal expression, involving a sum over all states $m$ and the inverted matrix $\big(B^{-1}\big)_{qm}$. In the weak binding limit $D_{np} \to 0$ and $\ket{\Omega}\to\ket{0}$. Then \eq{BBid} ensures that $(B^{-1})_{pn}=B_{np}$.

The coefficient $B_{pn}$ describes the ``valence'' electron momentum distribution in the bound state \eq{nstate}. For a confining potential it is expected to be limited to low $|\pv|$. In the next section I verify this for a linear potential in $D=1+1$ dimensions. This presumably also holds for the valence + sea electron distribution described by $\big(B^{-1}\big)_{pn}$ of \eq{nstate2}.

\subsection{Dirac states in $D=1+1$ dimensions \label{diracsubsec2}}

Physics in $D=1+1$ dimensions is often used as a model for confinement since the QED potential is linear. The Dirac equation can be solved analytically in terms of Confluent Hypergeometric functions. These turn out to have quite different properties compared to the Airy-function solutions of the Schr\"odinger equation with a linear potential. The difference stems from the $e^+e^-$ pairs that contribute to relativistic dynamics. 

\subsubsection{Wave function \label{d1p1subsec}}

The Dirac wave function for a linear potential in $D=1+1$ is known since a long time \cite{sauter}. The following discussion uses the notation and results of \cite{Dietrich:2012un}, as well as the relation between the wave function and the states presented in section \ref{dstatesec}.

Denoting the (positive or negative) eigenvalues by $M$ the Dirac equation in $D=1+1$ with $A^1 = 0$ is 
\beq\label{dir}
\big[-i\gz\go\partial_x+\gz m+ eA^0\big]\Phi(M,x)=M\Phi(M,x)
\eeq
The Dirac matrices can be represented in terms of Pauli matrices,
\beq\label{matdef}
\gz=\sigma_3 \hspace{1cm} \go=i\sigma_2 \hspace{1cm} \gz\go=\sigma_1
\eeq
The energy scale is set by the coupling $e=1$ and the potential is linear as in QED$_2$,
\beq\label{linpot}
V(x) \equiv eA^0(x) = V'|x|= \halft e^2|x| \equiv \halft |x|
\eeq

Let
\beq\label{potwf2}
\Phi(M,x) = \left[\begin{array}{c}  \vphi(x) \\ \chi(x) \end{array}\right] \hspace{1cm} \dsi(x) = (M-V)^2 \hspace{1cm} \partial_x = \varepsilon(-x)(M-V)\partial_\dsi
\eeq
where $\veps(x)=x/|x|$ is the sign function. The equations for the two components of the Dirac wave function are
\beqa\label{diraccomp2}
-i\partial_x\chi&=&(M-V-m)\vphi, \nn\\
-i\partial_x\vphi&=&(M-V+m)\chi.
\eeqa
They depend on $M$ and $x$ only in the combination $M-V$, motivating the introduction of $\dsi$ in \eq{potwf2}. The 2nd order equation for $\vphi(x)$ is
\beq\label{dirac2ord}
\partial_x^2\vphi(x) + \frac{\veps(x)}{2(M-V+m)}\,\partial_x\vphi(x)+\big[(M-V)^2-m^2\big]\vphi(x)=0
\eeq

Let $\vphi(x=0)$ be real and $\chi(x=0)$ imaginary. This reduces the four real parameters of the general solution to two real parameters.
The differential equations \eq{diraccomp2} ensure that the phases imposed at $x=0$ hold at all $x$,
\beq\label{phase}
\vphi^*(x) = \vphi(x) \hspace{2cm} \chi^*(x) = -\chi(x)
\eeq
For states of definite parity $\eta=\pm 1$ it suffices to find the wave function for $x\geq 0$ since
\beq\label{parity}
\vphi(-x) = \eta\vphi(x) \hspace{2cm} \chi(-x) = -\eta\chi(x)
\eeq
Continuity at $x=0$ requires that $\chi(0)=0$ for $\eta=+1$, while $\vphi(0)=0$ for $\eta=-1$.

The solution of the differential equations \eq{diraccomp2} are conveniently expressed in terms of the functions
\beqa\label{fgdef}
f(x) \equiv [\vphi(x)+\chi(x)]e^{i\dsi} \nn\\
g(x) \equiv [\vphi(x)-\chi(x)]e^{i\dsi}
\eeqa
Then for $x>0$,
\beqa\label{fgsol}
f(x)&=& N_M\Big[e^{i\beta} \kum(-i\halft m^2,\halft,2i\dsi)+2ime^{-i\beta} (M-V)\kum(\halft-i\halft m^2,\sfrac{3}{2},2i\dsi)\Big] \nn \\[2mm]
g(x) &=& N_M\Big[e^{-i\beta} \kum(\halft-i\halft m^2,\halft,2i\dsi)-2im e^{i\beta} (M-V)\kum(1-i\halft m^2,\sfrac{3}{2},2i\dsi)\Big]
\eeqa
The real normalization constant $N_M>0$ and the phase $\beta$ are defined in terms of the real constants $a,b$ of Eq. (2.15) in \cite{Dietrich:2012un} through $a+ib=N_M\,e^{i\beta}$. Since $\beta\to\beta+\pi$ only gives an overall sign change we may restrict the phase to $-\pi/2 \leq \beta < \pi/2$.

\subsubsection{Continuous spectrum \label{wcssubsubsec}}

The continuity condition at $x=0$ determines $M$ as a function of $\beta$. Since $\beta$ is a continuous variable also the eigenvalues $M$ take a {\em continuous range} of (positive and negative) values. This is in stark contrast to the discrete spectrum of the Schr\"odinger equation in $D=1+1$ with a linear potential. Already in the 1930's Plesset \cite{plesset} noted that the Dirac spectrum is continuous for positive and negative power-law potentials in any dimension. The only exception is the $1/r$ potential in $D=3+1$ dimensions.

The expression \eq{nstate} for the electron bound state has Fock components with any number of $e^+e^-$ pairs due to the structure \eq{vac2} of the vacuum $\ket{\Omega}$. The linear potential \eq{linpot} confines electrons but {\em repulses positrons}: the $e^+$ potential is {\em negative}, $-V(x)=-\halft |x|$. A positron can appear at any $x$ provided the sum of its kinetic and potential energies $E_p-V(x)$ is commensurate with the energy eigenvalue $M$ of the state (hence $|p| \simeq \halft |x|$ at large distances). This condition can be satisfied for any $M$, allowing a continuous spectrum.

The analytic solution \eq{fgsol} bears out these intuitive expectations. The parameter $\beta$ determines the relative amount of ``valence'' $e^-$ and ``sea'' $e^+e^-$ contributions. For finite fermion mass $m$ there is a specific value $\beta=\beta_{min}$ which corresponds to a minimal sea. This minimum is exponentially damped in $m^2$, similarly as the Schwinger pair production rate in a constant electric field \cite{Schwinger:1951nm}. Conversely, there is a value of $\beta$ that minimizes the valence $e^-$ contribution.

The large $x$ behavior of the wave functions in \eq{fgsol} is, with $\sigma \simeq \quart x^2-M|x|$,
\beq\label{fas}
\inv{N_M}\lim_{x\to\infty}e^{-i\dsi}f(x) = C_M\,e^{-i\dsi}(2\dsi)^{im^2/2}-\frac{m}{2\sqrt{\dsi}}C_M^*\,e^{i\dsi}(2\dsi)^{-im^2/2}+\morder{x^{-2}}
\eeq
where
\beqa
C_M &=& e^{i(\beta-\delta)}\frac{\Gamma(1-im^2/2)}{m\sqrt{2\pi}}e^{3\pi m^2/4} \sqrt{1-e^{-\pi m^2}}\Big[\sqrt{1+e^{-\pi m^2}}+e^{i(\delta-2\beta-\pi/4)}\sqrt{1-e^{-\pi m^2}}\Big] \label{cf}\\[2mm]
\delta &=& \arg\left[\frac{\Gamma(1-i\halft m^2)}{\Gamma(\halft-i\halft m^2)}\right]\ \ \buildrel m\to\infty\over{\rightarrow} -\frac{\pi}{4} \label{delta}
\eeqa
The corresponding result for the $g$-function is obtained from the relation $e^{-i\dsi}g(x) = \big[e^{-i\dsi}f(x)\big]^*$, which follows from \eq{fgdef} and the phase choice \eq{phase}. The wave functions \eq{fgsol} thus have constant asymptotic magnitude,
\beq\label{asnorm}
\lim_{x\to\infty}|f(x)| = \lim_{x\to\infty}|g(x)| = N_M\,|C_M|
\eeq

The continuous spectrum means that the orthonormality relation \eq{BBid} involves a Dirac $\delta$-function,
\beq\label{orno}
\int_{-\infty}^\infty dx\,\Phi_\eta^\dag(M,x)\Phi_{\eta'}(M',x)= \delta_{\eta\eta'}\delta(M-M')
\eeq
where $\eta,\eta'$ denote parities. As for plane waves the $\delta$-function must arise from the infinite range of the $x$-integral. For arbitrarily large $\Lambda$,
\beq\label{orno2}
\int_{\Lambda}^\infty dx\,\Phi_\eta^\dag(M,x)\Phi_{\eta}(M',x)= \halft\delta(M-M')+\ldots
\eeq
where $\ldots$ represents $\Lambda$-dependent finite contributions. Recalling that $\vphi(x)$ is real and $\chi(x)$ is imaginary, as well as the definition \eq{fgdef} of $f(x)$, \eq{orno2} requires
\beq\label{orno3}
\int_{\Lambda}^\infty dx\,\big[\vphi_M(x)\vphi_{M'}(x)-\chi_M(x)\chi_{M'}(x)\big] =
{\rm Re}\int_{\Lambda}^\infty dx\,\big[e^{-i\dsi}f_M(x)\big]\big[e^{-i\dsi'}f_{M'}(x)\big]^* =
 \halft\delta(M-M')+\ldots
\eeq
Using the asymptotic expression \eq{fas} allows to determine the normalization constant,
\beq\label{nm}
N_M=\inv{\sqrt{2\pi}\,|C_M|}
\eeq
According to \eq{asnorm} this implies that $|f(x\to\infty)|=1/\sqrt{2\pi}$ is independent of $\beta$ and $M$. The normalization \eq{nm} also gives the correct $\delta(x-y)$-function in the completeness condition \eq{comp2a}.

\subsubsection{Relative amount of valence and sea \label{vssubsubsec}}

Let us now consider how the parameter $\beta$ determines the relative amount of ``valence'' (small $|x|$) and ``sea'' (large $|x|$) contributions. Since $|f(x\to\infty)|$ is $\beta$-independent it suffices to consider the wave function in the region of small $|x|$. The comparison is most meaningful for large $m$, since the $V(x) \lsim 2m$ region is classically forbidden for the sea.

At large $m$ the dynamics is non-relativistic at small $|x|$. The Schr\"odinger equation with the linear potential \eq{linpot} is
\beq\label{seq}
 -\inv{2\moe}\partial_x^2 \rho(x)+\frac{1}{2} |x| \rho(x) = E_b\rho(x),
\eeq
where $E_b=M-m$ is the binding energy. Both the coordinate and binding energy scale as $x,E_b \propto m^{-1/3}$. The normalizable solutions are given by the Airy function,
\beq\label{seqAi}
\rho(x) = N\,\mathrm{Ai}[\moe^{1/3}(x-2E_b)] \qquad (x>0), 
\eeq
with $\rho(x)=\pm\rho(-x)$.

The non-relativistic limit of the wave function $f(x)$ in \eq{fgsol} is given in Eq. (2.21) of \cite{Dietrich:2012un}, and the derivation is detailed in App. B of \cite{Hoyer:2014gna},
\beq\label{slim}
\lim_{{\rule{0mm}{2mm}\scriptstyle m\to\infty\atop\scriptstyle {m^{1/3} x\ {\rm fixed}}}}e^{-i\dsi}f(x) = \sqrt{2\pi}N_M(\cos\beta+\sin\beta)m^{1/3}e^{\pi m^2/2}{\rm Ai}[m^{1/3}(x-2E_b)]
\eeq
The result agrees (up to the normalization) with the Schr\"odinger Airy-function solution. Determining $N_M$ from \eq{nm} and \eq{cf}, with $\delta=-\quart \pi$ as in \eq{delta}, we get
\beq\label{slim2}
\lim_{{\rule{0mm}{2mm}\scriptstyle m\to\infty\atop\scriptstyle {m^{1/3} x\ {\rm fixed}}}}e^{-i\dsi}f(x) = \inv{\sqrt{2}}\,\frac{(\cos\beta+\sin\beta)m^{1/3}{\rm Ai}[m^{1/3}(x-2E_b)]}{\left|\cos\big(\beta+\frac{\pi}{4}\big)+\frac{i}{2}e^{-\pi m^2}\sin\big(\beta+\frac{\pi}{4}\big)\right|}
\eeq

The value $\beta=-\quart\pi$ minimizes the ``valence'' wave function at low $|x|$. Conversely, $\beta=+\quart\pi$ suppresses the (relative) amount of sea by a factor $\exp(-\pi m^2)$, corresponding to the tunnelling rate from the classically forbidden region. As in the Schwinger mechanism \cite{Schwinger:1951nm}, there are always some virtual pairs in a non-vanishing electric field. 

\begin{figure}[h]
\includegraphics[width=1.\columnwidth]{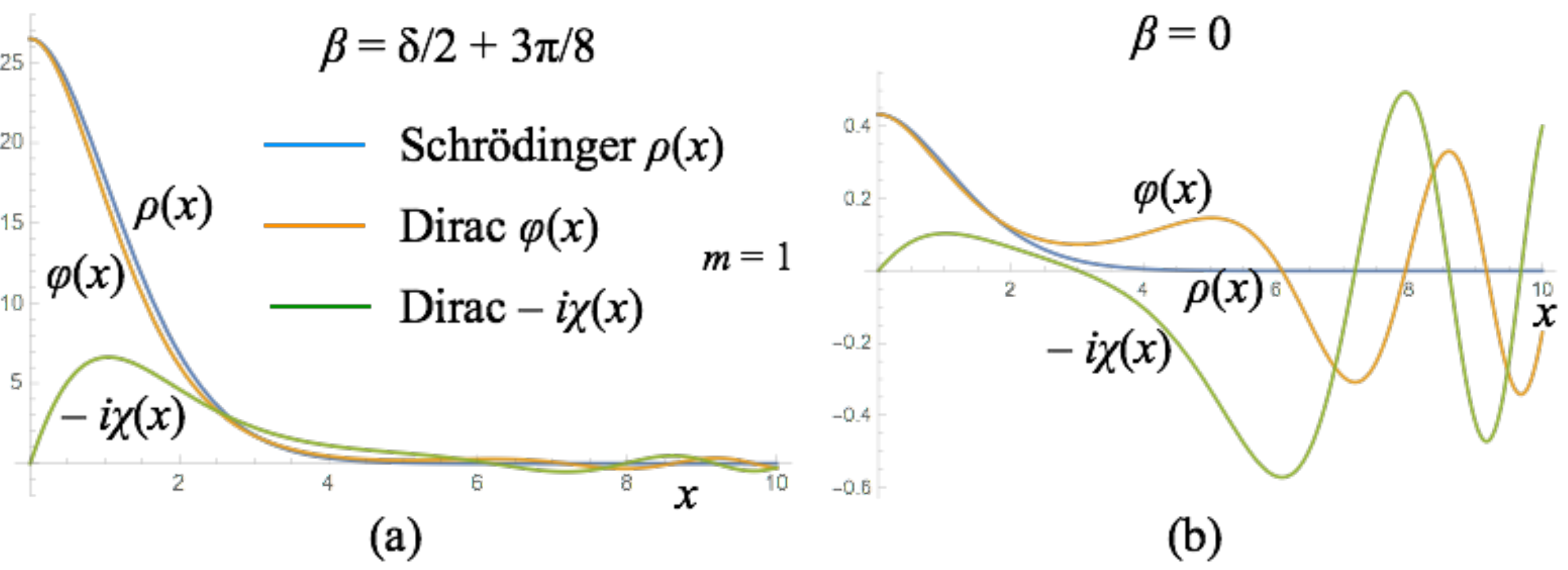}
\caption{The ground state Schr\"odinger wave function $\rho(x)$ for $m=1$ compared with the upper and lower components of the Dirac wave function \eq{potwf2} for (a) $\beta=\beta_{min}=0.925$ and (b) $\beta=0$. For ease of comparison the Schr\"odinger wave function is arbitrarily normalized by $\rho(0)=\vphi(0)$.
{\label{wfs-beta}}}
\end{figure}

\fig{wfs-beta} illustrates the $\beta$-dependence of the positive parity ground state wave function when $m=1$. In (a) $\beta=\beta_{min}= \halft \delta+\sfrac{3}{8}\pi$, the value which minimizes $C_M$ \eq{cf} and thus maximizes $\vphi(0)$. The relative magnitude of the oscillations at large $x$ are indeed seen to be smaller in (a) than in (b), where $\beta=0$. In both cases $\vphi(x)$ agrees in shape with the Schr\"odinger wave function $\rho(x)$ at low $x$.

\begin{figure}[h] \centering
\includegraphics[width=10cm]{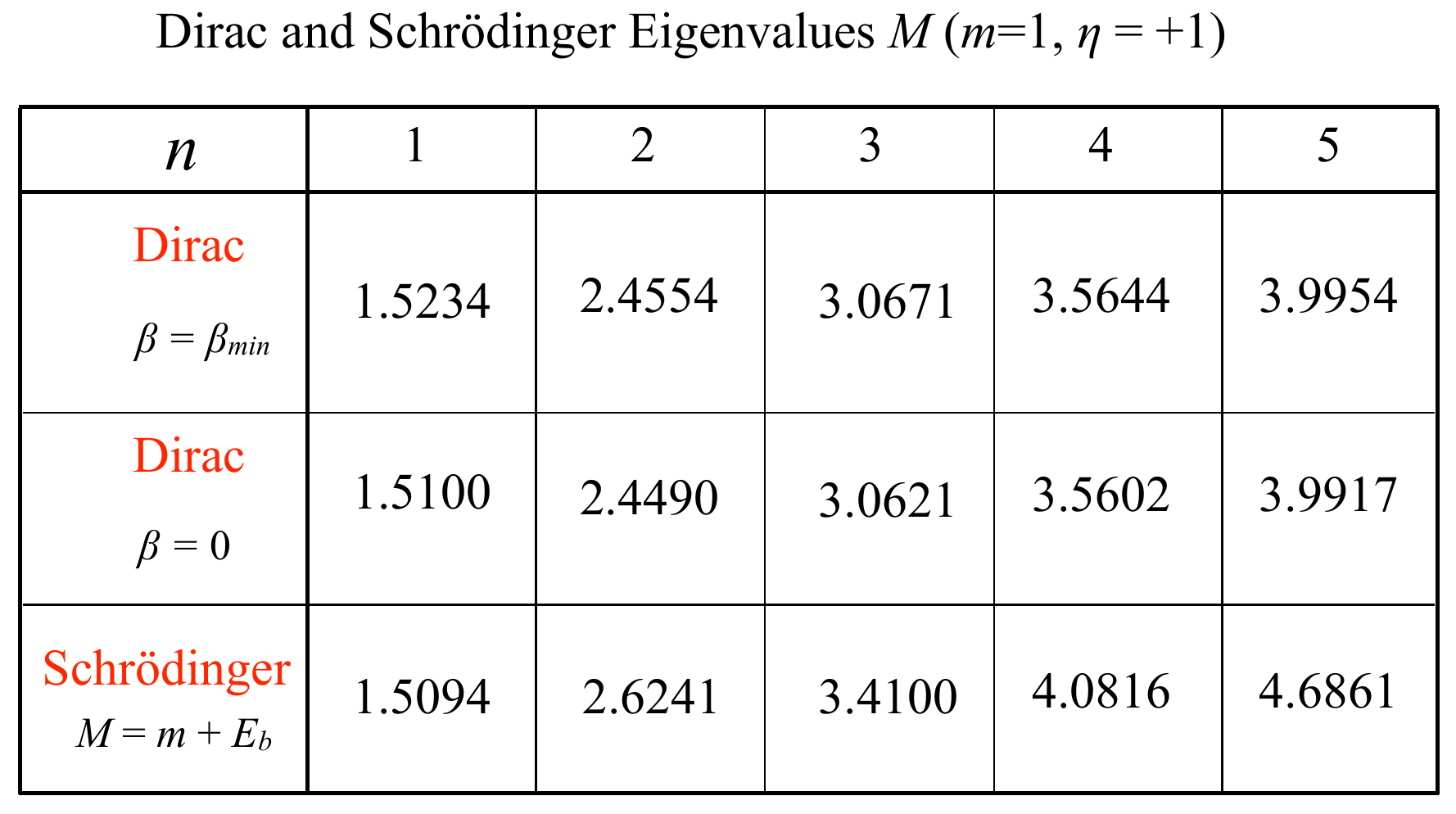}
\caption{The eigenvalues $M$ (in units of the electron charge $e$) of the five lowest (positive parity) states calculated from the Schr\"odinger and Dirac equations. In the Dirac case the fermion mass $m=1$, $\beta=0.925$ and $\beta=0$.\label{M-Table}}
\end{figure}

\fig{M-Table} compares the five first eigenvalues $M$ for the Dirac solutions with $\beta=\beta_{min}$ and $\beta=0$ with the Schr\"odinger values. The three solutions give nearly the same masses for the lowest excitations. Values of $M$ that differ widely from the Schr\"odinger ones occur only in a restricted range of $\beta \simeq -\quart\pi$, where the valence contribution is suppressed according to \eq{slim2}. That range moreover narrows quickly as $m$ increases. For the Dirac as well as the Schr\"odinger equation the quantization of $M$ is determined by the valence electron at low $|x|$. $M$ can therefore take a continuous range of values in the Dirac equation only when the valence contribution is suppressed.

The reduction of the Dirac wave function to the Schr\"odinger one is usually demonstrated by taking $m\to\infty$ with $x \propto m^{-1/3}$. From \eq{slim2} we see that this is not quite adequate in the case of a linear potential, since the relative size of the large $x$ oscillations decrease with $m$ only when $\beta=\beta_{min}$. A more proper limit is thus $m\to\infty$ with $\beta=\beta_{min}$.

\subsubsection{The electron and positron distributions \label{epsubsubsec}}

I have been rather loosely referring to the low and high $|x|$ regions as ``valence'' and ``sea'', respectively. The analysis in section \ref{dstatesec} allows to specify the $e^\pm$ distributions more precisely. The $M>0$ electron state \eq{nstate} is
\beq\label{wtBD}
\ket{M}=c_M^\dag\ket{\Omega} = \int_{-\infty}^\infty\frac{dp}{2\pi\,2E_p}\big[b_p^\dag u^\dag(p)+d_{-p}v^\dag(-p)\big]\Phi_M(p)\ket{\Omega}
\equiv \int_{-\infty}^\infty\frac{dp}{\sqrt{2\pi\,2E_p}}\big[b_p^\dag \wtB_{pM}+d_{p}\wtD_{pM}\big]\ket{\Omega}
\eeq
The $\wtB,\wtD$ matrices are defined to absorb the convention-dependent factor $1/\sqrt{2\pi\,2E_p}$ arising from the definition of the Fourier transform and the $u,v$-spinors,
\beqa
\wtB_{pM} &=& \inv{\sqrt{2\pi\,2E_p}}\,u^\dag(p)\Phi_M(p) \hspace{2cm}
u(p)=\inv{\sqrt{E_p+m}} \left(\begin{array}{c} E_p+m \\ p \end{array} \right) \label{wtB}\\[2mm]
\wtD_{pM} &=& \inv{\sqrt{2\pi\,2E_p}}\,v^\dag(p)\Phi_M(-p) \hspace{1.8cm}
v(p)=\inv{\sqrt{E_p+m}} \left(\begin{array}{c} p \\ E_p+m \end{array} \right) \label{wtD}
\eeqa
The $p$-depdendence of $\wtB_{pM}$ determines at what momenta an $e^-$ is added to the vacuum \eq{vac2}, whereas $\wtD_{pM}$ determines the $p$-distribution of the $e^+$ in the vacuum that is annihilated by $d_p$.

In terms of the solutions \eq{fgsol} of the Dirac equation, for $M>0$ positive parity states,
\beqa\label{bdexpr}
\wtB_{pM}^{\eta=+} &=& \inv{\sqrt{2\pi\,2E_p(E_p+m)}}\,{\rm Re}\int_0^\infty dx\,e^{-i\dsi}f(x)\big[(E_p+p+m)e^{-ipx}+(E_p-p+m)e^{ipx}\big] = \wtB_{-p,M}^{\eta=+}\nn\\[2mm]
\wtD_{pM}^{\eta=+} &=& \frac{-1}{\sqrt{2\pi\,2E_p(E_p+m)}}\,{\rm Re}\int_0^\infty dx\,e^{-i\dsi}f(x)\big[(E_p-p+m)e^{-ipx}-(E_p+p+m)e^{ipx}\big] = -\wtD_{-p,M}
\eeqa

For $p\to\infty$ the rapidly oscillating phases $\exp(\pm ipx)$ allows to approximate the $x$-integrals in \eq{bdexpr} using the stationary phase method,
\beq\label{statphase}
\int_{-\infty}^\infty dx F(x)e^{i\vphi(x)} \simeq e^{\veps(\vphi''(x_s)) i\pi/4}\sqrt{\frac{2\pi}{|\vphi''(x_s)|}}F(x_s)e^{i\vphi(x_s)}
\eeq
where $\veps(x)$ is the sign function and $\vphi(x) \simeq \vphi(x_s)+\halft \vphi''(x_s)(x-x_s)^2$ is a rapidly varying phase. The condition for the stationary phase is that $\partial_x\vphi(x_s)=0$.

As $x\to\infty$ the phase of the leading term in $e^{-i\dsi}f(x)$ \eq{fas} varies rapidly (the phase change of the power is subleading),
\beq\label{fphase} 
\arg\big[\lim_{x\to\infty}e^{-i\dsi}f(x)\big] =-\dsi = -M^2+Mx-\quart x^2
\eeq
Due to the $p \to -p$ symmetries in \eq{bdexpr} we need only consider $p\to +\infty$. Then \eq{fphase} can give a stationary phase at $x=x_s$ only with the Fourier phase $\exp(ipx)$ in \eq{bdexpr},
\beq\label{sphasecond}
\left.\frac{d}{dx}(-\dsi+px)\right|_{x=x_s}=M-\halft x_s+p=0 \hspace{2cm} 
\left\{\begin{array}{l} p-V(x_s)=-M \\[2mm] \dsi(x_s) = p^2 \\[2mm] px_s-\dsi(x_s)=p(p+2M)\end{array} \right.
\eeq 

In \eq{bdexpr} the factor multiplying $\exp(ipx)$ is of \order{p^{-1}} for $B_{Mp}$ and of \order{p^{0}} for $D_{Mp}$. Hence the oscillations of the wave function in \fig{wfs-beta} at large $x$ are due to positrons, not electrons. The positron kinetic energy $E_p \simeq p$ and potential energy $-V(x_s)$ sum to $-M$. The sea electron forming the pair with the positron is confined by the potential to low $x$, together with the valence electron created by $b_p^\dag\wtB_{pM}$. According to \fig{wfs-beta} the electrons are distributed similarly to the Schr\"odinger wave function. The two of them thus contribute $2M$ to the energy, so that the total energy of the $e^-(e^+e^-)$ Fock state adds up to the bound state energy $M$. 

The next-to-leading term in the asymptotic expansion \eq{fas} of $f(x)$ is suppressed by \order{\dsi^{-1/2}} = \order{p^{-1}} when $x \propto p$. It gives a stationary phase with the $\exp(-ipx)$ term in \eq{bdexpr}, whose coefficient in $\wtB_{pM}$ is of \order{p^0}. This contribution is of the same \order{p^{-1}} as the one from the $\exp(ipx)$ term in $\wtB_{pM}$. The two contributions cancel,
\beqa
\exp(ipx):&& \hspace{5mm} \frac{m\sqrt{2}}{p}\,N_M\,{\rm Re}\Big[C_M (2p^2)^{im^2/2}\,e^{-i\pi/4}e^{ip(p+2M)}\Big] \nn\\[2mm]
\exp(-ipx):&& \hspace{2mm} -\frac{m\sqrt{2}}{p}\,N_M\,\,{\rm Re}\Big[C_M^* (2p^2)^{-im^2/2}\,e^{i\pi/4} e^{-ip(p+2M)}\Big]
\eeqa
since the real part of a complex number and its conjugate are equal. Consequently 
\beq
\lim_{p\to\infty}\wtB_{pM}^+ \lsim\morder{p^{-2}}
\eeq

\begin{figure}[h] \centering
\includegraphics[width=1.\columnwidth]{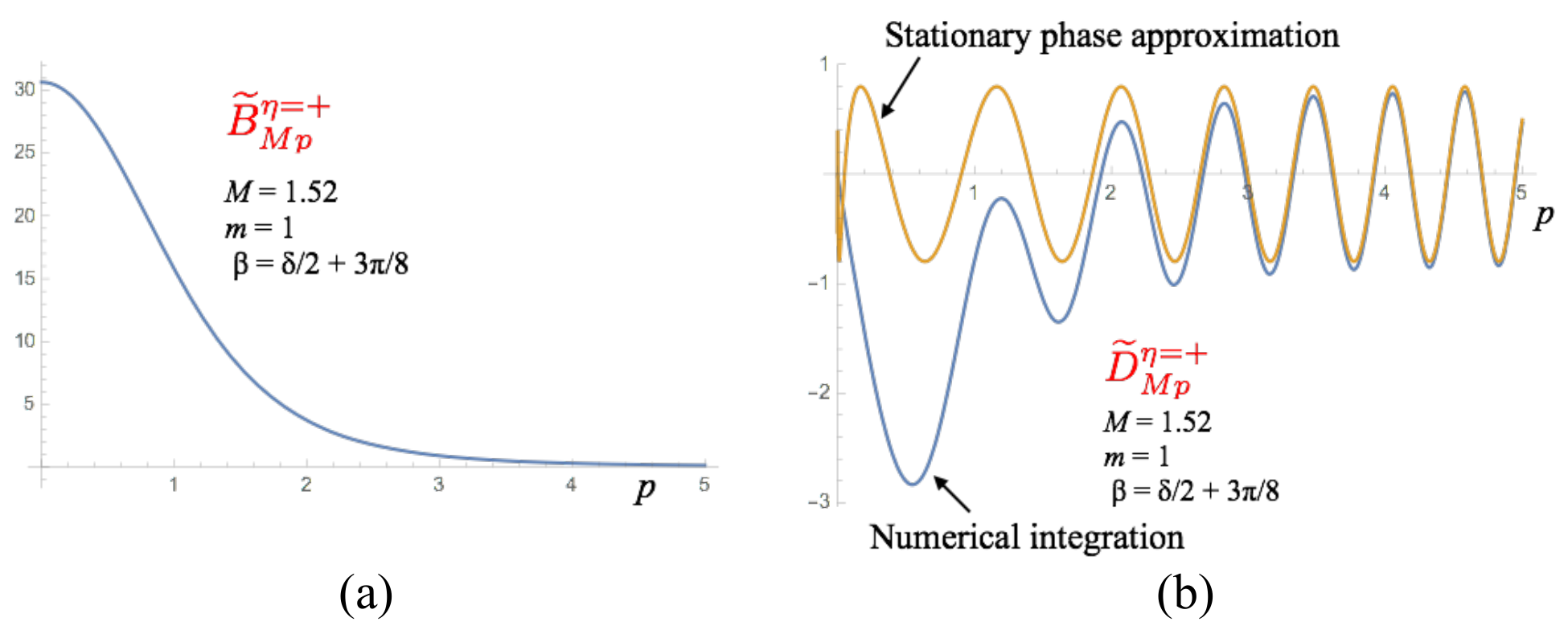}
\caption{Numerical results for the momentum dependence of the (a) $\wtB_{pM}^{\eta=+}$ and (b) $\wtD_{pM}^{\eta=+}$ matrices \eq{bdexpr} of the ground state, for $m=1$ and $\beta=\beta_{min}$. The stationary phase approximation for $\wtD_{pM}^{\eta=+}$ is indicated in (b).\label{BD-dist}}
\end{figure}

\subsubsection{Case of $m=0$ \label{m0subsubsec}}

The expressions of the previous section simplify considerably when the fermion mass $m=0$. Using
\beq
\kum(0,\halft,2i\sigma)=1 \hspace{2cm} \kum(\halft,\halft,2i\sigma)=e^{2i\sigma}
\eeq
and the parity $\eta$ convention \eq{parity} the components $\vphi_\eta(x), \chi_\eta(x)$ of Dirac wave function $\Phi(M,x)$ \eq{potwf2} are for any $x$,
\beqa\label{wfm0}
\vphi_+(x) &=& \frac{1}{\sqrt{2\pi}}\cos\big[x(M-\quart|x|)\big] \hspace{2cm}
\vphi_-(x) = \frac{1}{\sqrt{2\pi}}\sin\big[x(M-\quart|x|)\big] \label{phiexpr}\nn \\[2mm]
\chi_+(x) &=& \frac{i}{\sqrt{2\pi}}\sin\big[x(M-\quart|x|)\big] \hspace{2cm}
\chi_-(x) = -\frac{i}{\sqrt{2\pi}}\cos\big[x(M-\quart|x|)\big] \label{chiexpr}
\eeqa
where $M$ can be any real number. The relation between $\beta$ and $M$ is
\beq\label{beta0}
\beta = M^2-n\pi \ \ (\eta=+1) \hspace{3cm} \beta =  M^2-\frac{\pi}{2}-n\pi \ \ (\eta=-1)
\eeq
with the integer $n$ chosen so that $-\pi/2 \leq \beta < \pi/2$.

The simple form \eq{wfm0} of the wave function allows orthonormality and completeness to be explicitly verified,
\beq\label{orno4}
\int_{-\infty}^\infty dx\,\Phi_\eta^\dag(M,x)\Phi_{\eta'}(M',x)= \delta_{\eta\eta'}\delta(M-M')
\eeq
\beq\label{comp}
\sum_{\eta=\pm}\int_{-\infty}^\infty dM\,\Phi_\eta(M,x)\Phi_{\eta}^\dag(M,y)= 
\sum_\eta\int_{-\infty}^\infty dM\,\left(\begin{array}{cc}\vphi_\eta(x)\vphi_\eta(y) & -\vphi_\eta(x)\chi_\eta(y) \\[2mm] \chi_\eta(x)\vphi_\eta(y) & -\chi_\eta(x)\chi_\eta(y)\end{array}\right)=
\delta(x-y)\left(\begin{array}{cc}1 & 0 \\[2mm] 0 & 1\end{array}\right)
\eeq

With $m=0$ the matrix elements of the operators in the electron state \eq{wtBD} are, for $M>0$ and $p>0$,
\beqa\label{bdm0}
\wtB^{\eta=+}_{p>0,M} &=& \inv{\pi\sqrt{2}}\int_0^\infty dx\,\cos\big[x(M-p)-\quart x^2\big] \nn\\[2mm]
\wtD^{\eta=+}_{p>0,M} &=& \inv{\pi\sqrt{2}}\int_0^\infty dx\,\cos\big[x(M+p)-\quart x^2\big]
\eeqa
The ``valence'' electron distribution decreases with momentum as $1/p^3$,
\beq\label{Bas}
\lim_{p\to\infty}\wtB^{\eta=+}_{pM} = \inv{\pi\, 2\sqrt{2}}\,\inv{(p-M)^3}\left[1+\morder{p^{-1}}\right]
\eeq
while the stationary phase contribution to $\wtD^{\eta=+}_{pM}$ is of \order{p^0},
\beq\label{Dsp}
\lim_{p\to\infty}\wtD^{\eta=+}_{pM} = \frac{\sqrt{2}}{\sqrt{\pi}}\,\cos\big[(p+M)^{2}-\quart\pi\big]
\eeq

\section{Features of hadrons \label{secmotiv}}
\setcounter{subsubsection}{0}
\subsubsection{General remarks}

Quantum Chromodynamics accurately describes hard scattering using perturbation theory and universal parton distributions. Numerical lattice calculations support the correctness of QCD also for soft processes. Comparisons of QCD-inspired models with data have given a general understanding of hadron dynamics \cite{Pennington:2016dpj}. Color confinement implies that the constituents of hadrons (quarks and gluons) do not exist as free states. This is related to their relativistic binding energy, which enables pair production. The dynamical breaking of chiral symmetry leads, among other things, to the observed small mass of the pion. The vast literature on QCD is summarized in \cite{Kronfeld:2010bx}.

Analytic approaches to soft hadron processes complement numerical evaluations. This has only been possible at the expense of additional assumptions, outside the QCD field theory framework. It is desirable to derive from QCD those model features which successfully describe data. 

Perturbation theory is the main analytic, first-principles approach to QED and QCD. The perturbative characteristics of soft QCD dynamics has been underlined by Dokshitzer \cite{Dokshitzer:1998nz,Dokshitzer:2003bt,Dokshitzer:2010zza}.
It is not completely straightforward to apply perturbation theory even to QED bound states. Already the first approximation of a bound state requires summing an infinite number of Feynman diagrams. Reorderings of the perturbative expansion lead to different, and in principle equivalent formulations \cite{Lepage:1978hz,Siringo:2015jea}. The $\hbar$ expansion motivates a specific ``Born term'' for bound states characterized by the absence of loops: Only gauge fields which satisfy the classical field equations contribute. 

In section \ref{posisec} I derived the Born term for Positronium. In the rest frame it yields the standard Schr\"odinger equation with the classical potential $-\alpha/r$. It gives the correct dependence of the the Positronium energy on its momentum $\Pv$, and a wave function that Lorentz contracts as in classical physics. Higher order (loop) corrections are expected to be calculable from the $S$-matrix \eq{smatrix}, with the Born term used in the $in$ and $out$ states.

The loop expansion is a power series in $\hbar\alpha$ and as such converges only for small (perturbative) couplings. The binding energy $E_b=-\quart m_e\alpha^2$  of Positronium if of \order{m_e} only for $\alpha \gsim 1$. Hence strongly bound (relativistic) states cannot be accessed within perturbative QED. The situation for QCD may be different because the confining potential can cause relativistic binding even at small $\alpha_s$. This is the scenario that will be explored below.

In section \ref{diracsec} I discussed some properties of strongly bound states in the familiar setting of the Dirac equation, which describes the binding of an electron in a strong external field. Dirac states include virtual $e^+e^-$ pairs whose distributions are given by the negative energy components of the Dirac wave function. A potential which confines electrons repulses positrons, giving dramatic effects on the spectrum and wave functions.

In the rest of these lectures I consider whether the concept of Born term could be applicable to hadrons. The novel features of color confinement and chiral symmetry breaking then need to appear as a consequence of the classical gluon field and ground state. I begin by recalling some phenomenological features of hadron dynamics which hint at a perturbative framework.

\begin{figure}[h]
\includegraphics[width=1.\columnwidth]{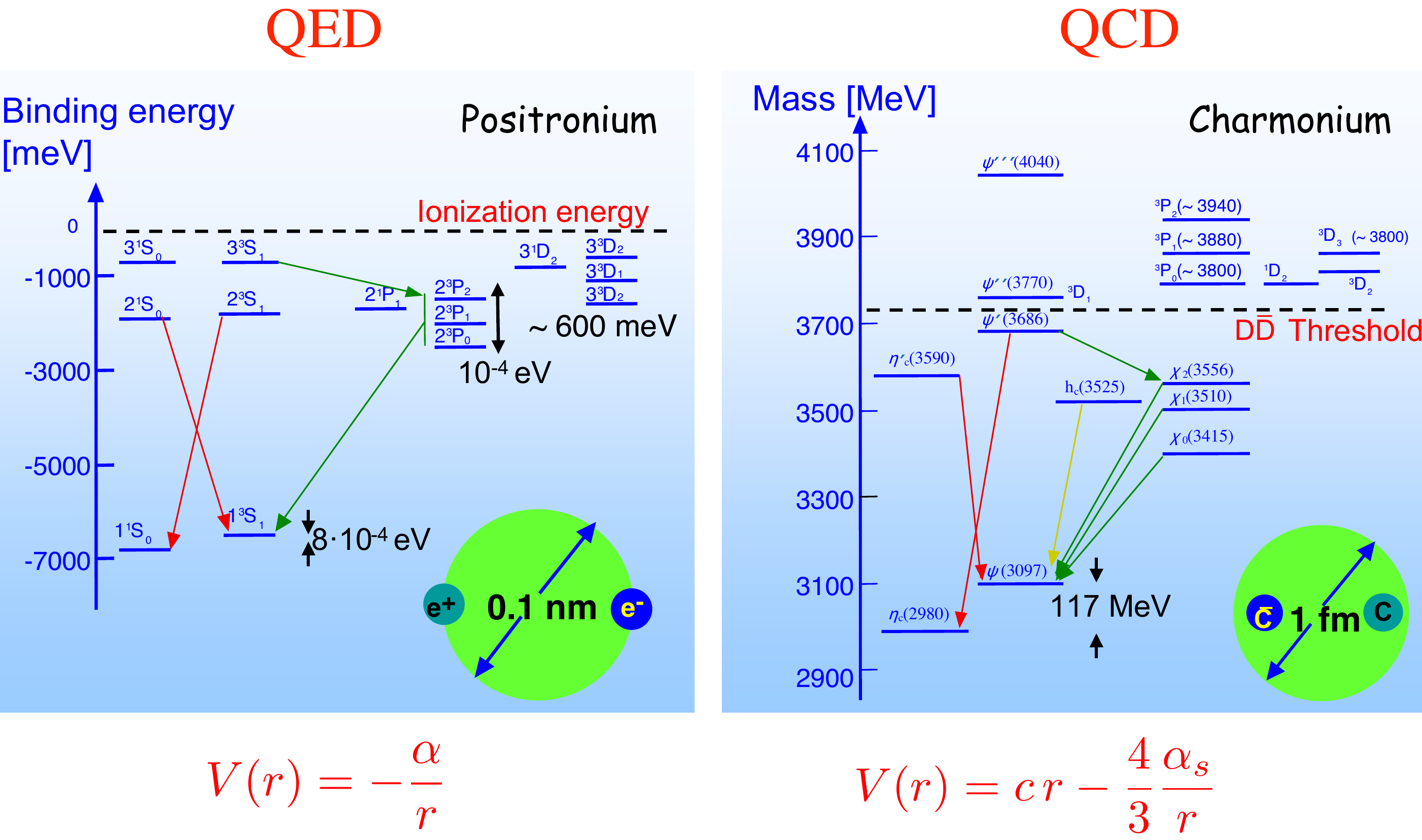}
\caption{Comparison of the Positronium and charmonium spectra. The overall features are surprisingly similar, considering that the hyperfine splittings differ by a factor $\gsim 10^{11}$ and that QCD is a confining, non-abelian theory.
{\label{atoms}}}
\end{figure}

\subsubsection{Perturbative aspects of the hadron spectrum}

The spectra of heavy quarkonia have features that are remarkably similar to the Positronium spectrum (\fig{atoms}). The $J^{PC}=1^{--}$ charmonium state $J/\psi$ was dubbed the ``Hydrogen atom of QCD'' because it was thought to provide the key to the deciphering of hadron structure. The quarkonium spectra are in fact well described \cite{Eichten:2007qx} by the Schr\"odinger equation, provided a linear term is added to the single gluon exchange potential,
\beq\label{qcdpot}
V_{QED}(r) = -\frac{\alpha}{r} \hspace{3cm} V_{QCD}(r) = V'r-\frac{4}{3}\frac{\as}{r}
\eeq 
The fine and hyperfine splittings agree with the perturbative corrections to the Schr\"odinger result.
This hint from Nature merits serious consideration. It suggests that the Born approximation of atoms discussed in section \ref{posisec} is applicable also in QCD. The QCD action has no dimensionful parameter corresponding to the slope $V'$ of the linear potential in \eq{qcdpot}. In the absence of loops (renormalization) $V'$ can only arise via a {\it boundary condition} in the solution of the classical gauge field equations.

The linear potential leads to color confinement and strong binding regardless of the size of $\as$. The ionization threshold of QED becomes a threshold for open charm ($D, \Lambda_c, \ldots$) pair production in QCD. Charmonium resonances of high mass are indeed observed to decay mainly into open charm states. In recent years narrow ``XYZ'' states have been discovered, often close to particle production thresholds \cite{Chen:2016qju}. As in QED, narrow widths and closeness to thresholds indicates weak coupling. The new states may be characterized as ``hadron molecules'' \cite{Voloshin:1976ap}, but a quantitative description awaits a better understanding of hadron dynamics.

Hadrons composed of light quarks ($u,d,s$) have quantum numbers that allow them to be classified as $q\bar q$ and $qqq$ states with specific values of quark spins and angular momenta. The bound state masses and couplings can be qualitatively modelled \cite{Godfrey:1985xj,Selem:2006nd}. The success of the quark description encourages a perturbative approach, since the degrees of freedom of strongly interacting constituents generally are not reflected in the spectrum. For example, the spectrum of QED in $D=1+1$ dimensions (the ``massive Schwinger model'') has the features of $e^+e^-$ ``atomic'' bound states only at weak coupling, $e/m \ll 1$. When the interaction is strong ($e/m \gg 1$) loops dominate and the spectrum has the characteristics of weakly bound scalars \cite{Coleman:1976uz}.

The binding energies (or mass splittings) of light hadrons are of the same order as the hadron masses. Hence the internal dynamics must be ultrarelativistic, in stark contrast to atoms. One would expect to see the degrees of freedom also of gluon constituents in the hadron spectrum. Searches for ``hybrid'' and ``glueball'' states, some of which would stand out because of their quantum numbers, have not led to definite results. Physical gluons appear mainly as radiative corrections to hard interactions. The gluon distribution in the proton is large at high $Q^2$ and low $x$, but strongly scale-dependent. At $Q \sim 1$ GeV the gluon distribution is insignificant, whereas the sea quark contribution persists at low $x$ \cite{CooperSarkar:2009xz}.

\subsubsection{Duality \label{dualsec}}

\begin{figure}[h]
\includegraphics[width=.8\columnwidth]{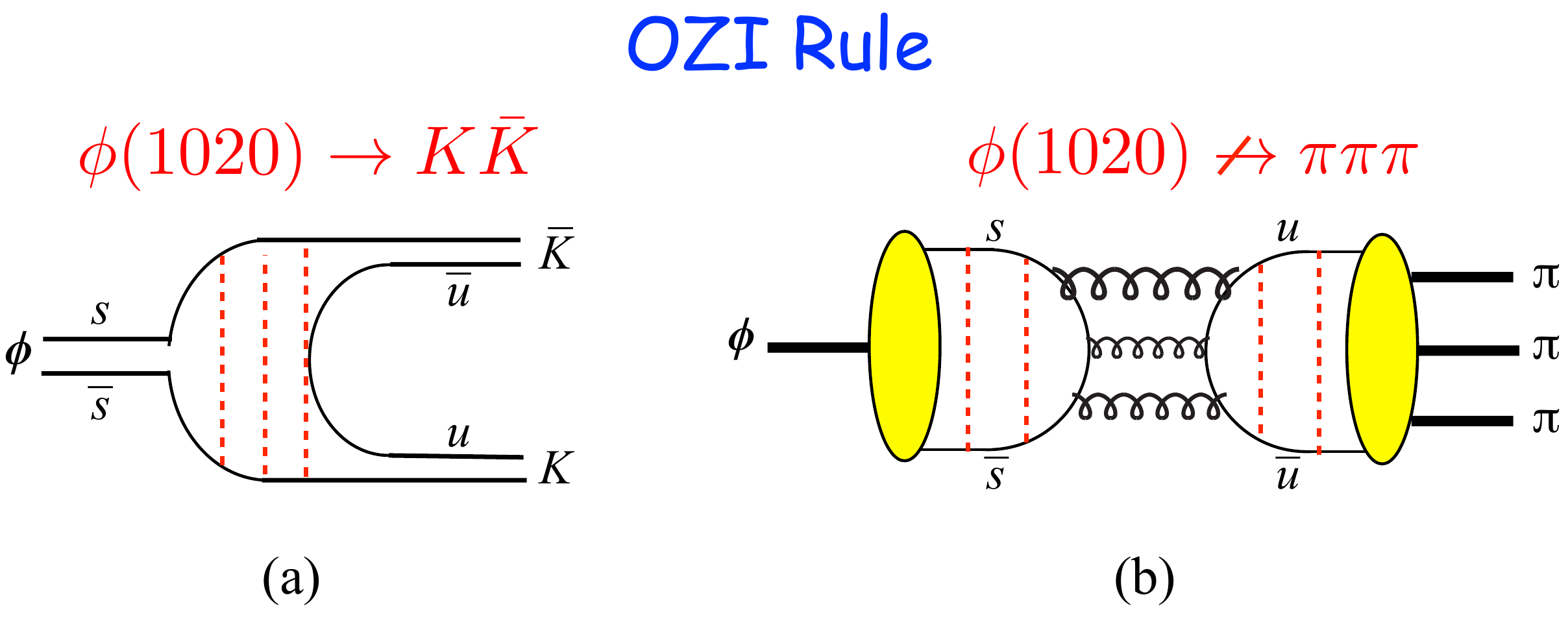}
\caption{According to the OZI rule, the connected quark diagram (a) dominates over the disconnected one in (b). In today's perspective the dashed lines in (a) represent the confining color field (the linear potential of \eq{qcdpot}). The $u\bar u$ pair is created by this field (``string breaking''). In (b) there is no confining field after the $s\bar s$ annihilates, and the transition can be mediated only by gluon exchange.
{\label{ozi}}}
\end{figure}

Hadrons couple selectively to each other. An extreme example is the X(3872) (molecular?) charmonium state, whose total width $\Gamma_{tot} < 1.2$ MeV is minimal despite apparently allowed, wide open decay channels. Light hadron dynamics is selective as well. Already at the dawn of the quark model Zweig \cite{Zweig:2015gpa} found it remarkable that the $\phi(1020)$ perfers to decay into $K\bar K$ (82\% of the time), even though kinematics favors $\pi\pi\pi$. \fig{ozi} illustrates Zweig's rule, which he formulated in terms of dual diagrams: The ``connected'' quark diagram (a) dominates the ``disconnected'' one in (b). This rule is applicable to other processes as well and today is known as the OZI rule, after Okuba, Zweig and Iizuka \cite{Okubo:1963fa}. A review of the status of the OZI rule may be found in \cite{Nomokonov:2002jb}. 

As indicated in \fig{ozi}(b) the $\phi(1020) \to \pi\pi\pi$ decay can be mediated by three (or more) gluons. The observed suppression of this process indicates that gluon exchange contributions are subleading in soft dynamics. This is consistent with the notion that the coupling $\as$ remains perturbative at low scales.

Quark diagrams like those of \fig{ozi}(a) illustrate the remarkable phenomenon of ``duality'' in hadron dynamics \cite{Harari:1981nn}. Phenomenological studies show that $2 \to 2$ hadron scattering amplitudes are described {\em either} by the resonances in the direct ($s$-) channel {\em or} by particle exchange in the crossed ($t$-) channel -- not by their sum. Resonances of different spins and masses must contribute in a coherent manner for them to mimic particle exchange. One aspect of this is that resonances are found to lie on approximately linear Regge trajectories \cite{Selem:2006nd}. A good illustration of duality is provided by the dual amplitude for $\pi^+\pi^- \to \pi^+\pi^-$ \cite{Lovelace:1969se}.

Another aspect of duality was observed by Bloom and Gilman in deep inelastic lepton scattering, $eN \to eX$ \cite{Bloom:1970xb}. Quite unexpectedly, the $x_{Bj}$-dependence of the $N^*$ contributions to the inclusive system $X$ agree (on average) with the quark distribution of the target proton.  A review of recent experimental results may be found in \cite{Niculescu:2015bla}. A further apparition is Local Parton Hadron Duality, which relates the perturbatively calculated inclusive {\it parton} production to the measured {\it hadron} distributions in high energy processes \cite{Dokshitzer:2003bt}. 

Duality is a pervasive and as yet poorly understood feature of hadron dynamics. It allows the average contribution of resonances to be described by a smoothly distributed scattering amplitude. A hint of how bound state and scattering features can mix is given by the Dirac bound states in D=1+1, discussed in section \ref{diracsubsec2}. The wave functions of hadrons found below have dual features such as indicated in \fig{dual}.

\subsubsection{$\as(Q^2)$ at low scales $Q$ \label{asfreeze}}

\begin{wrapfigure}[19]{r}{0.5\textwidth}
  \vspace{-30pt}
  \begin{center}
    \includegraphics[width=0.5\textwidth]{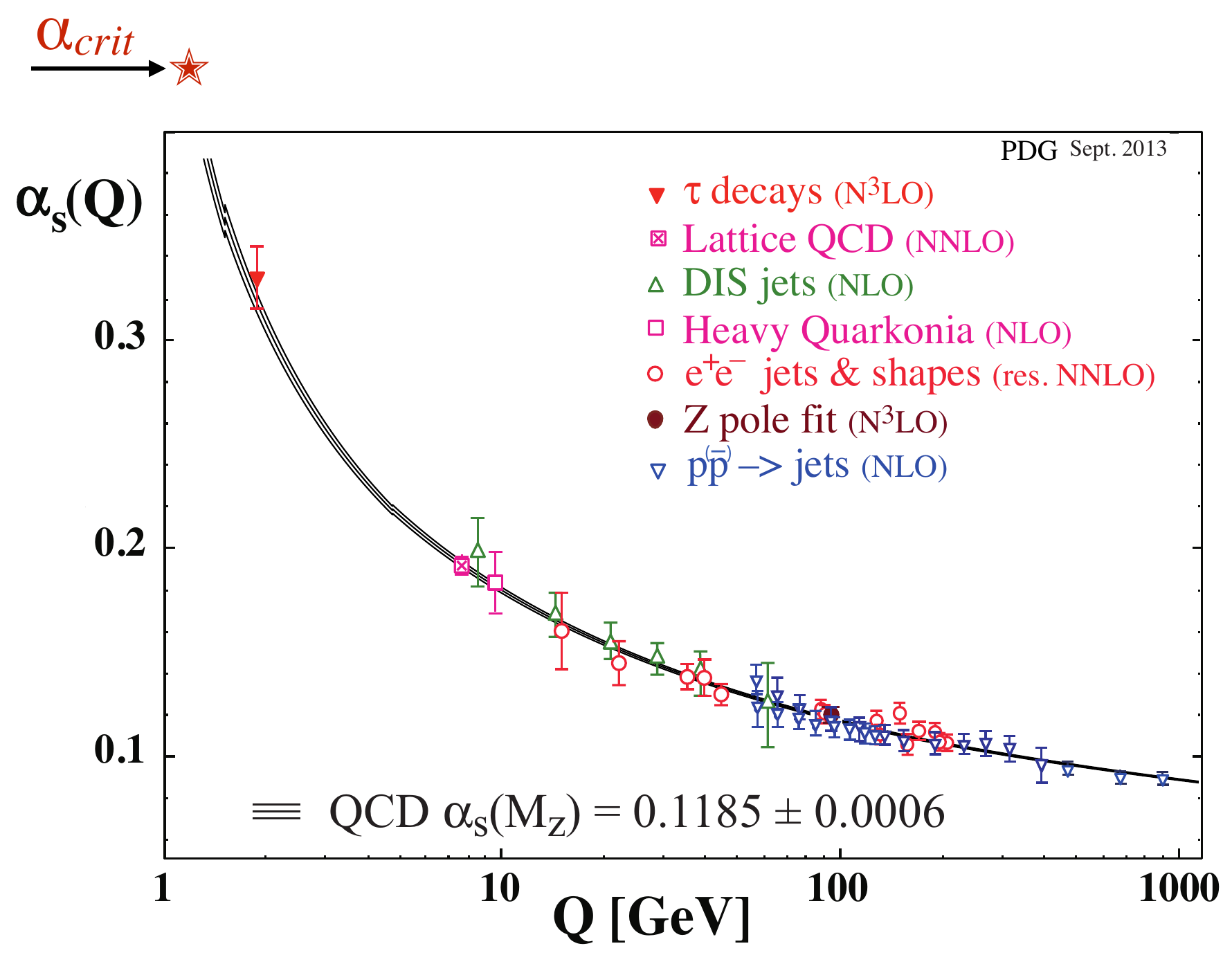}
  \end{center}
  \vspace{-15pt}
  \caption{The QCD coupling $\as$, as measured in several processes and calculated using lattice QCD \cite{Agashe:2014kda}. The point ``$\alpha_{crit}$'' refers to Eq. \eq{qcdcoup}.}
\label{alpha_s}
\end{wrapfigure}

A perturbative description of hadrons in QCD requires that the coupling $\as(Q^2)$ be of moderate size at low $Q$. 
Our knowledge \cite{Agashe:2014kda} of the perturbative coupling $\as$ is summarized in \fig{alpha_s}. At the lowest perturbative scale $\alpha_s^{\overline{MS}}(m_\tau^2) = 0.334\pm 0.014$ \cite{Bethke:2011tr}.

The coupling $\as$ is not directly measurable, and meaningful only in the context of a theoretical framework. Several analyses raise the possibility that $\as(Q^2)$ freezes around $Q \simeq 1$ GeV, at a value $\alpha_s^{\overline{MS}} \simeq 0.5$. In Gribov's picture of confinement \cite{Gribov:1999ui,Dokshitzer:2003bt} there is, in QCD as well as in QED, a critical value of the coupling where the vacuum changes character and new features emerge,
\beqa
\alpha^{crit}(QED)&=&\pi\left(1-\sqrt{\frac{2}{3}}\right) \simeq 0.58 \hspace{.5cm} \gg \inv{137}\nn \label{qedcoup}\\
\alpha_s^{crit}(QCD)&=& \frac{\pi}{C_F}\left(1-\sqrt{\frac{2}{3}}\right) \simeq 0.43 \hspace{.2cm} \gsim \as(m_\tau^2) \label{qcdcoup}
\eeqa

As indicated in \fig{alpha_s} the running QCD coupling reaches Gribov's critical value at $Q \simeq 1$ GeV. The value of the critical coupling is small enough that a an expansion in $\alpha_s^{crit}/\pi \simeq 0.14$ may be relevant.

\begin{figure}[h]
\includegraphics[width=1.\columnwidth]{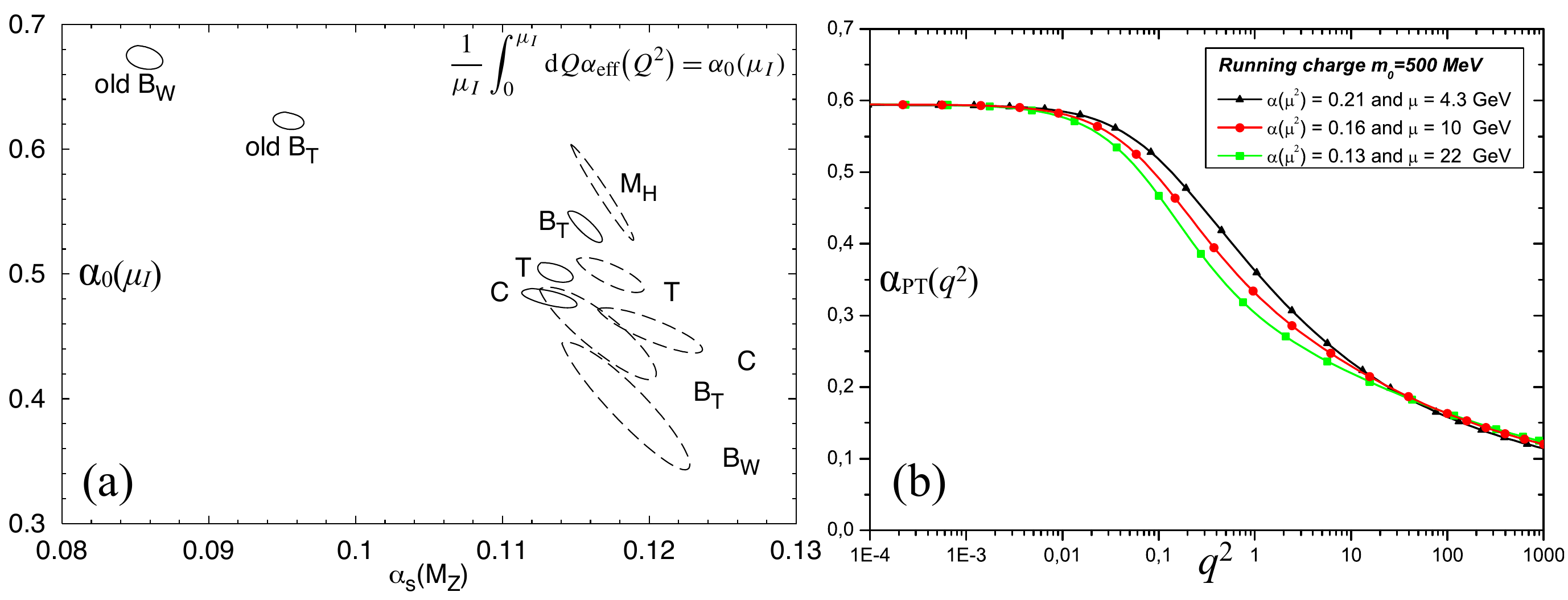}
\caption{(a) The average value of $\as$ for $Q \leq \mu_I = 2$ GeV based on several measures of event shapes \cite{Dokshitzer:1998qp}. (b) The effective coupling of the Pinch Technique, using Schwinger-Dyson equations \cite{Aguilar:2009nf}.
{\label{freeze}}}
\end{figure}

\fig{freeze}(a) shows estimates based on event shapes of the QCD coupling $\alpha_0(\mu_I)$, averaged over $0 \leq Q \leq \mu_I = 2$ GeV \cite{Dokshitzer:1998qp}. A similar analysis of event shapes in $e^+e^-$ annihilation using NNLO perturbative QCD gave $\alpha_0(2\ {\rm GeV}) = 0.5132 \pm 0.0115\,{\rm (exp)} \pm 0.0381\,{\rm (th)}$ \cite{Gehrmann:2009eh}. \fig{freeze}(b) shows the result of an analysis using the pinch technique \cite{Aguilar:2009nf}, in which the low scale coupling was estimated as $\alpha_{PT}(0)=0.7 \pm 0.3$. Further studies of the behaviour of the QCD coupling at low scales may be found in Refs. \cite{Brodsky:2002nb}.

The running of $\as$ originates from the renormalization of divergent loop integrals. The coupling does not run at the Born (no loop) level. If the Born approximation is adequate at low values of $Q$ the coupling necessarily freezes.

\section{QCD Bound States in the rest frame \label{qcdsec}}

\subsection{The meson and baryon states \label{statesubsubsec}}

Based on the previous study of Positronium \eq{bstate} and Dirac \eq{statedef} states I express the meson ($\mM$) and baryon ($\mB$) Born level states of QCD in the rest frame ($\Pv=0$) as
\beqa\label{mbstates}
\ket{\mM;M,\Pv=0} &=& \int d\xv_1\,d\xv_2\,\bar\psi_A(\xv_1)\Phi_\mM^{AB}(\xv_1-\xv_2)\psi_{B}(\xv_2)\ket{0} \label{meson}\\[2mm]
\ket{\mB;M,\Pv=0} &=& \int d\xv_1\,d\xv_2\,d\xv_3\,\psi_{A}^\dag(\xv_1)\psi_{B}^\dag(\xv_2)\psi_{C}^\dag(\xv_3) \Phi_{\mB}^{ABC}(\xv_1,\xv_2,\xv_3)\ket{0}\hspace{1cm} \label{baryon}
\eeqa
The states and fields are all evaluated at a common time $t$ (not shown). A sum over the repeated quark color indices ($A,B,C$) is understood, and Dirac indices suppressed. The baryon wave function needs to be translation invariant,
\beq\label{bartrans}
\Phi_{\mB}^{ABC}(\xv_1,\xv_2,\xv_3) = \Phi_{\mB}^{ABC}(\xv_1+\bs{a},\xv_2+\bs{a},\xv_3+\bs{a})
\eeq
for the state to carry momentum $\Pv=0$. 

Similarly as for Positronium, there are no gauge links connecting the fields. The meson state is invariant under a local gauge transformation $U(\xv)$ provided the wave function transforms accordingly,
\beq\label{gaugetransform}
\psi(\xv) \to U(\xv)\psi(\xv) \hspace{2cm} \Phi_{\mM}(\xv_1-\xv_2) \to U^\dag(\xv_1)\Phi_\mM(\xv_1-\xv_2)U(\xv_2)
\eeq
The baryon wave function $\Phi_{\mB}$ is similarly gauge dependent. The usual ``color singlet'' wave functions,
\beqa
\Phi_\mM^{AB}(\xv_1-\xv_2) &=& \inv{\sqrt{N_C}}\delta^{AB}\Phi_\mM(\xv_1-\xv_2) \label{mcolor} \\[-1mm]
&&\hspace{6cm} (N_C=3) \nn\\[-1mm]
\Phi_\mB^{ABC}(\xv_1,\xv_2,\xv_3) &=& \inv{\sqrt{N_C!}}\epsilon^{ABC}\Phi_\mB(\xv_1,\xv_2,\xv_3) \label{bcolor}
\eeqa
are in fact invariant only under global gauge transformations.

\subsection{Symmetries: Poincar\'e, Parity and Charge Conjugation for Mesons \label{symsubsubsec}}

The QCD action is invariant under Poincar\'e transformations (time and space translations, rotations and boosts) as well as under the discrete parity and charge conjugation transformations. The ``kinetic'' transformations (space translations, rotations, parity and charge conjugation) do not change the time coordinate of the field operators. 
The transformation of the states under kinetic transformations is explicitly determined by the representation of the corresponding subgroup to which the wave function belongs. 

Time translations are generated by the Hamiltonian, which includes the interaction terms of the action. Consequently this is called a ``dynamic'' transformation. In order that a state be an eigenstate of the Hamiltonian its wave function should satisfy the bound state equation (section \ref{bsesubsec}). Also boosts are dynamical, and moreover change the definition of simultaneity. Equal-time states in one frame are non-trivially related to equal-time states in another frame. In section \ref{scheq} we found the frame dependence of the Positronium wave function. In section \ref{framesec} I discuss how relativistic, Born level bound states transform under boosts.

\subsubsection{Space translations \label{spacetrans}}

Under space translations $\xv \to \xv+\bs{a}$ the quark fields are transformed by the operator
\beq\label{transop}
U(\bs{a}) = \exp[-i\bs{a}\cdot\bs{\mP}] \hspace{1cm} {\rm where} \hspace{1cm} \bs{\mP} = \int d\xv\,\psi^\dag(\xv)(-i\nv)\psi(\xv)
\eeq 
(I suppress the common time $t$ in all operators). The momentum operator satisfies
\beq\label{momcom}
\com{\bs{\mP}}{\psi(\xv)} = i\nv \psi(\xv)  \hspace{2cm}  \com{\bs{\mP}}{\bar\psi(\xv)} = i\nv \bar\psi(\xv)
\eeq
For finite translations,
\beq
U(\bs{a})\psi(\xv)U^\dag(\bs{a})=\psi(\xv+\bs{a})
\eeq
Using \eq{momcom} and $\bs{\mP}\ket{0}=0$ we may verify that the state \eq{meson} is indeed at rest,
\beq\label{restcheck}
\bs{\mP}\ket{\mM;M,\Pv=0} =0
\eeq

\subsubsection{Rotations \label{rotations}}

A rotation $\xv \to R\,\xv$, where $R$ is an orthogonal $3\times 3$ matrix, should be represented by an operator $U(R)$ which transforms the quark momentum states accordingly,
\beq\label{rotmom}
U(R)b(\pv,\lambda)U^\dag(R) = b(R\pv,\lambda) \hspace{2cm} U(R)d(\pv,\lambda)U^\dag(R) = d(R\pv,\lambda)
\eeq
where the quantization axis for the spin component $\lambda$ is understood to be rotated. I next verify that this implies
\beq\label{Udef}
U(R)\psi(\xv)U^\dag(R) = S^{-1}(R)\psi(R\xv)
\eeq
For a rotation by an angle $\theta$ around the unit vector $\hat{\bs{n}}$ the Dirac matrix $S(R)$ is 
\beq\label{Sdef}
S(R) = \exp\big(i\halft\theta\,\hat{\bs{n}}\cdot\bs{\Sigma}\big) \hspace{1cm} {\rm where}  \hspace{1cm}  \bs{\Sigma} = \gamma_5\gz\gv = 
\left(\begin{array}{cc} \bs{\sigma} & 0 \\ 0 & \bs{\sigma} \end{array} \right)
\eeq
$S(R)$ has the property
\beq\label{Sgamrot}
S(R)\,\gv\,S^{-1}(R) = R^{-1}\gv
\eeq
which transforms the $u$-spinor,
\beq\label{urot}
S(R)u(\pv,\lambda) = S(R)\frac{E\gz-\pv\cdot\gv+m}{\sqrt{E+m}}
\left(\begin{array}{c} \vphi_\lambda \\ 0  \end{array} \right) = 
\frac{E\gz-\pv\cdot R^{-1}\gv+m}{\sqrt{E+m}} S(R)
\left(\begin{array}{c} \vphi_\lambda \\ 0  \end{array}\right) \equiv u(R\pv,\lambda)
\eeq
In the last step I used $\pv\cdot R^{-1}\gv = R\,\pv\cdot \gv$ and defined the rotated $u$-spinor by
\beq\label{defrotspinor}
u(R\pv,\lambda) = \frac{E\gz-R\pv\cdot\gv+m}{\sqrt{E+m}}
\left(\begin{array}{c} \exp\big(i\halft\theta\,\hat{\bs{n}}\cdot\bs{\sigma}\big)\vphi_\lambda \\ 0  \end{array} \right)
\eeq
The $v$-spinor is rotated in the same way since $\bs{\Sigma}$ in \eq{Sdef} is block diagonal. Using the expression \eq{eop} for the field,
\beq\label{fieldrotcheck}
S(R)U(R)\psi(\xv)U^\dag(R) = \int\frac{d\pv}{(2\pi)^3 2E}\sum_\lambda\Big[u(R\,\pv,\lambda)e^{iR\pv\cdot R\xv}b(R\,\pv,\lambda)+v(R\,\pv,\lambda)e^{-iR\pv\cdot R\xv}d^\dag(R\,\pv,\lambda)\Big] =\psi(R\,\xv)
\eeq
In the last step I changed the integration measure $d\pv = d(R\,\pv)$.
Rotations are generated by the angular momentum operator $\bs{\mJ}$,
\beq\label{rotop}
U(R) = \exp[i\theta\,\hat{\bs{n}}\cdot\bs{\mJ}]
\eeq 
given by
\beq\label{rotgen}
\bs{\mJ} = \int d\xv\,\psi^\dag(\xv)\,\bs{J}\,\psi(\xv) \hspace{3cm} \bs{J} = \xv\times(-i\nv)+\halft\bs{\Sigma}
\eeq
Using
\beq\label{angmomcom}
\com{\bs{\mJ}}{\psi(\xv)}=-\big[\xv\times(-i\rnab)+\halft\bs{\Sigma}\big]\psi(\xv) \hspace{2cm} 
\com{\bs{\mJ}}{\bar\psi(\xv)}= \bar\psi(\xv)\big[\xv\times(i\lnab)+\halft\bs{\Sigma}\big]
\eeq
and $\bs{\mJ}\ket{0}=0$ we can determine the condition that the meson state \eq{meson} is an eigenstate of some component of the angular momentum operator $\bs{\mJ}$,
\beqa\label{angstate}
\bs{\mJ}\ket{\mM;M,\Pv=0} &=& \int d\xv_1\,d\xv_2\,\bar\psi(\xv_1)\Big\{\big[\xv_1\times(-i\rnab_1)+\halft\bs{\Sigma}\big]\Phi_\mM(\xv_1-\xv_2)  \nn\\[2mm]
&& \hspace{2cm} -\Phi_\mM(\xv_1-\xv_2)\big[\xv_2\times(i\lnab_2)+\halft\bs{\Sigma}\big]\Big\}\psi(\xv_2)\ket{0}
\eeqa
Thus
\beq\label{angwf}
\mJ^z\ket{\mM;M,\Pv=0}=\lambda\ket{\mM;M,\Pv=0}\hspace{1cm} {\rm provided}
\hspace{1cm} \com{J^z}{\Phi_\mM(\xv)} = \lambda\,\Phi_\mM(\xv)
\eeq

A finite rotation transforms the meson state as
\begin{align}\label{finiterot}
U(R)\ket{\mM;M,\Pv=0} &=\int d\xv_1\,d\xv_2\,\bar\psi(R\xv_1)S(R)\Phi_\mM(\xv_1-\xv_2)S^{-1}(R)\psi(R\xv_2)\ket{0} \nn\\[2mm]
&= \int d\xv_1\,d\xv_2\,\bar\psi(\xv_1)S(R)\Phi_\mM\big[R^{-1}(\xv_1-\xv_2)\big]S^{-1}(R)\psi(\xv_2)\ket{0}
\end{align}
The wave function in the rotated coordinate system is thus given by
\beq\label{rotwf}
\Phi_\mM^{(R)}(\xv) = S(R)\Phi_\mM\big(R^{-1}\xv\big)S^{-1}(R)
\eeq

\subsubsection{Parity \label{paritytrans}}

The parity operator $\bP$ reverses 3-momenta $\pv$ but leaves spins $\lambda$ invariant:
\beq\label{parop}
\bP b(\pv,\lambda)\bP^\dag = b(-\pv,\lambda) \hspace{2cm} \bP d(\pv,\lambda)\bP^\dag = -d(-\pv,\lambda)
\eeq
We could add an ``intrinsic'' parity for the quarks, but it is irrelevant for $q\bar q$ states so I omit it. The {\em relative} intrinsic parity $-1$ of quarks and antiquarks in \eq{parop} ensures that the field transforms as
\beq\label{parop1}
\bP \psi(t,\xv)\bP^\dag = \gz\psi(t,-\xv) \hspace{2cm} \bP \bar\psi(t,\xv)\bP^\dag = \bar\psi(t,-\xv)\gz
\eeq
That in turn allows to consider the condition for the (rest frame) meson states to be eigenstates of parity,
\beq\label{parop2}
\bP\ket{\mM;M,\Pv=0} = \int d\xv_1\,d\xv_2\,\bar\psi(\xv_1)\gz\Phi_\mM(-\xv_1+ \xv_2)\gz \psi(\xv_2)\ket{0} = \eta_P \ket{\mM;M,\Pv=0}
\eeq
The wave function of a parity eigenstate with eigenvalue $\eta_P = \pm 1$ satisfies
\beq\label{parop3}
\gz\Phi_\mM(-\xv)\gz = \eta_P \Phi_\mM(\xv)
\eeq

\subsubsection{Charge conjugation \label{chargeconj}}

The charge conjugation operator $\bC$ transforms particles into antiparticles,
\beq\label{ccop}
\bC b(\pv,\lambda)\bC^\dag = d(\pv,\lambda) \hspace{2cm} \bC d(\pv,\lambda)\bC^\dag = b(\pv,\lambda)
\eeq
In the standard Dirac matrix representation this implies ($T$ indicates transpose)
\beq\label{ccop1}
\bC \psi(t,\xv)\bC^\dag = i\gz\gamma^2\bar\psi^T(t,\xv) \hspace{2cm} \bC \bar\psi(t,\xv)\bC^\dag = i\psi^T(t,\xv)\gz\gamma^2
\eeq
For a meson state to be an eigenstate of charge conjugation with eigenvalue $\eta_C=\pm 1$,
\beq\label{ccop2}
\bC\ket{\mM;M,\Pv=0} = \int d\xv_1\,d\xv_2\,\bar\psi(\xv_1)\gz\gamma^2\Phi_\mM^T(\xv_2- \xv_1)\gz\gamma^2 \psi(\xv_2)\ket{0} = \eta_C \ket{\mM;M,\Pv=0}
\eeq
its wave function should satisfy
\beq\label{ccop3}
\gz\gamma^2\Phi_\mM^T(-\xv)\gz\gamma^2 = \eta_C \Phi_\mM(\xv)
\eeq

\subsection{The confining field $\mA^\mu$ \label{confsubsec}}

\subsubsection{Requirements on a viable solution \label{confsubsubsec}}

I now consider whether there is a classical field $\mA^\mu$ which could be relevant for hadrons. Recall that in Positronium the classical field \eq{qedfield} depends on the positions $\xv_1,\xv_2$ of the $e^-$ and $e^+$. In QCD the field will also depend on the quark colors $A,B,C$. The total color field generated by the hadron (which would be felt by an external color probe, excluding \order{\as} gluon exchange) is the coherent sum of the fields of all quark color components. The total field vanishes for the meson and baryon states, see \eq{mfieldsum} and \eq{bfieldsum3} below. Hence the confining field $\mA^\mu$ referred to below concerns only the field between quarks with specific colors.

To make my arguments concerning the confining field transparent I list them as separate items. The outcome is a field with a single dimensionful parameter $\la$. Possibly this is the only acceptable scenario.

\begin{enumerate}

\item[\#1.] There is no dimensionful constant $\lqcd$ in the classical field equations. It can only appear in their solution via a boundary condition.

\item[\#2.] Rotational symmetry requires that $\mA^\mu$ depend on the positions of the quarks, \cf\ the QED case \eq{qedfield}.

\item[\#3.] The quark positions determine $\mA^0$ for all $\xv$ at an instant of time.  The propagating components $\bs{\mA}$ respond to the quark positions with a delay, which would be infinite at the spatial boundary, $|\xv| \to \infty$.

\item[\#4.] I choose $\bs{\mA}=0$ in the hadron rest frame ($\Pv=0$), in the gauge with wave functions \eq{mcolor}, \eq{bcolor}. The field felt by a moving hadron ($\Pv\neq 0$) is then given by a Lorentz boost of $\mA^0$ as in \eq{Aboost}.

\item[\#5.] Each quark interacts with the confining field independently of the others. The color structures \eq{mcolor}, \eq{bcolor} are maintained only if $\mA_a^0 T^{AB}_a \propto \delta^{AB}$. Hence $\mA_a^0 \neq 0$ only for $a=3,8$, in the standard notation where
\beq\label{colgen}
T_3=\inv{2}\left(\begin{array}{ccc} 1 & 0 & 0 \\ 0 & -1 & 0 \\ 0 & 0 & 0 \end{array} \right) \hspace{3cm}
T_8=\inv{2\sqrt{3}}\left(\begin{array}{ccc} 1 & 0 & 0 \\ 0 & 1 & 0 \\ 0 & 0 & -2 \end{array} \right)
\eeq

\item[\#6.] In the rest frame the confining field should be a homogeneous solution of Gauss' law,
\beq\label{glawqcd}
\nv^2 \mA_a^0=0 \hspace{1cm} (a=3,8)
\eeq

\item[\#7.] The field energy density is, including a convenient normalization factor,
\beq\label{confen}
\sum_{a=3,8} \big[\nv \mA_a^0(\xv)\big]\cdot \big[\nv \mA_a^0(\xv)\big] \equiv \la^4_3+\la^4_8 \equiv 12\la^4
\eeq
Translation invariance requires that $\la$ be $\xv$-independent,  $\la \neq \la(\xv)$.

\item[\#8.] The infinite field energy $\int d\xv\,\sum_{a=3,8} \big[\nv \mA_a^0(\xv)\big]^2 = 12\la^4\int d\xv$ must be universal. Hence $\la$ is independent of the quark positions and colors.

\item[\#9.] The simplest field $\mA_a^0$ which satisfies the above requirements is linear in $\xv$. Color and Poincar\'e invariance imposes a scalar product between $\xv$ and a vector formed from the quark positions $\xv_i$ weighted by their color charges. This implies a specific classical field for each quark component of a hadron with the color structure of \eq{mcolor}, \eq{bcolor} (no sum on colors):
\beqa
\bar\psi_A(\xv_1)\psi_{A}(\xv_2)\ket{0}:&& \mA_a^0(\xv)=\xv\cdot\frac{T_a^{AA}\,(\xv_1-\xv_2)}{|T_a^{AA}\,(\xv_1-\xv_2)|}\,\la_a^2(\xv_i,A) \label{mfield}\\
&& \hspace{8cm} (a=3,8) \nn\\
\psi_A^\dag(\xv_1)\psi_B^\dag(\xv_2)\psi_C^\dag(\xv_3)\ket{0}:&& \mA_a^0(\xv)=\xv\cdot\frac{T_a^{AA}\,\xv_1+T_a^{BB}\,\xv_2+T_a^{CC}\,\xv_3}{|T_a^{AA}\,\xv_1+T_a^{BB}\,\xv_2+T_a^{CC}\,\xv_3\big|}\,\la_a^2(\xv_i,A,B,C) \label{bfield}
\eeqa
As indicated, $\la_3$ and $\la_8$ may depend on the quark positions and colors, but not on $\xv$. \eq{confen} requires that their 4th powers sum to the universal scale $12\la^4$. 

\item[\#10.] The ratio $\la_3/\la_8$ is determined by stationarity of the field energy density \eq{confen}.  

\end{enumerate}

The parameter $\la$ \eq{confen} is of \order{g^0} and determines the energy scale through the strength of the confining field $\mA^0_a$. The quark interactions with the confining field are $\propto g\mA^0_a$, \ie, of \order{g\la^2}.

\subsubsection{Field energy of the meson component \eq{mfield} \label{mfieldsubsubsec}}

The field energy $\int d\xv\,\halft F_{j0}F^{j0}$ has a universal \order{\la^4} contribution (\#8) and an \order{g\la^2} contribution from the interference of the confining field $\mA^0_a$ with the standard Coulomb field $A^0_a$ due to the quark color charges. The field energy  needs to be evaluated to the same \order{g\la^2} as the quark interactions.

The standard Coulomb field $A^0$ generated by the $q_A(\xv_1)\bar q_A(\xv_2)$ component of the meson state (no sum on color $A$) is an inhomogeneous solution of Gauss' law,
\beq\label{glawm}
-\nv^2 A_a^0(\xv) = g T_a^{AA} \big[\delta^3(\xv-\xv_1)-\delta^3(\xv-\xv_2)\big]
\eeq
When the quarks move relativistically they will generate also a vector field $\Av_a$. The vector field does not interfere with $\mA_a^0$ and so contributes only at \order{g^2}. To \order{g\la^2},
\beq\label{fenm1}
E_\mM^{AA}(\xv_1,\xv_2) = -\halft\sum_{a=3,8}\int d\xv\,\nv (\mA_a^0+A_a^0) \cdot \nv (\mA_a^0+A_a^0) \to \halft\sum_{a=3,8}\int d\xv\,\mA_a^0\,\nv^2 A^0_a
\eeq
In the second step I omitted the \order{\la^4} and \order{g^2} contributions, as well as the insertion at spatial infinity from the partial integration. The integral of the \order{g\la^2} contribution is marginally convergent, its definition affects the coefficient of the result. This scheme dependence is absorbed in the definition of $\la$. From \eq{mfield} and \eq{glawm} we have
\beq\label{fenm2}
E_\mM^{AA}(\xv_1,\xv_2) = -\halft\sum_{a=3,8}gT_a^{AA}\big[\mA_a^0(\xv_1)-\mA_a^0(\xv_2)\big] = -\halft\sum_{a=3,8}g\la_a^2\,|T_a^{AA}(\xv_1-\xv_2)|
\equiv E_\mM(\xv_1,\xv_2)
\eeq
Let us verify that the energy is independent of the quark color $A=1,2,3$. Since $\la_3^4+\la_8^4 = 12\la^4$ I parametrize 
\beq\label{betapar}
\la_3^2=2\sqrt{3}\,\la^2 \sin\beta \hspace{2cm} \la_8^2=2\sqrt{3}\,\la^2 \cos\beta
\eeq

\vspace{.5cm}
\underline{A=1}:\ \ 
\beq\label{fenm11a}
E_\mM^{11}(\xv_1,\xv_2) = -\quart g\Big(\la_3^2+\inv{\sqrt{3}}\la_8^2\Big)|\xv_1-\xv_2| = -\frac{\sqrt{3}}{2} g\la^2\Big(\sin\beta+\inv{\sqrt{3}}\cos\beta\Big)|\xv_1-\xv_2|
\eeq
From $dE_\mM^{11}/d\beta=0$ (\#10 above) follows $\sin\beta=\sqrt{3}/2$ and $\cos\beta=1/2$. Thus
\beq\label{fenm11b}
E_\mM^{11}(\xv_1,\xv_2) = - g\la^2|\xv_1-\xv_2| = E_\mM(\xv_1,\xv_2)
\eeq

\vspace{.5cm}
\underline{A=2}:\ \ $E_\mM^{22}=E_\mM$ follows trivially since $|T_a^{11}|=|T_a^{22}|$ implies the same values for $\la_3^2$ and $\la_8^2$ as for $A=1$.

\vspace{.5cm}
\underline{A=3}:\ \ Since $T_3^{33}=0$ stationarity gives $\la_3^2=0$ and $\la_8^2=2\sqrt{3}\,\la^2$,
\beq\label{fenm11c}
E_\mM^{33}(\xv_1,\xv_2) = -g\la^2|\xv_1-\xv_2| = E_\mM(\xv_1,\xv_2)
\eeq
This completes the demonstration that the meson field energy is independent of the quark color $A$.

An external probe at $\xv$ would sense the confinement fields generated by the coherent sum over of all quark colors $A=B$ in the meson state \eq{meson}, with each color equally weighted due to the singlet wave function \eq{mcolor}. Let us verify that a color singlet meson does not generate a color field for any quark positions $\xv_1,\xv_2$. From \eq{mfield},
\beq\label{mfieldsum}
\sum_{A=1,2,3} \mA_{a}^0(\xv;\xv_1,\xv_2,A) = \frac{\xv\cdot(\xv_1-\xv_2)}{|\xv_1-\xv_2|}\sum_A \frac{T_a^{AA}}{|T_a^{AA}|}\,\la_a^2(\xv_i,A)=0
\eeq
The sum vanishes for $a=3$ since $T_3^{11}=-T_3^{22}$ and $\la_3^2(A=1)=\la_3^2(A=2),\ \la_3^2(A=3)=0$. For $a=8$ we had $\la_8^2(A=1,2)= \sqrt{3}\,\la^2$ while $\la_8^2(A=3)= 2\sqrt{3}\,\la^2$. Weighting with the signs of $T_8^{AA}$ makes \eq{mfieldsum} vanish also for $a=8$.

\subsubsection{Field energy of the baryon component \eq{bfield} \label{bfieldsubsubsec}}

The color singlet baryon wave function \eq{bcolor} assigns the quark colors $1,2,3$ to quarks at $\xv_1,\xv_2,\xv_3$ in all permutations, with even and odd permutations having a opposite signs. Consider first the case where $A,B,C$ in \eq{bfield} is $1,2,3$. Gauss' law is
\beq\label{glawb}
-\nv^2 A_a^0 = g \big[T_a^{11}\delta^3(\xv-\xv_1)+T_a^{22}\delta^3(\xv-\xv_2)+T_a^{33}\delta^3(\xv-\xv_3)\big]
\eeq
As for mesons, the field energy of \order{g\la^2} arises from the interference between the confining field $\mA_a^0$ of \eq{bfield} and the inhomogeneous solution $A_a^0$ of \eq{glawb},
\beqa\label{fenb1}
E_\mB^{123}(\xv_1,\xv_2,\xv_3) &=& \halft\sum_{a=3,8}\int d\xv\,\mA_a^0\,\nv^2 A^0_a = -\halft\sum_{a=3,8}g \big[T_a^{11}\mA_a^0(\xv_1)+T_a^{22}\mA_a^0(\xv_2)+T_a^{33}\mA_a^0(\xv_3)\big] \nn \\[2mm]
&=& -\halft\sum_{a=3,8}g \big|T_a^{11}\xv_1+T_a^{22}\xv_2+T_a^{33}\xv_3\big|\, \la_a^2(\xv_i,1,2,3) \equiv E_\mB(\xv_1,\xv_2,\xv_3)
\eeqa
Let us verify that the \order{g\la^2} field energy is independent of the color permutation. With the parametrization \eq{betapar},
\beq\label{fenb2}
E_\mB^{123}(\xv_1,\xv_2,\xv_3) = -\sqrt{3}\, g\la^2 \Big[\halft\sin\beta\,|\xv_1-\xv_2|+ \sfrac{1}{2\sqrt{3}}\cos\beta\,|\xv_1+\xv_2-2\xv_3|\Big]
\eeq
The stationarity condition $dE_\mB^{123}/d\beta=0$ gives
\beq\label{betab}
\sin\beta= \frac{\sqrt{3}\,|\xv_1-\xv_2|}{2d(\xv_1,\xv_2,\xv_3)} \hspace{2cm} 
\cos\beta= \frac{|\xv_1+\xv_2-2\xv_3|}{2d(\xv_1,\xv_2,\xv_3)} 
\eeq
where
\beq\label{ddef}
d(\xv_1,\xv_2,\xv_3) \equiv \inv{\sqrt{2}}\sqrt{(\xv_1-\xv_2)^2+(\xv_2-\xv_3)^2+(\xv_3-\xv_1)^2}
\eeq
Using this in \eq{fenb2} gives
\beq\label{fenb3}
E_\mB^{123}(\xv_1,\xv_2,\xv_3) = -\,g\la^2\,d(\xv_1,\xv_2,\xv_3)
\eeq
The field components \eq{bfield} with any other permutation of the colors $A,B,C$ will give the same field energy since $d(\xv_1,\xv_2,\xv_3)$ is independent of the permutations of $\xv_1,\xv_2,\xv_3$. This justifies the definition of $E_\mB$ in \eq{fenb1}.

It remains to consider the total color field of a baryon, given by the sum over colors $A,B,C$ in \eq{bfield} weighted by the color dependence $\veps^{ABC}$ of the wave function \eq{bcolor}. As for mesons, we may keep the quark positions $\xv_1,\xv_2,\xv_3$ fixed. For $a=3$ and $A,B,C=1,2,3$ we have using \eq{betab},
\beq\label{bfield3}
\mA_{a=3}^0(\xv;123)=\frac{\xv\cdot(\xv_1-\xv_2)}{|\xv_1-\xv_2|}\,\frac{\sqrt{3}\,|\xv_1-\xv_2|}{2d(\xv_1,\xv_2,\xv_3)}\,2\sqrt{3}\,\la^2 =3\la^2\,\frac{\xv\cdot(\xv_1-\xv_2)}{d(\xv_1,\xv_2,\xv_3)}
\eeq
Since $T_3^{11}=-T_3^{22}$ and $\veps^{123}=-\veps^{213}$ we have $\mA_{a=3}^0(\xv;123) = \mA_{a=3}^0(\xv;213)$. The switch $1 \leftrightarrow 2$ will similarly not change the color field for any other permutation of $1,2,3$. Multiplying by a factor 2 we may thus restrict the sum to the even permutations,
\beq\label{bfieldsum1}
\sum_{A,B,C}\veps^{ABC}\mA_{a=3}^0(\xv;ABC)= \frac{6\la^2}{d(\xv_1,\xv_2,\xv_3)}\,\xv\cdot\big[(\xv_1-\xv_2)+(\xv_2-\xv_3)+(\xv_3-\xv_1)\big] = 0
\eeq

For $a=8$ and $A,B,C=1,2,3$ we have
\beq\label{bfield8}
\mA_{a=8}^0(\xv;123)=\frac{\xv\cdot(\xv_1+\xv_2-2\xv_3)}{|\xv_1+\xv_2-2\xv_3|}\,\frac{|\xv_1+\xv_2-2\xv_3|}{2d(\xv_1,\xv_2,\xv_3)}\,2\sqrt{3}\,\la^2 =\sqrt{3}\,\la^2\,\frac{\xv\cdot(\xv_1+\xv_2-2\xv_3)}{d(\xv_1,\xv_2,\xv_3)}
\eeq
Since $T_8^{11}=T_8^{22}$ and $\veps^{123}=-\veps^{213}$ the sum $\mA_{a=8}^0(\xv;123) +\mA_{a=8}^0(\xv;213) =0$. The two amplitudes related by a switch $1 \leftrightarrow 2$ will similarly cancel each other for any other permutation of $1,2,3$. Thus an external color probe will not detect the baryon via its confining field,
\beq\label{bfieldsum3}
\sum_{A,B,C}\veps^{ABC}\mA_{a}^0(\xv;ABC)=  0
\eeq

The vanishing of the total confining field $\mA_{a}^0$ for mesons and baryons means that this field does not transmit interactions between hadrons, only between quarks of specific colors in the same hadron. This is a non-abelian feature. In a $U(1)$ gauge theory the confinement field corresponding to \eq{mfield} would be
\beq\label{u1field}
\mA^0(\xv;\xv_1,\xv_2) = \frac{\xv\cdot(\xv_1-\xv_2)}{|\xv_1-\xv_2|}\,\la^2
\eeq
An external observer would see an electric field of magnitude $\la^2$, aligned with $\xv_1-\xv_2$. The sum over $\xv_1,\xv_2$ weighted by the wave function need not vanish, giving rise to long-range correlations of undiminished strength. The vanishing of the field in non-abelian theories for each $\xv_1,\xv_2$ avoids this feature.

Interactions between hadrons can occur through annihilations between a quark in one hadron and an antiquark in another. This is related to string breaking as in \fig{ozi}(a). There is also \order{\as} perturbative gluon exchange which is not considered here.

\subsection{The bound state equations ($\Pv=0$) \label{bsesubsec}}

\subsubsection{The meson confining potential \label{mpotsubsubsec}}

The meson state \eq{meson} should be an eigenstate of the QCD Hamiltonian,
\beq\label{hmeson}
H_{QCD}(\mA)\ket{\mM;M,\Pv=0} = M\ket{\mM;M,\Pv=0}
\eeq
At \order{\alpha_s^0} (more precisely, \order{g\la^2}) only the classical confining field $\mA$ determined in section \ref{confsubsubsec} appears in the Hamiltonian. Since the field was determined for a system at rest we have $\Pv=0$ and $E=M$ in \eq{hmeson}. The Hamiltonian is assumed to annihilate the vacuum, $H_{QCD}\ket{0}=0$, since the color summed confining field vanishes for each hadron. String breaking effects before color summation is treated iteratively, as discussed in sections \ref{tree} and \ref{discsubsec6}. Quark and gluon production is possible via perturbative contributions which are not considered here. 

The classical field $\mA_a^0$ is specific for each color and spatial configuration of the quarks. The quark interaction term in $H_{QCD}$ gives for the component \eq{mfield} of the meson state (no sum on $A$)
\beqa\label{qintm}
\Big[\int d\xv\sum_{a,C,D}\psi_C^\dag(\xv)g\mA_a^0(\xv)T^{CD}_a \psi_D(\xv)\Big]
\bar\psi_A(\xv_1)\psi_{A}(\xv_2)\ket{0} && \nn\\[2mm] && \hspace{-8cm}
=\Big[\sum_{a,C}\bar\psi_C(\xv_1)g\mA_a^0(\xv_1)T^{CA}_a \psi_A(\xv_2)
-\sum_{a,D}\bar\psi_A(\xv_1)g\mA_a^0(\xv_2)T^{AD}_a \psi_D(\xv_2)\Big]\ket{0} \nn\\[2mm] && \hspace{-8cm}
=\sum_{a=3,8} gT_a^{AA}\big[\mA_a^0(\xv_1)-\mA_a^0(\xv_2)\big]\bar\psi_A(\xv_1)\psi_{A}(\xv_2)\ket{0} = -2E_\mM(\xv_1,\xv_2) \bar\psi_A(\xv_1)\psi_{A}(\xv_2)\ket{0}
\eeqa
In the final equality I noted that the dependence on the confining field is the same as in \eq{fenm2}. This combination was found to be independent of the quark color $A$, and given by \eq{fenm11b}.

Since $\bar\psi_A(\xv_1)\psi_{A}(\xv_2)\ket{0}$ is an eigenstate of the quark interactions this is also the case for the meson potential, obtained by adding the field energy $E_\mM$, 
\beq\label{mpot}
V_\mM(\xv_1-\xv_2) = -2E_\mM(\xv_1,\xv_2)+E_\mM(\xv_1,\xv_2)=g\la^2\,|\xv_1-\xv_2|
\eeq
The fact that the potential is linear is obviously welcome from a phenomenological point of view. It appears to be the only alternative in the framework I am describing here.

\subsubsection{The baryon confining potential \label{bpotsubsubsec}}

The quark interaction term in $H_{QCD}$ gives, operating on the component \eq{bfield} of the baryon wave function,
\beqa\label{qintb}
\Big[\int d\xv\sum_{a,D,E}\psi_D^\dag(\xv)g\mA_a^0(\xv)T^{DE}_a \psi_E(\xv)\Big]
\psi_A^\dag(\xv_1)\psi_B^\dag(\xv_2)\psi_C^\dag(\xv_3)\ket{0} && \nn\\[2mm] && \hspace{-8cm}
=\sum_{a=3,8} g\big[T_a^{AA}\mA_a^0(\xv_1)+T_a^{BB}\mA_a^0(\xv_2)+T_a^{CC}\mA_a^0(\xv_3)\big]\psi_A^\dag(\xv_1)\psi_B^\dag(\xv_2)\psi_C^\dag(\xv_3)\ket{0}
\eeqa
From \eq{fenb1} we see that the eigenvalue is $-2E_\mB(\xv_1,\xv_2,\xv_3)$. The baryon potential is obtained by adding the field energy to the quark interaction energy,
\beq\label{bpot}
V_\mB(\xv_1,\xv_2,\xv_3) = -E_\mB(\xv_1,\xv_2,\xv_3)=g\la^2\,d(\xv_1,\xv_2,\xv_3)
\eeq
with $d(\xv_1,\xv_2,\xv_3)$ defined in \eq{ddef}. When two quarks in the baryon are at the same position, $\xv_2=\xv_3$, we have $d(\xv_1,\xv_2,\xv_2)=|\xv_1-\xv_2|$. Consequently the baryon potential reduces to the meson potential,
\beq\label{mbpot}
V_\mB(\xv_1,\xv_2,\xv_2)=V_\mM(\xv_1-\xv_2)
\eeq

The translation invariance ($\xv_i \to \xv_i+\bs{a}$) of the meson and baryon potentials \eq{mpot} and \eq{bpot} is a consequence of the states being color singlets. Postulating analogous confining fields for single quark states would break Poincar\'e invariance. 

Furthermore, as may be surmised from the expression \eq{fenb2} for the baryon field energy, the three quarks are confined along $|\xv_1-\xv_2|$ by the $\mA_3^0$ potential and along $|\xv_1+\xv_2-2\xv_3|$ by the $\mA_8^0$ potential. In \#5 of section \ref{confsubsubsec} I noted that only the $a=3,8$ color fields, which have diagonal color generators, preserve the color structure \eq{bcolor} of the wave function. These two diagonal generators of $SU(3)$ allow to confine at most three quarks (having two independent separations). Thus an analogous confining potential could not be constructed for, \eg, $(q\bar q)(q\bar q)$ states, even if they were color singlets. In the present scenario the $XYZ$ multi-quark states \cite{Chen:2016qju} seem best understood as hadron molecules.

\subsubsection{Meson bound state equation \label{bsemsubsubsec}}

In section \ref{mpotsubsubsec} we saw that the component \eq{mfield} of the meson state is an eigenstate of the interaction Hamiltonian $H_{int}(\mA^0)$, with the potential $V_\mM$ \eq{mpot} as eigenvalue. For the meson to be an eigenstate of the full Hamiltonian as in \eq{hmeson} that component must be an eigenstate also of the free Hamiltonian $H_0$,
\beq\label{hmeson2}
\big[H_0+V_\mM(\xv_1-\xv_2)-M\big]\,\bar\psi_A^{(1)}(\xv_1)\Phi_\mM(\xv_1-\xv_2)\psi_A^{(2)}(\xv_2)\ket{0} = 0 \hspace{2cm} ({\rm no\ sum\ on\ }A)
\eeq
The color reduced wave function \eq{mcolor} is included in \eq{hmeson2} to indicate the Dirac matrix structure. A flavor index $f=1,2$ is added to the quark field $\psi^{(f)}$ to allow for the case of unequal masses. The free Hamilonian is
\beq\label{hfree}
H_0=\sum_{f,C} \int d\xv\,\bar\psi_C^{(f)}(\xv)\big(-i\rnab\cdot\gv+m_f\big)\psi_C^{(f)}(\xv)
\eeq
where the arrow on $\nv$ indicates the direction of differentiation. Recalling that the vacuum $\ket{0}$ is annihilated by $H_{QCD}$ \eq{hmeson2} becomes
\beqa\label{hmeson3}
\Big\{\bar\psi_A^{(1)}(\xv_1)\big(i\lnab_1\cdot\gv+m_1\big)\gamma^0\Phi_\mM(\xv_1-\xv_2)\psi_A^{(2)}(\xv_2)-\,\bar\psi_A^{(1)}(\xv_1)\Phi_\mM(\xv_1-\xv_2)\gamma^0\big(-i\rnab_2\cdot\gv+m_2\big)\psi_A^{(2)}(\xv_2)\Big]\ket{0}\hspace{1cm}&& \nn\\[2mm]
&& \hspace{-12cm}=\big[M-V_\mM(\xv_1-\xv_2)\big] \bar\psi_A^{(1)}(\xv_1)\Phi_\mM(\xv_1-\xv_2)\psi_A^{(2)}(\xv_2)\ket{0} = 0
\eeqa
where $\nv_j \equiv \partial/\partial\xv_j$. Since the state \eq{meson} involves an integral over $\xv_1,\xv_2$ we may partially integrate $\lnab_1 \to -\rnab_1$ so that the derivative acts on the wave function, and similarly $\rnab_2 \to -\lnab_2$. Identifying the coefficients of the quark fields gives the bound state equation
\beq\label{mbse}
i\nv\cdot\acom{\gz\gv}{\Phi(\xv)}+m_1\gz\Phi(\xv)-m_2\Phi(\xv)\gz = \big[M-V_\mM(\xv)\big]\Phi(\xv)
\eeq
where I denoted $\xv_1-\xv_2 = \xv$.

The separation of angular variables in the rest frame and the derivation of radial equations for the meson wave function may be found in \cite{Geffen:1977bh} for the equal-mass case ($m_1=m_2$). I review this in the section \ref{mes31subsec}.

\subsubsection{Baryon bound state equation \label{bsebsubsubsec}}

The derivation of the bound state equation for the baryon wave function $\Phi_\mB$ \eq{bcolor} of the state \eq{baryon} is similar to that for mesons. For generality I shall assume that all three quarks have distinct flavors $f=1,2,3$. The condition corresponding to \eq{hmeson2} gives
\beq\label{bbse}
\sum_{j=1}^3 \big[-i\rnab_j\cdot(\gz\gv)_j+m_j\gz_j\big]\Phi_\mB(\xv_1,\xv_2,\xv_3)= \big[M-V_\mB(\xv_1,\xv_2,\xv_3)\big]\Phi_\mB(\xv_1,\xv_2,\xv_3)
\eeq
where $\nv_j \equiv \partial/\partial\xv_j$ and the subscript $j$ on the Dirac matrices indicates which (suppressed) index on $\Phi_\mB$ they are contracted with. The expression for the baryon potential $V_\mB(\xv_1,\xv_2,\xv_3)$ is given in \eq{bpot}.

The separation of variables in the baryon equation \eq{bbse} and the frame dependence of its wave function remain to be addressed.

\section{Rest frame meson wave functions \label{messec}}

In this section I discuss some properties of the solutions of the meson bound state equation \eq{mbse} with the linear potential \eq{mpot}. I begin with the case of $D=1+1$ dimensions, which allows to address several of the novel properties in a simple setting with analytical solutions \cite{Dietrich:2012un}. I then consider $D=2+1$ and $D=3+1$. For simplicity I assume the quark masses to be equal, $m_1=m_2=m$. For $D=1+1$ the solution with unequal masses may be found in \cite{Dietrich:2012un}. Solutions in frames with non-vanishing bound state momentum $\Pv$ are discussed in section \ref{framesec}.

\subsection{$D=1+1$ dimensions \label{mes11subsec}}

\subsubsection{Analytic solution \label{wfsubsubsec}}

I represent the Dirac matrices with Pauli matrices as in \eq{matdef} for the Dirac equation. The bound state equation \eq{mbse} with $m_1=m_2=m$ then reads
\beq\label{mbse11}
i\partial_x\acom{\dsi_1}{\Phi(x)}+m\com{\dsi_3}{\Phi(x)}=[M-V(x)]\Phi(x)
\eeq
where\footnote{For convenience in comparison with earlier results the notation in section \ref{diracsec} uses $e=1$ in \eq{linpot}, corresponding to a scale in which $V'=\halft$. From now on I show the scale $V'$ explicitly.}
\beq\label{potdef}
V(x) = g\la^2\,|x| \equiv V'\,|x|
\eeq
Since $\Phi(x)$ is a $2\times 2$ matrix it can be expanded in Pauli matrices,
\beq\label{wf11}
\Phi(x) = \phi_0(x)\,\mathbb{1}+\phi_1(x)\,\dsi_1+\phi_2(x)\,i\dsi_2+\phi_3(x)\,\dsi_3
\eeq
The bound state equation \eq{mbse11} for the wave function of the $q\bar q$ state \eq{meson} has common features, but also differences, compared to the Dirac equation \eq{dir} for the single fermion state \eq{statedef}.
\begin{itemize}

\item[(a)] The Dirac coordinate $x$ refers to the fixed external field $A^0(x)$. For the meson case $x=x_1-x_2$ is the distance between the quark and the antiquark, and $V(x)=V(x_1-x_2)$. Thus the meson equation is translation invariant.

\item[(b)] The meson wave function \eq{wf11} has four scalar components {\it vs.} the Dirac wave function's two components \eq{potwf2}. However, $\phi_2$ and $\phi_3$ do not contribute to the anti-commutator in \eq{mbse11} and may be solved algebraically:
\beq\label{phi23}
\phi_2(x)=\frac{2m}{M-V(x)}\phi_1(x) \hspace{2cm} \phi_3(x)=0
\eeq

\item[(c)] The coupled differential equations for $\phi_0$ and $\phi_1$,
\beqa\label{m1stord}
2i\partial_x\phi_1(x) &=& (M-V)\phi_0(x) \label{m1stord1} \\[2mm] 
2i\partial_x\phi_0(x) &=& (M-V)\Big[1-\frac{4m^2}{(M-V)^2}\Big]\phi_1(x) \label{m1stord2}
\eeqa
reduce to the single 2nd order equation
\beq\label{m2ndord}
\partial^2_x \phi_1(x)+\frac{\veps(x)V'}{M-V}\,\partial_x\phi_1(x)+\big[\quart(M-V)^2-m^2\big]\phi_1(x) =0
\eeq
where $\veps(x)$ is the sign function. This may be compared with \eq{dirac2ord} in the Dirac case.

\item[(d)] A phase convention analogous to the Dirac \eq{phase} may be imposed at $x=0$ and then holds for all $x$,
\beq\label{mphase}
\phi_1^*(x)=\phi_1(x)  \hspace{2cm} \phi_0^*(x)=-\phi_0(x)
\eeq

\item[(e)] Solutions of parity $\eta=\pm 1$ may be defined analogously\footnote{$\eta=-\eta_P$ in \eq{parop3} since I follow the convention of \cite{Dietrich:2012un}. In $D=1+1$ charge conjugation does not provide an independent constraint.} to the Dirac \eq{parity},
\beq\label{mparity}
\phi_1(-x)=\eta\,\phi_1(x) \hspace{2cm} \phi_0(-x)=-\eta\,\phi_0(x)
\eeq
Consequently the differential equations may be solved for $x \geq 0$ with the constraint
\beq\label{mbc}
\phi_0(0)=0 \hspace{.5cm} (\eta=+1) \hspace{2cm} \phi_1(0)=0 \hspace{.5cm} (\eta=-1)
\eeq

\item[(f)] The wave functions depend on $M$ and $x$ only through the dimensionless variable\footnote{This corresponds to the variable $\sigma(x)$ in \eq{potwf2} for the Dirac equation. Since the expressions in section \ref{diracsec} assume $V'=\halft$ the definition \eq{msig} of $\tau(x)$ differs from the definition of $\sigma(x)$ by a factor of 2.} $\tau(x)$ 
\beq\label{msig}
\tau(x) = (M-V)^2/V' \hspace{2cm} \partial_x = -2\,\veps(x)(M-V)\partial_\tau
\eeq
The 2nd order equation \eq{m2ndord} is in terms of $\tau$,
\beq\label{msig2ndord}
16\,\partial_\tau^2\,\phi_1(\tau) + \Big[1-\frac{4m^2}{V'\tau}\Big]\phi_1(\tau)=0
\eeq
which has the general solution
\beq\label{mgensol}
\phi_1(\tau)=\tau\,\exp(-i\quart\tau)\big[a\kum(1-i\halft\,\tm^2,2,i\halft\tau)+b\,U(1-i\halft\,\tm^2,2,i\halft\tau)\big]
\eeq
where $\kum$ and $U$ are the confluent hypergeometric functions of the first and second kind, respectively. The parameters $a$ and $b$ are constants and the dimensionless mass $\tm$ is
\beq\label{tmdef}
\tm^2 \equiv \frac{m^2}{V'}
\eeq

\end{itemize}

\subsubsection{Non-relativistic and relativistic limits \label{limitsubsubsec}}

(a) $\tm \to \infty$: {\em Non-relativistic limit}

In the non-relativistic limit $M=2m+E_b$, where the fermion mass $m$ is much larger than the binding energy $E_b$ and the potential $V(x)$. The Schr\"odinger equation \eq{seq} specifies the mass dependence $E_b \sim V \propto m^{-1/3}$. Taking $\tm\to\infty$ at fixed $\tm^{1/3}V(x)$ in \eq{mgensol} gives \cite{Dietrich:2012un,Hoyer:2014gna}
\beqa
\tau\,e^{-i\tau/4}\kum(1-i\halft\,\tm^2,2,i\halft\tau) &=& 4\,\tm^{-2/3}e^{\pi\tm^2/2}{\rm Ai}\big[\tm^{1/3}(V-E_b)/\sqrt{V'}\,\big] \Big[1+{\cal O}\big(\tm^{-4/3}\big)\Big] \label{nr1} \\[2mm]
\tau\,e^{-i\tau/4}\;U(1-i\halft\,\tm^2,2,i\halft\tau) &=& -2\,\tm^{4/3}
\frac{\pi e^{-\pi\tm^2/2}}{\Gamma(1-i\halft\,\tm^2)}\Big\{{\rm Ai}\big[\tm^{1/3}(V-E_b)/\sqrt{V'}\,\big]+i\,{\rm Bi}\big[\tm^{1/3}(V-E_b)/\sqrt{V'}\,\big]\Big\} \hspace{1cm}\nn\\
&&\times \Big[1+{\cal O}\big(\tm^{-4/3}\big)\Big] \label{nr2}
\eeqa
The Airy Ai-solution \eq{nr1} agrees (up to the normalization) with the Schr\"odinger wave function \eq{seqAi}.
The Airy Bi-function is not normalizable, indicating that we need to take $b=0$ in \eq{mgensol}. In the next section \ref{discretesubsubsec} I discuss why this is necessary even in the relativistic case.

\vspace{.5cm}
(b) $\tau \to \infty$: {\em Large $M$ or large $V(x)$} 

Substituting the ansatz $\phi_1(\tau\to\infty)=N\,{\rm Re}\big[\tau^{i\alpha} \, e^{i\beta\tau+\gamma}\big]$ into the differential equation \eq{msig2ndord} and comparing the terms of \order{\tau^n} with $n=0,-1$ determines the parameters $\alpha=-\halft\tm^2$ and $\beta=\quart$. Thus 
\beq\label{11asymp1}
\phi_1(\tau\to\infty)=N\cos\big[\quart\tau-\halft\tm^2 \log(\halft\tau)+\gamma\big]
\eeq
The analytic solution \eq{mgensol} with $b=0$ specifies 
\beq
N=\frac{4a}{\tm\sqrt{\pi}}\sqrt{\exp(\pi\tm^2)-1} \hspace{2cm} \gamma=\arg\Gamma(1+i\halft\tm^2)-\halft\pi 
\eeq
From \eq{m1stord1},
\beq
\phi_0(\tau\to\infty)= -4i\partial_\tau\phi_1(\tau\to\infty)=i\,N\sin\big[\quart\tau-\halft\tm^2 \log(\halft\tau)+\gamma\big]
\eeq
The result may be summarized as
\beq\label{siglim}
\lim_{\tau\to\infty} \big[\phi_1(\tau)+\phi_0(\tau)\big]= \lim_{\tau\to\infty} \big[\phi_1(\tau)-\phi_0(\tau)\big]^* =N\exp\big[i\quart\tau-i\halft\tm^2\log(\halft\tau)+i\gamma\big]
\eeq 

The absolute value of \eq{siglim} is independent of $\tau$ as in the Dirac case \eq{fas}. There we saw that the wave function at large $|x|$ was due to the positron component of the virtual $e^+e^-$ pairs, since positrons were repulsed by the linear potential. In the present case the constant norm of the wave function might be interpreted as being dual to pair production. The real pairs would be produced when string breaking is taken into account. I discuss this further below.

\vspace{.5cm}
(c) $m \to 0$: {\em Massless quarks} 

The bound state equation \eq{msig2ndord} is trivial for $m=0$. The $\tau\to\infty$ solution \eq{siglim} is then exact for all values of $\tau$,
\beq\label{m0lim}
\lim_{m\to 0} \big[\phi_1(\tau)+\phi_0(\tau)\big]= \lim_{m\to 0} \big[\phi_1(\tau)-\phi_0(\tau)\big]^* =N\exp\big[i\quart\tau+i\gamma\big]
\eeq

\subsubsection{Discrete spectrum \label{discretesubsubsec}}

An essential difference between the meson and Dirac equations arises due to the algebraic conditions \eq{phi23}: $\phi_2(x)$ is singular at $M-V(x)=0$ unless $\phi_1(\tau=0)=0$. The differential equation \eq{msig2ndord} allows $\phi_1(\tau\to 0) \propto \tau^\alpha$ with $\alpha=0,\ 1$. In order for $\phi_2(x)$ to be regular we must choose $\phi_1(\tau\to 0) \propto \tau^1$, which implies $b=0$ in the solution \eq{mgensol}. The same constraint ensures the normalizability of the wave function in the non-relativistic limit \eq{nr2}. The $x=0$ continuity constraint \eq{mbc} can then be satisfied only for {\em discrete} bound state masses $M$.

Meson states with distinct eigenvalues $M_a \neq M_b$ are orthogonal, as seen from
\beq\label{mortho1}
\bra{M_b}H\ket{M_a}= M_a \bra{M_b}{M_a}\rangle = M_b \bra{M_b}{M_a}\rangle
\eeq 
The bound state equations imply wave function orthogonality \cite{Dietrich:2012un},
\beq\label{mortho2}
\int_{-\infty}^\infty dx\, \tr\big[\Phi_b^\dag(x)\Phi_a(x)\big] = 0
\ \ \ {\rm for}\ M_a \neq M_b 
\eeq
which is equivalent to the orthogonality of the states,
\beqa\label{mortho3}
\bra{M_b}{M_a}\rangle &=& \int_{-\infty}^\infty\Big[\prod_{i=1,2}dx_{bi}dx_{ai}\Big]
\bra{0}\psi^\dag(x_{b2})\Phi_b^\dag(x_{b1}-x_{b2})\gz\psi(x_{b1})
\bar\psi(x_{a1})\Phi_a(x_{a1}-x_{a2})\psi(x_{a1})\ket{0}\nn\\[2mm]
&=& \int_{-\infty}^\infty dx_{a1}dx_{a2}\,\tr\big[\Phi_b^\dag(x_{a1}-x_{a2})\Phi_a(x_{a1}-x_{a2})\big]
=2\pi\delta(0)\int_{-\infty}^\infty dx\, \tr\big[\Phi_b^\dag(x)\Phi_a(x)\big] = 0
\eeqa
where $\delta(0)$ represents momentum conservation: both states are defined in the rest frame. For $a=b$ the normalization integral diverges due to the constant norm \eq{siglim} of the meson wave function at large $|x|$. A normalization to $\delta(M_a-M_b)$ as in \eq{orno} is not possible when $M_a,M_b$ are discrete.

\begin{wrapfigure}[15]{r}{0.3\textwidth}
  \vspace{-30pt}
  \begin{center}
    \includegraphics[width=0.3\textwidth]{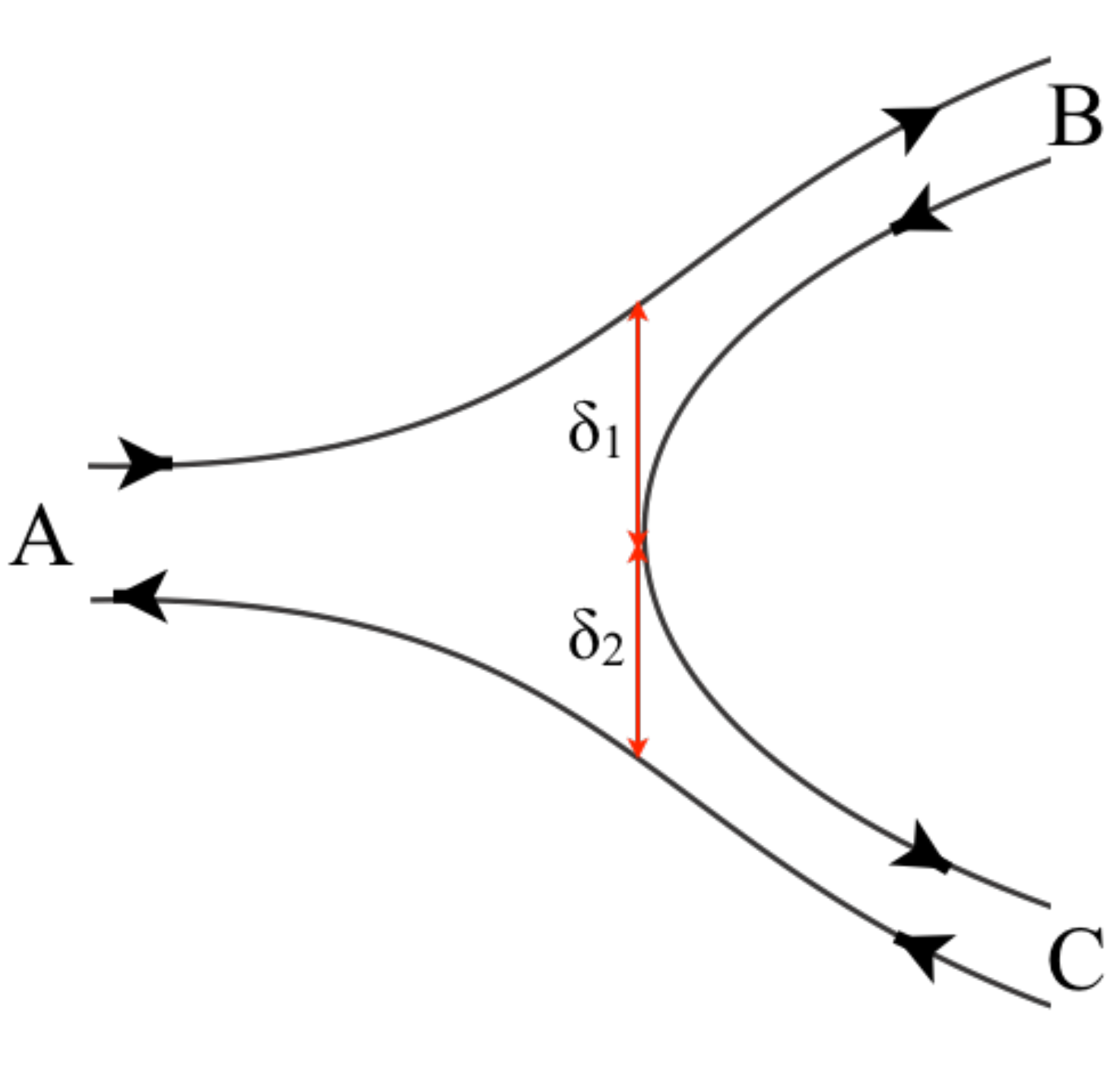}
  \end{center}
  \vspace{-15pt}
  \caption{Dual diagram generated by the annihilation of an antiquark in meson $B$ with a quark in meson $C$.}
\label{dualdiag}
\end{wrapfigure}

To illuminate this puzzle, consider the $m \to 0$ analytic solution \eq{m0lim}. For any finite $m$, however small, the regularity of $\phi_2(\tau)$ requires $\phi_1(\tau = 0)$ = 0. Imposing also the continuity condition \eq{mbc} at $x=0$ (where $\tau=M^2/V'$) gives for $x\geq 0$,
\beqa
M_n^2(m\to 0) &=& 2V'\big[2n-\halft(1+\eta)\big]\pi \label{m0spect} \\[2mm]
\phi_1(x) &=& N\cos\big[-\halft M_nx+\quart V'x^2+\quart(1-\eta)\pi\big] \label{m0wf}
\eeqa
where the parity $\eta$ is defined in \eq{mparity} and $n$ is an integer. The trigonometric functions \eq{m0wf} do not form a complete functional basis when the masses are restricted to the discrete values \eq{m0spect}. 

These issues do not arise for the Dirac states. The Dirac wave functions \eq{fgsol} are regular for all values of $\sigma$, allowing a continuous spectrum. This reflects the unconfined positron component of the Dirac state which can have any energy. The continuous mass spectrum ensures orthonormality and completeness, as was verified in \eq{orno4} and \eq{comp} for the $m\to 0$ wave functions \eq{phiexpr}.

The Dirac equation is derived under the assumption that the electron interacts only with the external field, not with any other bound electron nor with the $e^+e^-$ pairs in the higher Fock states.  In mesons the $q\bar q$ pairs of a given color mutually interact via the classical field \eq{mfield} that they generate. Since the classical field vanishes when summed over quark colors \eq{mfieldsum} the field does not induce interactions between mesons (neglecting \order{\alpha_s} gluon exchange). However, a quark in one meson can annihilate an antiquark in another meson if their space-time coordinates coincide (\cf\ the double annihilation in \eq{mortho3}). This gives rise to ``string breaking'', as depicted by the dual diagram in \fig{dualdiag}. The amplitude for this diagram may be calculated in terms of the meson wave wavefunctions obtained from a ``tree-level'' bound state equation such as \eq{mbse11} (more precisely, \eq{bse3} for bound state momentum $P$). For a linear potential $V(\delta_1+\delta_2) = V(\delta_1) + V(\delta_2)$, so there is no instantaneous change in the potential energy. The amplitude of \fig{dualdiag} is of \order{1/\sqrt{N_c}} compared to the tree-level states \eq{meson}, $N_c$ being the number of colors.

The process of \fig{dualdiag} implies that the meson states \eq{meson} are not orthogonal to two-meson states. The splitting and fusion process can be repeated any number of times, leading to overlaps with multi-meson states and hadron loop corrections to the tree-level wave functions. A new basis of ``dressed'' states will be obtained by diagonalizing the interactions induced by \fig{dualdiag}. Such corrections are necessary, and hopefully also sufficient, to generate a complete orthonormal set of normalizable states with unitary scattering amplitudes at \order{\alpha_s^0}. Verifying this is an import avenue for future research.

\begin{figure}[h] \centering
\includegraphics[width=.6\columnwidth]{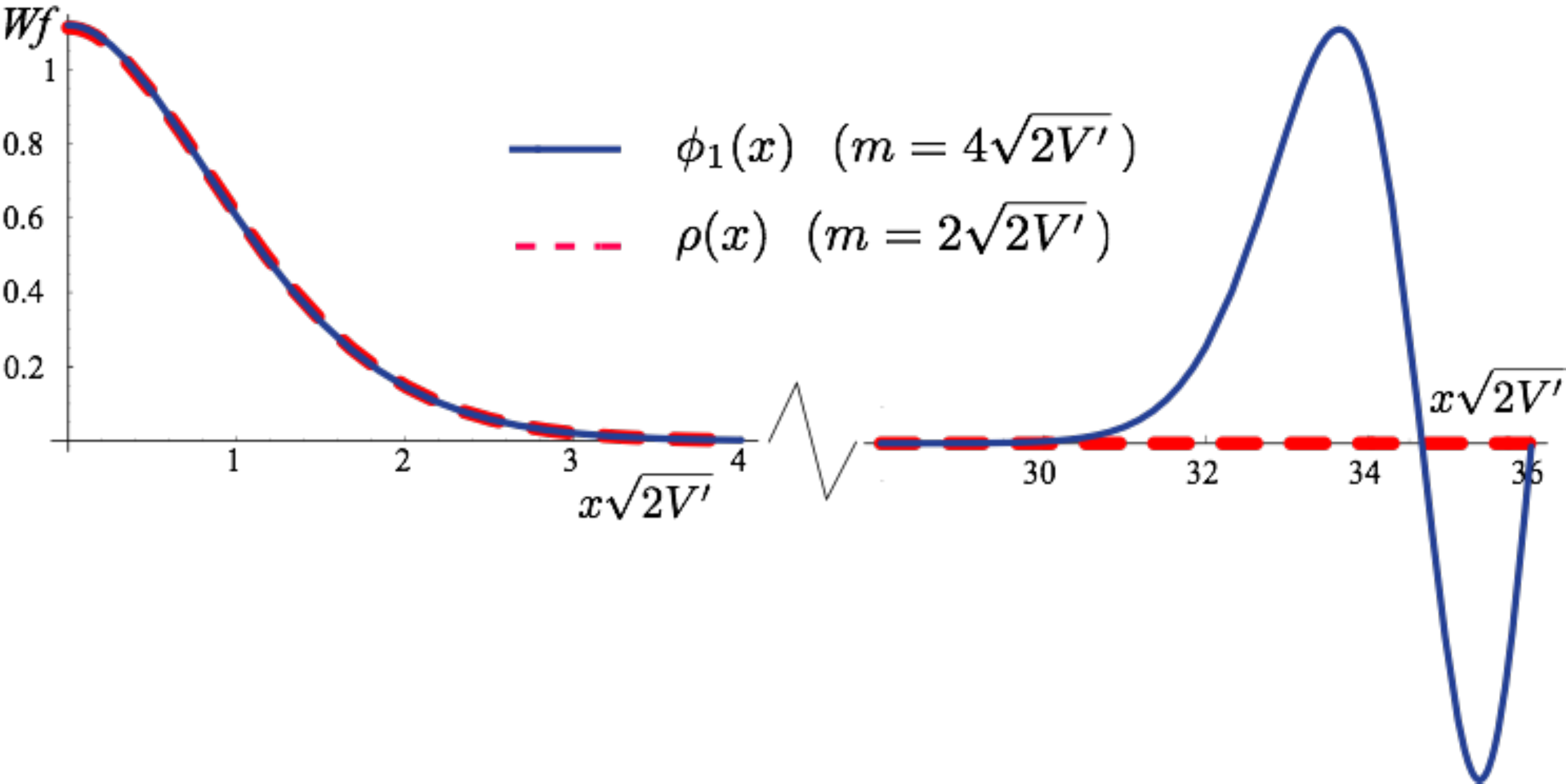}
\caption{The $\phi_1(x)$ wave function \eq{mgensol} (with $b=0$) of the ground state meson wave function  (solid blue line) compared to the solution $\rho(x)$ of the Schr\"odinger equation \eq{seq} with the reduced mass (dashed red line). The meson mass is $8.4100\sqrt{2V'}$ while the binding energy of the Schr\"odinger solution is $E_b=0.4043\sqrt{2V'}$. Both wave functions are normalized to unity in the region $0 \leq x\sqrt{2V'}\leq 5$. Figure from \cite{Dietrich:2012un}. \label{ffwf}}
\end{figure}

\subsubsection{Duality \label{dualitysubsubsec}}

\fig{ffwf} shows the $\phi_1(x)$ wave function of the ground state ($\eta=+1$) meson. The quark mass $m= 4\sqrt{2V'}$ is large enough to ensure good agreement with the Schr\"odinger wave function $\rho(x)$ in the non-relativistic region where $V(x) \ll 2m$. In the region where $V(x) \gg 2m$ the wave function oscillates with constant amplitude \eq{siglim}. This behavior may be compared with that of the Dirac wave function in \fig{wfs-beta} (where $m=\sqrt{2V'}$). A comparison of Figs.\ \ref{wfs-beta}a,b shows that the magnitude of the oscillations relative to the wave function at $x=0$ depends sensitively on the parameter $\beta$ of the Dirac wave function \eq{fgsol}. Mesons with non-singular wave functions have $b=0$ in \eq{mgensol}, leaving only the overall normalization $a$ as a free parameter: the relative magnitude of the oscillations cannot be adjusted. Since $\phi_1$ depends on $x$ via the function $\tau(x)=(M-V)^2/V'$ we have $\phi_1(x=0) = \phi_1(x=2M/V')$, as seen in \fig{ffwf}.

The Dirac state \eq{nstate} has a ``valence'' component $\propto b^\dag_p$ and a ``sea'' component $\propto d_p$. The sea reflects the momentum distribution of the positron component in the state. Eq. \eq{Dsp} shows that at large momenta (related to the rapid oscillations at large $|x|$ in coordinate space) the wave function has only a sea component. The valence component vanishes for $p\to\infty$, as seen in \fig{BD-dist}(a) and Eq. \eq{Bas}.

The meson state \eq{meson} has four components\footnote{The color field causes a Bogoliubov transformation analogous to \eq{cndef1}, so that $b$ and $d$ do not annihilate the vacuum.}, $b^\dag b,\,b^\dag d^\dag,\,db \text{ and }d d^\dag$. For high excitations $M \gg V(x)$ and in the region of the oscillations where $V(x) \gg M$ only one component dominates. Both cases correspond to the $\tau\to\infty$ limit given in \eq{siglim}. In order to make contact with the parton picture I consider the ultrarelativistic limit where the quark mass $m$ can be neglected.

Expressing the quark fields in the meson state \eq{meson} in terms of the $b$ and $d$ operators as in \eq{eop}, with the $u,\,v$ spinors of \eq{wtB} and \eq{wtD} evaluated for $m=0$, and Fourier transforming the wave function as in \eq{fourierdef} we get (suppressing colors),
\beq\label{mexpand}
\ket{M} = \int_{-\infty}^\infty\frac{dp}{2\pi\,2|p|}\Big\{\veps(p)\phi_2(p)b_p^\dag b_p + \big[\phi_1(p)-\veps(p)\phi_0(p)\big]b_p^\dag d^\dag_{-p} + \big[\phi_1(p)+\veps(p)\phi_0(p)\big]b_p d_{-p} + \veps(p)\phi_2(p)d_{-p}^\dag d_{-p}\Big\}\ket{0}
\eeq
where the $\phi_i(p)$ are the momentum space versions of the wave function components $\phi_i(x)$ defined in \eq{wf11}. According to \eq{phi23} $\phi_2(x) =2m\phi_1(x)/(M-V)$  vanishes for $m=0$ (and contributes at non-leading order even for finite $m$ in the $\tau\to\infty$ limit). Using the parity relations \eq{mparity} the Fourier transforms of $\phi_1(x)$ and $\phi_0(x)$ are
\beqa\label{ftrans}
\phi_1(p) &=& \int_{-\infty}^{\infty} dx\,\phi_1(x)\,e^{-ipx} = 
\int_{0}^{\infty} dx\,\phi_1(x)\big(e^{-ipx}+\eta e^{+ipx}\big) = \eta\,\phi_1(-p) \nn \\[2mm]
\phi_0(p) &=& \int_{-\infty}^{\infty} dx\,\phi_0(x)\,e^{-ipx} = 
\int_{0}^{\infty} dx\,\phi_0(x)\big(e^{-ipx}-\eta e^{+ipx}\big) = -\eta\,\phi_0(-p)
\eeqa
The combinations appearing in the state \eq{mexpand} are thus
\beq\label{ftrans1}
\phi_1(p)\pm \phi_0(p) = 
\int_{0}^{\infty} dx\Big\{\big[\phi_1(x)\pm \phi_0(x)\big]e^{-ipx}+\eta \big[\phi_1(x) \mp \phi_0(x)\big]e^{ipx}\Big\}
\eeq

From \eq{m0wf} we find
\beq\label{m0phicomb}
\phi_1(x)+ \phi_0(x)=\big[\phi_1(x)- \phi_0(x)\big]^* = N\exp\big[-i\halft Mx+i\quart V'x^2+i\quart(1-\eta)\pi\big]
\eeq
Using $\eta\exp[-i(1-\eta)\pi/4] = \exp[i(1-\eta)\pi/4]$ gives
\beqa
\phi_1(p)+ \phi_0(p) &=& N\,e^{i(1-\eta)\pi/4}\int_0^\infty dx\Big\{\exp\big[-ix(p+\halft M)+i\quart V'x^2\big]+\exp\big[ix(p+\halft M)-i\quart V'x^2\big]\Big\} \label{m0phicomb1} \\[2mm]
\phi_1(p)- \phi_0(p) &=& N\,e^{-i(1-\eta)\pi/4}\int_0^\infty dx\Big\{\exp\big[-ix(p-\halft M)-i\quart V'x^2\big]+\exp\big[ix(p-\halft M)+i\quart V'x^2\big]\Big\} \label{m0phicomb2}
\eeqa
I now consider the two limits mentioned above.

\vspace{.5cm}
(a) $M \to \infty,\ V(x)$ fixed

We may drop the \order{x^2} term in the exponents: $\quart xV(x) \ll \halft xM$. The two terms may be combined into a single integral,
\beqa\label{m0phicomb3}
\phi_1(p)+ \phi_0(p) &=& N\,e^{i(1-\eta)\pi/4}\int_{-\infty}^\infty dx\,\exp\big[-ix(p+\halft M)\big] = N\,e^{i(1-\eta)\pi/4}\,2\pi\,\delta(p+\halft M)\nn\\[2mm]
\phi_1(p)- \phi_0(p) &=& N\,e^{-i(1-\eta)\pi/4}\int_{-\infty}^\infty dx\,\exp\big[-ix(p-\halft M)\big] = N\,e^{-i(1-\eta)\pi/4}\,2\pi\,\delta(p-\halft M)
\eeqa
The expression \eq{mexpand} for the state reduces to
\beq\label{mexpand1}
\ket{M\gg V}=\inv{M}\,N\,e^{i(1-\eta)\pi/4}\Big[b_{M/2}^\dag d_{-M/2}^\dag + \eta\, b_{-M/2}^\dag d_{M/2}^\dag\Big]
\eeq
The wave function of a highly excited state (large $M$), at separations $x$ between the quarks where the potential $V(x)$ is small compared to $M$, has only the valence $b^\dag d^\dag$ quark component, as in the parton picture. The bound state wave function reduces to a plane wave $(p=\pm \halft M)$ due to its oscillations at large $\tau$ and the linearity of the potential. 

\begin{wrapfigure}[8]{r}{0.5\textwidth}
  \vspace{-25pt}
  \begin{center}
    \includegraphics[width=0.5\textwidth]{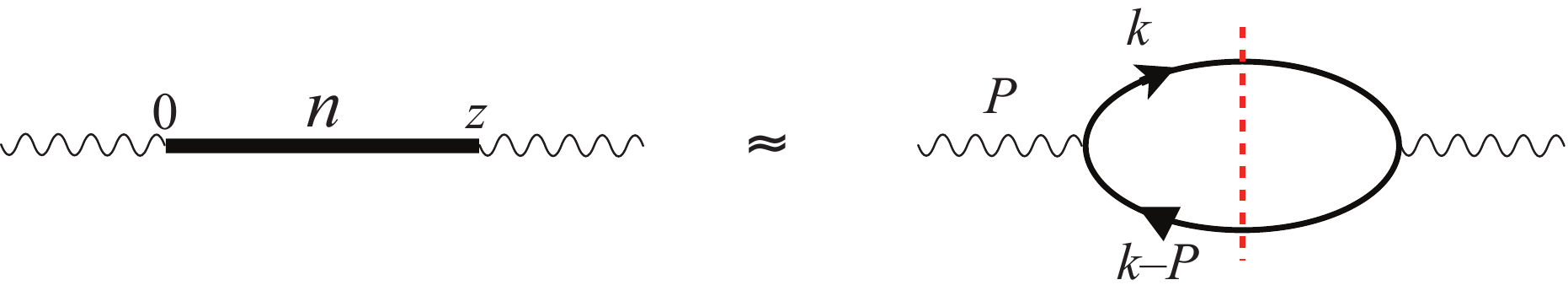}
  \end{center}
  \vspace{-15pt}
  \caption{Duality between a resonance $n$ and a quark-loop contribution to the imaginary part of a current propagator \cite{Dietrich:2012un}.}
\label{dual}
\end{wrapfigure}

Quark-Hadron duality is a remarkable feature of data on hadron scattering (section \ref{dualsec}). In $e^+e^- \to X$ (with $X$ an inclusive hadron state) direct channel resonances build the cross section at low energies. At high energies the total cross section for hadron production can be calculated in the quark and gluon basis. Even the distribution of individual hadrons (mostly pions) follows the distribution of the partons (mostly gluons).

The Bloom-Gilman duality of Deep Inelastic electron Scattering $e+M \to e+X$ expresses the fact that the resonance contributions to the final state $X$ agree with the scaling parton contribution at high values of $M_X$. This may be another implication of the valence quark dominance in \eq{mexpand1}. 

\vspace{.5cm}
(b) $M$ fixed, $V(x) \to \infty$

For $p\to +\infty$ the phase in the exponent of \eq{m0phicomb1} has a stationary phase at
\beq\label{statphase1}
\frac{d}{dx}\big[-x(p+\halft M)+\quart V'x^2\big]=0 \hspace{2cm} M=V(x)-2p
\eeq
The stationary phase approximation \eq{statphase} gives
\beqa\label{VggM}
\lim_{p\to +\infty}\big[\phi_1(p)+\phi_0(p)\big] &=& 4Ne^{i(1-\eta)\pi/4}\sqrt{\frac{\pi}{V'}}\,\cos\Big[\inv{V'}(p+\halft M)^2-\frac{\pi}{4}\Big] \\[2mm]
\lim_{p\to +\infty}\big[\phi_1(p)-\phi_0(p)\big] &=& 0
\eeqa
where the latter result follows because the stationary phase in the exponent of \eq{m0phicomb2} is at $p<0$. On the other hand, from \eq{ftrans}
\beq\label{VggM1}
\lim_{p\to +\infty}\big[\phi_1(-p)-\phi_0(-p)\big]= \lim_{p\to +\infty}\eta\,\big[\phi_1(p)+\phi_0(p)\big]
\eeq
We may combine the nonvanishing $p\to\pm\infty$ limits as
\beq\label{VggM2}
\lim_{p\to \infty}\big[\phi_1(p)+\veps(p)\phi_0(p)\big]=
4Ne^{\veps(p)i(1-\eta)\pi/4}\sqrt{\frac{\pi}{V'}}\,\cos\Big[\inv{V'}(|p|+\halft M)^2-\frac{\pi}{4}\Big]
\eeq
The stationary phase approximation \eq{statphase1} implies that $x\propto p$. Comparing with \eq{mexpand} we find
\beq\label{mexpand2}
\ket{M \ll V} = \int_{|p|\gg M}\frac{dp}{2\pi\,2|p|}\big[\phi_1(p)+\veps(p)\phi_0(p)\big]b_p d_{-p}\ket{0}
\eeq
Only the $b_p d_{-p}\ket{0}$ component of a meson state survives in the region of large $|x|$ where the wave function is oscillating. The rapid oscillations of the coefficient \eq{VggM2} will suppress this component when it is convoluted with a function that has a smooth $p$-dependence. The result \eq{mexpand1} is analogous to the one for the Dirac state \eq{wtBD}, where the $d_p$ contribution dominates the $b_p^\dag$ one at large $|p|$ according to \eq{Bas} and \eq{Dsp}.

\subsection{Mesons in $D=2+1$ dimensions \label{21subsec}}

\subsubsection{Bound state equation \label{21bsesubsubsec}}

In $D=2+1$ dimensions there is one transverse direction, but the Dirac matrices may still be represented using the $2\times 2$ Pauli matrices,
\beq\label{21dirdef}
\gz=\sigma_3 \hspace{1cm} \gz\gamma^1=\sigma_1 \hspace{1cm} \gz\gamma^2=\sigma_2
\eeq
The arguments for the classical confining potential in section \ref{confsubsec} are independent of the dimension. The bound state equation for mesons at rest has the same form as in \eq{mbse}. In the representation \eq{21dirdef} and assuming $m_1=m_2=m$,
\beq\label{bse2}
i\acom{\sv\cdot\nv}{\Phi(\xv)}+\moe\com{\sigma_3}{\Phi(\xv)} = \big[M-V(\xv)\big]\Phi(\xv)
\hspace{2cm} V(\xv) = V'|\xv|
\eeq
with $V'=g\la^2$ and the $D=2+1$ vector notation illustrated by $\sv=(\sigma_1,\sigma_2)$,
Since the potential depends only on $|\xv|$ there is rotational symmetry in the $(1,2)$-plane. In terms of the cylindrical $(r,\vphi)$ coordinates
\bal
x^1&=r\cos\vphi \hspace{2cm} \partial_1=\cos\vphi\,\partial_r-\frac{\sin\vphi}{r}\partial_\vphi\nn\\
x^2&=r\sin\vphi \hspace{2cm} \partial_2=\sin\vphi\,\partial_r+\frac{\cos\vphi}{r}\partial_\vphi
\end{align}
the single ``$z$'' component of the angular momentum $\bs{J}$ defined in \eq{rotgen} is
\beq\label{jzdef}
J^z = -i(x^1\partial_2-x^2\partial_1)+\sfrac{i}{2}\gamma^1\gamma^2=-i\partial_\vphi+\halft\sigma_3
\eeq

The $2\times 2$ wave function $\Phi(\xv)$ may be expanded so that each of the four components transforms covariantly under rotations \eq{rotwf} in the plane,
\beq\label{wfdef1}
\Phi^\jz(\xv) = \big[F_a^\jz(r) +\frac{\sv\cdot\xv}{r} F_b^\jz(r)+\frac{\sv\times\xv}{r}F_c^\jz(r)+\sigma_3\,F_d^\jz(r)\big]e^{i\jz\vphi}
\eeq 
The coefficients of the radial functions are invariant under rotations due to \eq{Sgamrot}: 
$S(R)\gv S^{-1}(R) = R^{-1}\gv$,
\beq\label{coeffinv}
S(R)\gv S^{-1}(R)\cdot R^{-1}\xv= (R^{-1}\gv)\cdot(R^{-1}\xv)= \gv\cdot\xv \hspace{2cm} S(R)\gv S^{-1}(R)\times R^{-1}\xv= \gv\times\xv
\eeq
The cross product has only a $z$-component, which is preserved by rotations around the $z$-axis. Consequently a rotation by $\theta$ only changes $\vphi \to \vphi-\theta$ in the wave function  \eq{wfdef1}.

Equivalently, we may easily check that
\beq\label{coeffinv1}
\com{J^z}{\sv\cdot\xv}= \com{J^z}{\sv\times\xv}= \com{J^z}{\sigma_3}=0
\eeq
Hence the wave function \eq{wfdef1} is an eigenfunction of $J^z$ with eigenvalue $\jz$,
\beq
\com{J^z}{\Phi^\jz(\xv)}=\jz\,\Phi^\jz(\xv)
\eeq
which as in \eq{angwf} ensures that the meson state \eq{meson} is an eigenstate of the angular momentum operator $\mJ^z$ with eigenvalue $\jz$.

In the parity condition \eq{parop3} $\xv\to-\xv$ implies $\vphi\to\vphi+\pi$. From \eq{wfdef1} we see that
\beq
\sigma_3\Phi^\jz(-\xv)\sigma_3 = (-1)^\jz\Phi^\jz(\xv) 
\eeq
which implies that the parity of the state is $\eta_P=(-1)^\jz$. The charge conjugation condition \eq{ccop3} 
\beq
\sigma_2\Phi^T(-\xv)\sigma_2 = (-1)^\jz\Phi(\xv) 
\eeq
gives the same result since $F_d(r)=0$ (see \eq{bsd} below). Thus also $\eta_C=(-1)^\jz$.

Substituting the expansion \eq{wfdef1} into the bound state equation \eq{bse2} gives the constraints
\beqa
1: \hspace{1.cm} 2i\Big(\frac{F_b^\jz}{r}+\partial_r F_b^\jz\Big)+\frac{2\jz}{r}F_c^\jz &=& (M-V)F_a^\jz \label{bsa}\\[2mm]
\sv\cdot\xv: \hspace{2.25cm} 2i\partial_r F_a^\jz+2i\moe F_c^\jz &=& (M-V)F_b^\jz \label{bsb}\\[2mm]
\sv\times\xv: \hspace{2.4cm} \frac{2\jz}{r} F_a^\jz-2i\moe F_b^\jz &=& (M-V)F_c^\jz \label{bsc}\\[2mm]
\sigma_3: \hspace{4.65cm} 0 &=& (M-V)F_d^\jz \label{bsd}
\eeqa
Eliminating $F_c^\jz$ we get two coupled first order radial equations,
\begin{align}\label{21coupled}
2r(M-V)\partial_r F_a^\jz &= -4m\jz F_a^\jz-i\big[(M-V)^2-4m^2\big]rF_b^\jz \nn \\[2mm]
2r(M-V)\partial_r F_b^\jz &= -2\big[(M-V)-2\jz m\big]F_b^\jz -i\big[(M-V)^2-\frac{4\jz^2}{r^2}\big]rF_a^\jz 
\end{align}

\subsubsection{Wave function in the $r\to\infty$ limit \label{21asympsubsubsec}}

In $D=1+1$ dimensions the bound state equation \eq{msig2ndord} depends on the coordinate $x$ and the bound state mass $M$ only via the variable $\tau$ of \eq{msig}. In $D=2+1$ this no longer holds, as the radius $r$ appears explicitly. In studying the $r\to\infty$ limit it is nevertheless helpful to introduce
\beq\label{msig21}
\tau(r) = \big[M-V(r)\big]^2/V' \hspace{2cm} \partial_r = -2(M-V)\partial_\tau
\eeq
The bound state equations can then be expressed as
\begin{align}\label{21coupled1}
-4V'\tau\partial_\tau F_a^\jz &= -\frac{4m\jz}{r} F_a^\jz-i\big[V'\tau-4m^2\big]F_b^\jz \\[2mm]
-4V'\tau\partial_\tau F_b^\jz &= -\frac{2}{r}\big[(M-V)-2\jz m\big]F_b^\jz -i\big[V'\tau-\frac{4\jz^2}{r^2}\big]rF_a^\jz 
\end{align}
Neglecting terms of \order{r^{-1}} on the rhs. this simplifies to
\begin{align}\label{21coupled2}
\partial_\tau iF_a^\jz &= -\inv{4}\Big(1-\frac{4m^2}{V'\tau}\Big)F_b^\jz \\[2mm]
\partial_\tau F_b^\jz &= -\frac{1}{2\tau}F_b^\jz +\inv{4}iF_a^\jz 
\end{align}
Eliminating $F_a^\jz$ and neglecting \order{\sigma^{-2}} leaves
\beq\label{21asymp}
\partial_\tau^2 F_b^\jz+\inv{2\tau}\partial_\tau F_b^\jz+\inv{16}\Big(1-\frac{4m^2}{V'\tau}\Big)F_b^\jz = 0  \hspace{2cm} (\tau \to \infty)
\eeq

As in section \ref{limitsubsubsec}(b) \eq{21asymp} implies the leading behavor
\beq\label{21asymp1}
\lim_{r\to\infty}F_a^\jz = \lim_{r\to\infty}F_b^\jz \propto\inv{\tau^{1/4}}\,\tau^{-im^2/2V'}e^{i\tau/4} \propto \frac{1}{\sqrt{r}}\,r^{-im^2/V'}\exp\big[i(M-V)^2/4V'\big]
\eeq
Apart from the factor $\tau^{-1/4} \simeq 1/\sqrt{r}$ this result is similar to the behavior of the $D=1+1$ wave function \eq{siglim}. From \eq{bsc} follows that $F_c^\jz$ is suppressed by a factor $1/r$ compared to $F_{a,b}^\jz$.

As in \eq{mortho3} the normalization integral is proportional to
\beq\label{21density}
\int d\xv\,\tr\big[{\Phi^\jz}^\dag(\xv)\Phi^\jz(\xv)\big] = 4\pi\int_0^\infty dr\,r\big[|F_a^\jz|^2+|F_b^\jz|^2+|F_c^\jz|^2\big]
\eeq
The norm of the wave function for large $r$ is constant, since the phase space factor $r$ compensates for the $1/\sqrt{r}$ fall-off of the wave function in \eq{21asymp1}. Hence an interpretation of the asymptotic wave function as representing, via duality, the constant density of hadrons produced via string breaking remains viable.

\subsubsection{Spectrum in $D=2+1$ \label{21discsubsubsec}}

Eq. \eq{bsc} implies that $F_c^\jz$ is singular at $M=V(r)$ unless the combination of $F_{a,b}^\jz$ appearing on the lhs. of the equation vanishes at this point. This condition, which ensures that the norm \eq{21density} of the wave function is locally finite, is satisfied only for discrete values of the bound state mass $M$. The demonstration is quite analogous to the $D=1+1$ case discussed in section \ref{discretesubsubsec}. For simplicity I discuss here only two special cases, where the coupled equations \eq{21coupled} are easily reduced to a single second order equation.

\vspace{.5cm}
(a) $m=0$, $\jz \neq 0$

Eliminating $F_b^\jz$ gives for  a second order equation for $F_a^\jz$
\beq\label{21adouble}
\partial_r^2 F_a^\jz+\Big[\inv{r}+\frac{V'}{M-V}\Big]\partial_r F_a^\jz +\Big[\quart(M-V)^2-\frac{\jz^2}{r^2}\Big]F_a^\jz =0 \hspace{1cm} (m=0)
\eeq
implying $F_a^\jz(r\to 0) \propto r^{\pm\jz}$. A finite density at $r=0$ requires $F_a^\jz(r\to 0) \propto r^{|\jz|}$. For $\jz \neq 0$ this fixes $F_a^\jz(r= 0)=0$ and $\partial_r F_a^\jz(r= 0) \neq 0$, with the slope of the derivative defining the absolute normalization of the wave function.

Similarly at $M-V=0$, $F_a^\jz(r\to M/V') \propto (M-V)^\beta$ with $\beta=0,2$. As already remarked, \eq{bsc} requires us to choose $F_a^\jz(r\to M/V') \propto (M-V)^2$ for $F_c^\jz(r\to M/V')$ to be finite. Since the boundary condition at $r=0$ already determined the wave function (up to an overall normalization) the condition at $M-V=0$ can only be fulfilled by adjusting the bound state mass $M$ accordingly.

\vspace{.5cm}
(b) $\jz=0$, any $m$

Eliminating $F_a^{\jz=0}$ we get a second order equation for $F_b^{\jz=0}$,
\beq\label{21bdouble}
\partial_r^2 (rF_b^0) + \Big[-\inv{r}+\frac{V'}{M-V}\Big]\partial_r (rF_b^0) +\big[\quart(M-V)^2-m^2\big](rF_b^0) =0 \hspace{1cm} (\jz=0)
\eeq
Now $F_b^0(r\to 0) \propto r^{\pm 1}$, and we need to choose $F_b^0(r\to 0) \propto r$ for a finite density at the origin. This determines the solution up to its overall normalization.

For $F_c^0$ to be finite at $M-V=0$ we must have $F_b^0(r\to M/V') \propto (M-V)^2$. This allows only discrete values of $M$. Note that this condition is absent if $m=0$, but does persist as $m\to 0$, as long as $m \neq 0$. In $D=1+1$ dimensions the bound state equation \eq{msig2ndord} similarly has singular solutions at $\tau=0$ only when $m\neq 0$.

\vspace{.5cm}
(c) Numerical example 

\begin{figure}[h] \centering
\includegraphics[width=1.\columnwidth]{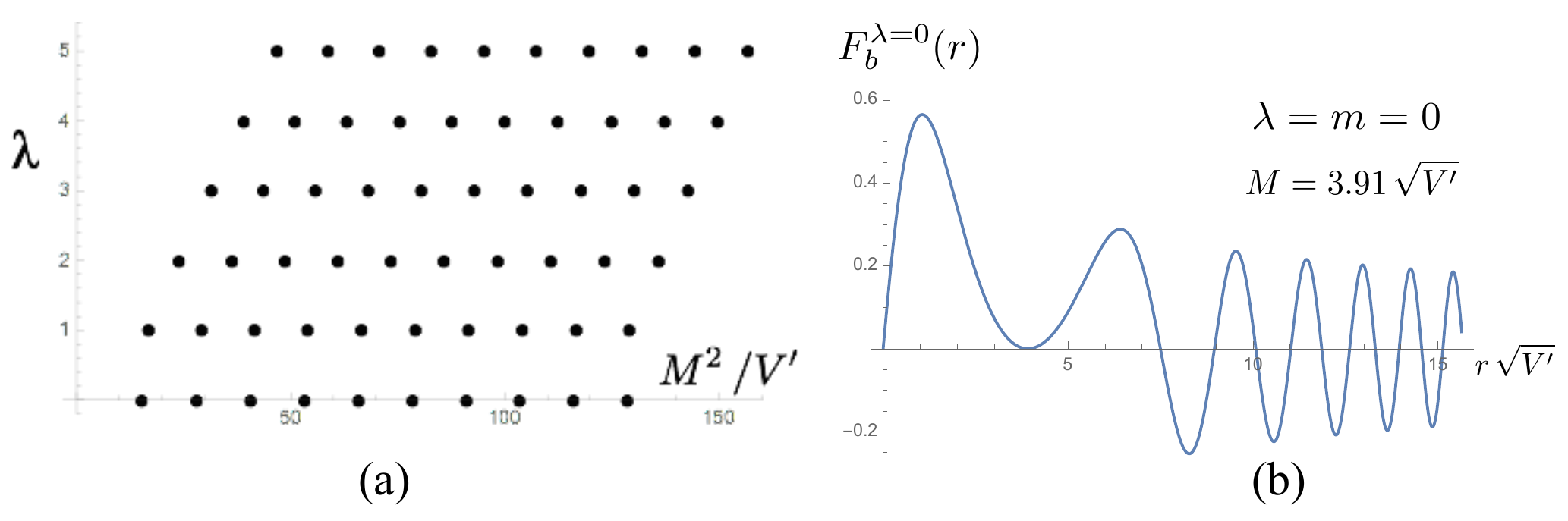}
\caption{(a) Mass spectrum of the $D=2+1$ mesons for $m=0$, when the only energy scale is $\sqrt{V'}$. (b) Plot of the (arbitrarily normalized) radial wave function $F_b^{\jz=0}(r)$ of the ground state. \label{2+1_CF+Wf}}
\end{figure}

The radial bound state equations \eq{21adouble}, \eq{21bdouble} can be readily solved numerically, and the value of $M$ determined which makes the wave function vanish at $M-V=0$. In \fig{2+1_CF+Wf} I show some examples for the case of vanishing quark mass, $m=0$. The Chew-Frautshi type plot in \fig{2+1_CF+Wf}(a) demonstrates that the spectrum resembles that of the dual model, with a linear dependence of $M^2$ on $J^z=\jz$ for the leading trajectory, and parallel daughter trajectories corresponding to radial excitations. The $\jz=0$ states appear to lie on trajectories with smaller slope. The $F_b^{\jz=0}(r)$ wave function in \fig{2+1_CF+Wf}(b) is seen to vanish at $M=V(r)= 3.91\sqrt{V'}$, and oscillate at larger radii according to \eq{21asymp1}. 

\begin{figure}[h] \centering
\includegraphics[width=.6\columnwidth]{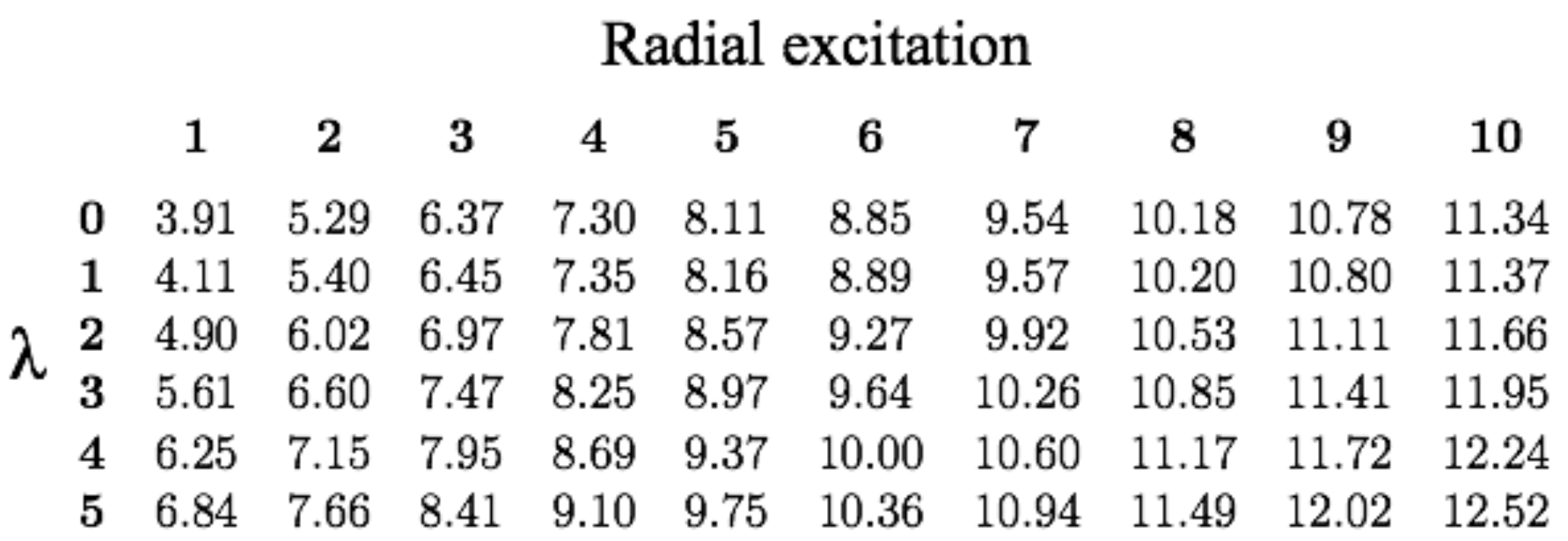}
\caption{Masses $M$ of the $D=2+1$ mesons for $m=0$, in units of $\sqrt{V'}$. \label{2+1_Table}}
\end{figure}

The values of the bound states masses plotted in \fig{2+1_CF+Wf}(a) are given in \fig{2+1_Table}.

\subsection{Mesons in $D=3+1$ dimensions \label{mes31subsec}}

\subsubsection{Quantum numbers \label{31qnsubsubsec}}

The separation of variables in the meson bound state equation \eq{mbse} is less straightforward in $D=3+1$ than in lower dimensions. However, limiting ourselves to the equal mass case $m_1=m_2=m$ we can use the results of Geffen and Suura \cite{Geffen:1977bh}, who considered the same equation (multiplied by $\gz$ from the left). They studied the spectrum for a phenomenological potential $V(r)$. For future reference I  keep $V''=\partial_r^2 V(r)$ in the radial equations, although $V''=0$ for the linear potential \eq{mpot}.

Considering the angular momentum $(J,J^z=\jz)$, parity $\eta_P$ and charge conjugation $\eta_C$, similarly to section \ref{symsubsubsec}, the states can be grouped into three ``Regge trajectories'' \cite{Geffen:1977bh},
\beq\label{31traj}
\begin{array}{lll}\pi\ trajectory: & \eta_P=(-1)^{J+1} &\hspace{.5cm} \eta_C=(-1)^{J} \\[2mm]
a_1\ trajectory: & \eta_P=(-1)^{J+1} &\hspace{.5cm} \eta_C=(-1)^{J+1} \\[2mm]
\rho\ trajectory: & \eta_P=(-1)^{J} &\hspace{.5cm} \eta_C=(-1)^{J} 
\end{array} 
\eeq
The names associate to well-known mesons with the corresponding quantum numbers. However, we are considering a single flavor so there is no isospin. It should be easy to added more flavors, whereas the case of $m_1 \neq m_2$ may require more effort.
Since we are considering relativistic dynamics, the orbital angular momentum $L$ and spin $S$ are not separately conserved. In the quark model $\eta_P=(-1)^{L+1}$ and $\eta_C=(-1)^{L+S}$, which implies that the combination $\eta_P=(-1)^{J},\ \eta_C=(-1)^{J+1}$ should not exist. The quantum numbers \eq{31traj} which are allowed by the meson bound state equation \eq{mbse} agree with the quark model in this respect. Furthermore, the quark model does not allow $J^{PC} = 0^{--}$, and the $J=0$ state is indeed missing on the $a_1$ trajectory.

I next quote the results of \cite{Geffen:1977bh} for each of the above trajectories in terms of the standard spherical coordinates $(r,\theta,\vphi)$ and the spherical harmonics $Y_{J\jz}$. 

\subsubsection{$\pi$ trajectory \label{31pisubsubsec}}

The $4\times 4$ wave functions on the pion trajectory with angular momentum $J$ and $J^z=\lambda$ have the form
\beq\label{pidirstr}
\Phi_\pi(\xv)=\Big[-\frac{2im}{M-V}\,\gz\gff-i\gff + \frac{2}{M-V}\,\gff\,\alv\cdot\nv\Big]F_2(r)Y_{J\lambda}(\theta,\vphi)
\eeq

The radial function satisfies the second order differential equation
\beq\label{pirad}
F_2''(r) + \Big(\,\frac{2}{r}+\frac{V'}{M-V}\Big)F_2'(r) + \Big[\quart(M-V)^2-m^2-\frac{J(J+1)}{r^2}\Big]F_2(r)=0
\eeq
which has a structure similar to \eq{21adouble} in the $D=2+1$ case and \eq{m2ndord} in $D=1+1$.

The behavior of $F_2(r)$ at large radii may be found similarly as in section \ref{limitsubsubsec}(b). For $V(r)=V'r$,
\beq\label{piasr}
\lim_{r\to\infty} F_2(r) = N\,\inv{r}\,r^{-im^2/V'}\exp\left[i\,\frac{(M-V)^2}{4V'}\right]
\eeq
where $N$ is a constant. The factor $1/r$ compensates for phase space so that the local density is constant,
\beq\label{pinorm}
\int d\xv\,\tr\big[\Phi^\dag(\xv)\Phi(\xv)\big] \sim \int dr\,r^2 \frac{8|N|^2}{r^2} \sim 8|N|^2\int dr
\eeq
As in $D=1+1$ and $D=2+1$ this allows an interpretation of the constant density as representing a constant rate of string breaking per unit length. The precise meaning and correctness of such an interpretation requires further study.

The radial equation \eq{pirad} allows $F_2(r\to 0) \propto r^\alpha$ with $\alpha=J$  and $\alpha=-J-1$. We must choose $\alpha=J$ for the norm in \eq{pinorm} to be integrable at $r=0$. Similarly $F_2(r\to M/V') \propto (M-V)^\beta$ imposes $\beta=0,\ 2$ and integrability of the norm at $r= M/V'$ requires $\beta=2$. Together with the condition at $r=0$ this can only be satisfied for discrete values of $M$. 

\begin{figure}[h] \centering
\includegraphics[width=1.\columnwidth]{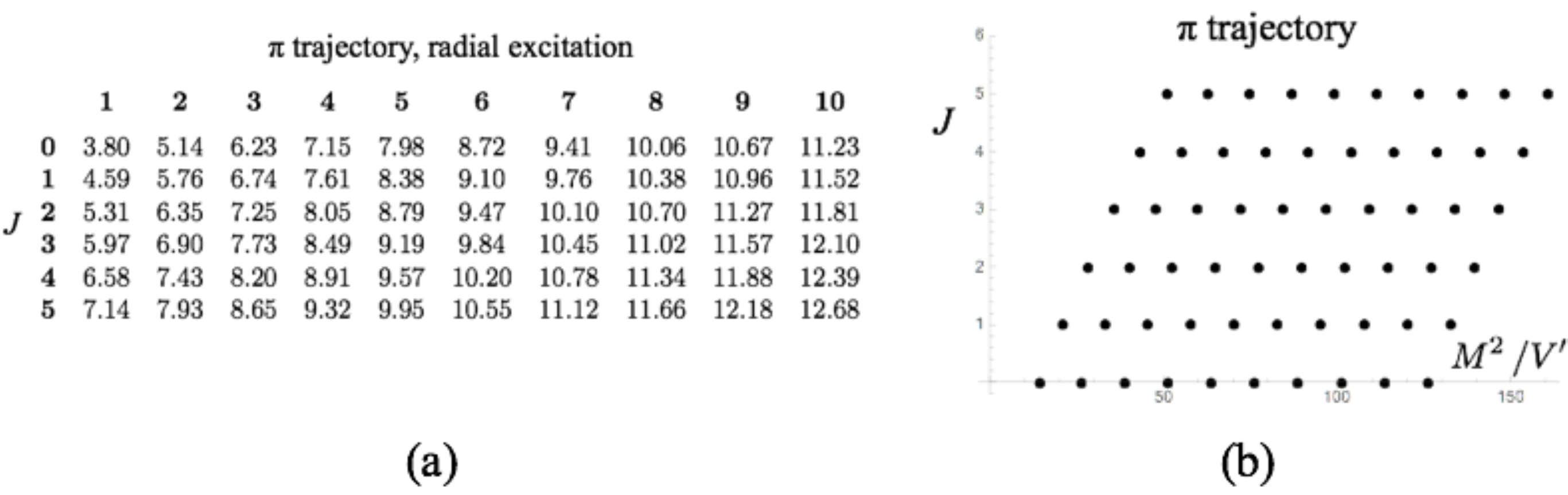}
\caption{(a) Masses $M$ of the mesons on the $\pi$ trajectory for $m=0$, in units of $\sqrt{V'}$. (b) Plot of the spin $J$ {\em vs.} $M^2/V'$ for the states listed in (a). \label{3+1_pion}}
\end{figure}

\fig{3+1_pion}(a) lists the masses $M/\sqrt{V'}$ of the mesons on the $\pi$ trajectory when the quark mass $m=0$. The states lie on nearly linear Regge trajectories as shown in (b).

The radial equation \eq{pirad} allows a locally normalizable solution also for $M=0$, in which case the potentially singular points $r=0$ and $r=M/V'$ coincide. Massless solutions in the rest frame have four-momenta $P=0$ in all frames. The $M=0,\ J^{PC}=0^{-+}$ state on the pion trajectory might take the role of a Goldstone boson in chiral symmetry breaking. I return to this issue in section \ref{csbsubsubsec}. 

\subsubsection{$a_1$ trajectory \label{31a1subsubsec}}

The wave functions on the $a_1$ trajectory have the form
\beq\label{a1dirstr}
\Phi_{a1}(\xv)=\Big[\gv\cdot \bs{L}+\frac{2m}{M-V}\,\alv\cdot \bs{L} - \frac{2}{M-V}\,\gff\,\gv\cdot\nv\times\bs{L} \Big]F_1(r)Y_{J\lambda}(\theta,\vphi)
\eeq
Since the orbital angular momentum 
\beq\label{Ldef}
\bs{L} = -i\,\xv\times\nv
\eeq
differentiates only wrt. $\theta$ and $\vphi$ the $a_1$ wave functions \eq{a1dirstr} are non-vanishing only for $J \geq 1$.
The radial function satisfies the second order differential equation
\beq\label{a1rad}
F_1''(r) + \Big(\,\frac{2}{r}+\frac{V'}{M-V}\Big)F_1'(r) + \Big[\quart(M-V)^2-m^2-\frac{J(J+1)}{r^2}+\frac{V'}{r(M-V)}\Big]F_1(r)=0
\eeq

\begin{figure}[h] \centering
\includegraphics[width=1.\columnwidth]{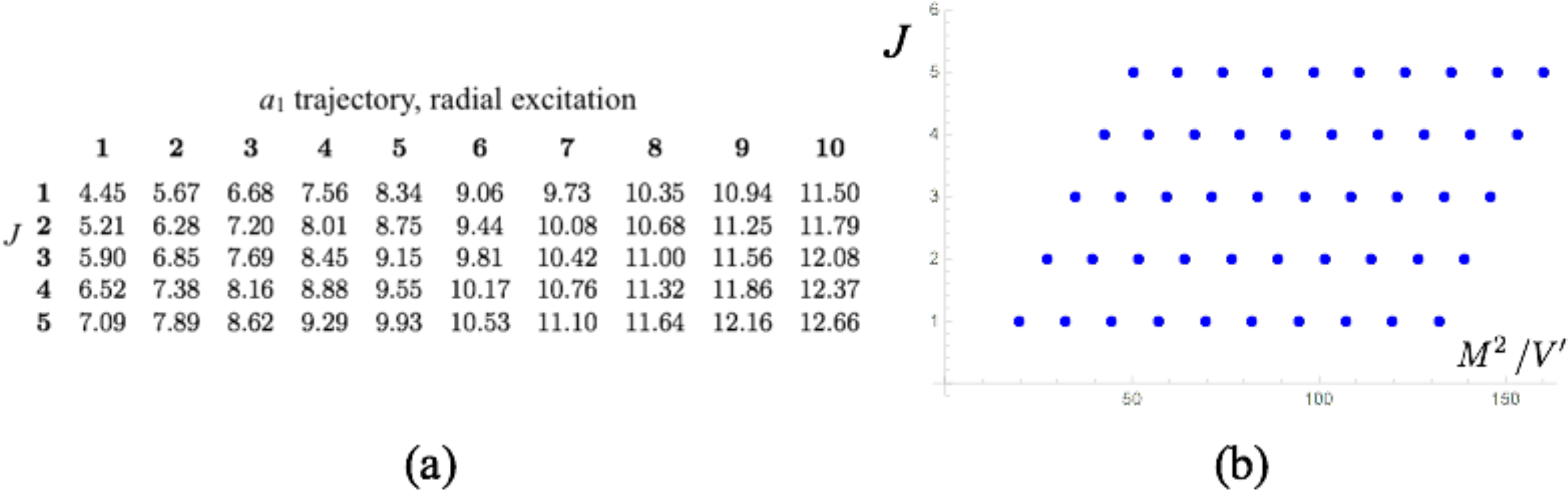}
\caption{(a) Masses $M$ of the mesons on the $a_1$ trajectory for $m=0$, in units of $\sqrt{V'}$. (b) Plot of the spin $J$ {\em vs.} $M^2/V'$ for the states listed in (a). \label{3+1_a1}}
\end{figure}

The behavior of this equation for large $r$ is the same as the pion equation \eq{pirad}. Hence the asymptotic limit $F_1(r\to\infty)$ is the same as that of $F_2(r)$ in \eq{piasr}. The leading behaviors at $r=0$ and $M-V=0$ are also the same as for the pion trajectory. The numerical results for $m=0$ are given in \fig{3+1_a1}.

\subsubsection{$\rho$ trajectory \label{31rhosubsubsec}}

The wave functions of states on the $\rho$ trajectory have the form
\beqa\label{rhodirstr}
\Phi_{\rho}(\xv) &=&  \gv\cdot\nv\left\{{\displaystyle \frac{4}{M-V}}\Big[-i\halft mF_2(r)+\Big({\displaystyle \frac{rG_1(r)}{M-V}}\Big)'\, \Big]\, Y_{J\jz}(\theta,\vphi)\right\} + \gv\cdot\xv\,G_1(r)\, Y_{J\jz}(\theta,\vphi)\nn \\[6mm] 
&& -i \alv\cdot\nv\big[ F_2(r)\, Y_{J\jz}(\theta,\vphi)\big] + {\displaystyle \frac{\alv\cdot\xv}{M-V}\Big[\frac{iV'}{r}F_2(r)+2mG_1(r)\Big]}Y_{J\jz}(\theta,\vphi) \nn\\[6mm]
&& +\,\mathbb{1}\, \Big[{\displaystyle -\frac{(M-V)^2-4m^2}{2(M-V)}\,F_2(r)+ \frac{4im}{M-V}\Big(\frac{rG_1(r)}{M-V}\Big)'}\, \Big]\, Y_{J\jz}(\theta,\vphi)\nn \\[6mm]
&& +\,\gff\,\gv\cdot\bs{L} {\displaystyle\frac{2i}{M-V}}\,G_1(r)\, Y_{J\jz}(\theta,\vphi)
\eeqa
where the radial function $F_2(r)$ is unrelated to the $F_2$ of the $\pi$ solution \eq{pidirstr}.
The radial functions satisfy coupled second order equations,
\beqa
G_1''(r)+\Big(\frac{2}{r}+\frac{3V'}{M-V}\Big) G_1'(r)+\Big[\quart(M-V)^2-m^2-\frac{J(J+1)}{r^2} +\frac{{3V'}^2}{(M-V)^2}+\frac{3V'}{r(M-V)}+\frac{V''}{M-V}\Big]G_1(r)&& \nn\\[2mm]
 = \frac{imV'}{r}\,F_2(r) \hspace{1cm} &&  \label{rhoG1} \\[4mm]
F_2''(r)+\Big(\frac{2}{r}-\frac{V'}{M-V}\Big) F_2'(r)+\Big[\quart(M-V)^2-m^2-\frac{J(J+1)}{r^2} -\frac{{V'}^2}{(M-V)^2}-\frac{2V'}{r(M-V)}-\frac{V''}{M-V}\Big]F_2(r)&&  \nn\\[2mm]
 = -\frac{4im}{M-V}\,G_1(r) \hspace{1cm} &&  \label{rhoF2}
\eeqa
where $V''=0$ for a linear potential. The $\rho$ solutions have twice as many degrees of freedom as the $\pi$ or $a_1$ trajectories.

The identification of the physical, locally normalizable solutions at $r=0$ and $M-V=0$ requires care due to the coupled equations. I only summarize the discussion in \cite{Geffen:1977bh}.

\vspace{.5cm}
(a) $r = 0$
\beq\label{rhor0}
G_1(r\to 0) = \morder{r^J} \hspace{3cm} F_2(r\to 0) = \morder{r^J}
\eeq
This ensures that the most singular terms of \order{r^{J-2}} vanish separately in \eq{rhoG1} and \eq{rhoF2}, and that the norm of the wave function is locally normalizable at $r=0$. The ratio $\lim_{r\to 0}G_1(r)/F_2(r)$ remains a free parameter (in addition to the overall normalization). Together with the bound state mass $M$ this allows to choose normalizable solutions for $G_1$ and $F_2$ at $M-V=0$.

\vspace{.5cm}
(b) $M-V(r)= 0$

Assuming $G_1 \sim (M-V)^\alpha$ and $F_2 \sim (M-V)^\beta$ we may first conclude that the terms of ${\cal O}\big[(M-V)^{\alpha-2}\big]$ and ${\cal O}\big[(M-V)^{\beta-2}\big]$ on the lhs. of \eq{rhoG1} and \eq{rhoF2}, respectively, must vanish. This allows $\alpha=1,3$ and $\beta=\pm 1$. Local normalizability of the norm of $\Phi_{\rho}$ imposes $\beta=1$. Comparing the powers of $M-V$ on the lhs. and rhs. of \eq{rhoG1} excludes $\alpha=3$, leaving $\alpha=\beta=1$. The normalizability of the $G_1$ contributions to $\Phi_{\rho}$ imposes a specific behavior of $G_1$ at $r=R=M/V'$,
\beqa\label{rhoMV}
G_1(r\to R) &=& \frac{M-V}{r}\Big[A+B(M-V)^2\big]+{\cal O}\big[(M-V)^4\big] \label{rhoMVG} \\[2mm] 
F_2(r\to R) &=& -\frac{F_2'(R)}{V'}(M-V)+ {\cal O}\big[(M-V)^2\big] \label{rhoMVF}
\eeqa
We may verify that the contributions of ${\cal O}\big[(M-V)^0\big]$ to \eq{rhoG1} vanish. At ${\cal O}\big[(M-V)^1\big]$ we find a relation between the coefficient $A$ and $F_2'(R)$,
\beq\label{Aconstr}
A = \frac{im}{m^2+J(J+1)/R^2}\,F_2'(R)
\eeq

\vspace{.5cm}
(c) $m \to 0$, $J\geq 1$

The limit $m\to 0$ is interesting because of the chiral invariance of the action when $m=0$. When $J\geq 1$ we have $A \to 0$ in \eq{Aconstr}, and the rhs. of \eq{rhoG1} and \eq{rhoF2} vanish. With the redefinitions
\beqa\label{redef}
G_1(r) &=& (M-V)H_1(r) \nn\\[2mm]
F_2(r) &=& \inv{M-V}\,H_2(r)
\eeqa
the equation \eq{rhoG1} for $H_1$ becomes the equation \eq{a1rad} for the $a_1$ trajectory (with $m=0$), and  \eq{rhoF2} for $H_2$ becomes the equation \eq{pirad} for the pion trajectory. Thus the spectra are parity degenerate in the $m\to 0$ limit when $J\geq 1$ \cite{Geffen:1977bh}.

\vspace{.5cm}
(d) $J=0$ state on $\rho$ trajectory

The relation \eq{Aconstr} is singular when $m\to 0$ and $J=0$. It is therefore easier to consider the spinless case separately.
The $J=0$ wave functions on the $\rho$ trajectory take the form
\beq\label{rhodirstr0}
\Phi_{\rho}^{J=0}(\xv) = -\frac{2im}{M-V}\,\gv\cdot\xv\,G_2(r)-i\alv\cdot\xv\,G_2(r)+\frac{2\big[r^3G_2(r)\big]'}{r^2(M-V)}\,\mathbb{1}
\eeq

The radial function $G_2(r)$ satisfies
\beq\label{rhorad0}
G_2''(r) + \Big(\,\frac{4}{r}+\frac{V'}{M-V}\Big)G_2'(r) + \Big[\quart(M-V^2)-m^2+\frac{3V'}{r(M-V)}\Big]G_2(r)=0
\eeq
Since this equation differs from the $J=0$ pion equation \eq{pirad} the $J^{PC}=0^{++}$ and $J^{PC}=0^{-+}$ mesons do not have the same mass. The absence of parity degeneracy arises from the requirement of local normalizability and implies the breaking of chiral invariance \cite{Geffen:1977bh}. I return to this issue in section \ref{cstranssubsubsec1}.

The behavior of $G_2(r\to\infty)$ may be analyzed as for the pion, giving $|G_2(r\to\infty)| \sim r^{-2}$. It implies a constant asymptotic norm of the wave function, as in \eq{pinorm}.

\begin{figure}[h] \centering
\includegraphics[width=.6\columnwidth]{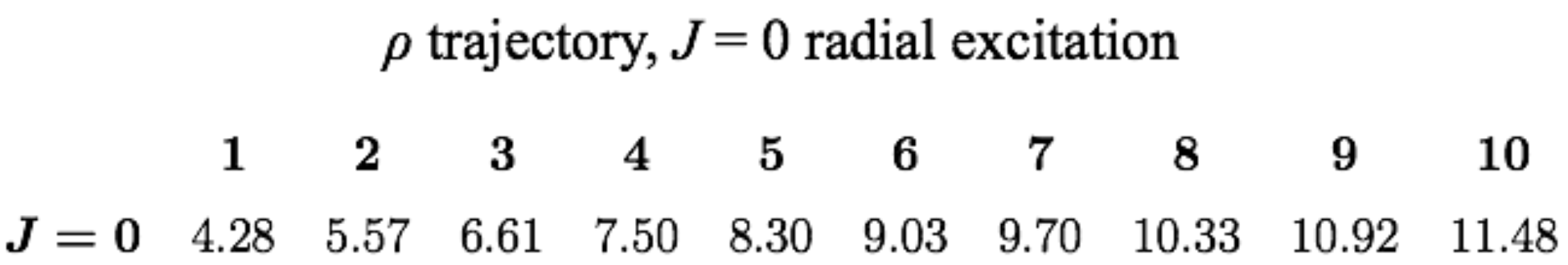}
\caption{Masses $M$ of the $J=0$ mesons on the $\rho$ trajectory for $m=0$, in units of $\sqrt{V'}$. The $\rho$ trajectory states with $J\geq 1$ are degenerate with those of the same spin but opposite parity on the $\pi$ and $a_1$ trajectories. \label{3+1_rho}}
\end{figure}

The requirement of local normalizability implies
\beq
G_2(r \to 0) \sim r^0 \hspace{2cm} G_2(M-V \to 0) \sim (M-V)^2
\eeq 
The numerical values of the masses $M$ for $J=m=0$ are given in \fig{3+1_rho}.

\subsection{Chiral symmetry \label{cs31subsec}}

\subsubsection{Chiral transformations \label{cstranssubsubsec}}

For vanishing quark mass $m=0$ the QCD action is invariant under (global) chiral transformations\footnote{I disregard the $U(1)$ anomaly. The present discussion for a single quark flavor is illustrative and should be easily generalized to the case of two light flavors.}. The chiral transformation operator is
\beq\label{chiop}
U_\chi(\alpha) = \exp\Big[i\alpha\int d\xv\,\psi^\dag(\xv)\gff\psi(\xv)\Big]
\eeq
where $\gff$ anticommutes with all the Dirac matrices. The operator \eq{chiop} transforms the fields as
\beq\label{psichi}
U_\chi(\alpha)\psi(\xv)U_\chi^\dag(\alpha) = \exp(-i\alpha\gff)\psi(\xv) \hspace{2cm}
U_\chi(\alpha)\bar\psi(\xv)U_\chi^\dag(\alpha) = \bar\psi(\xv)\exp(-i\alpha\gff)
\eeq

For a chiral invariant the vacuum, $U_\chi(\alpha)\ket{0}=\ket{0}$, the meson state \eq{meson} is transformed as
\beq\label{mesonchi}
U_\chi(\alpha)\ket{M,\Pv=0} = \int d\xv_1\,d\xv_2\,\bar\psi(\xv_1)\exp(-i\alpha\gff)\Phi(\xv_1-\xv_2)\exp(-i\alpha\gff)\psi(\xv_2)\ket{0}
\eeq
Thus the wave function is transformed into
\beq\label{wfchi}
\Phi_\chi(\xv) \equiv \exp(-i\alpha\gff)\Phi(\xv)\exp(-i\alpha\gff)
\eeq
Multiplying the bound state equation \eq{mbse} by $\exp(-i\alpha\gff)$ from both the left and the right we note that if $m_1=m_2=0$ then $\Phi_\chi(\xv)$ is a solution with the same eigenvalue $M$ as $\Phi(\xv)$. 

The $m=0$ states may be classified as chiral even or odd \cite{Geffen:1977bh} according to the (anti)commutation of their wave function with $\gff$,
\beq\label{chisign}
\acom{\gff}{\Phi^+}=0 \hspace{3cm} \com{\gff}{\Phi^-}=0
\eeq
Chiral invariance implies that $\Phi^\pm$ and $\gff\Phi^\pm$ have the same mass eigenvalues $M$. If $\Phi^\pm$ has parity $\eta_P$ and charge conjugation $\eta_C$ we find from \eq{parop3} and \eq{ccop3} that 
\beq\label{pcrel}
\gz\gff\Phi^\pm(-\xv)\gz = -\eta_P\,\gff\Phi^\pm(\xv) \hspace{2cm}
\gz\gamma^2\big[\gff\Phi^\pm(-\xv)\big]^T\gz\gamma^2 = \mp\eta_C\,\gff\Phi^\pm(\xv)
\eeq
Hence chiral invariance implies the existence of mass degenerate pairs of bound states of opposite parity and opposite (equal) charge conjugation for chiral even (odd) states. Let us see how this is realized for the solutions found above.

\subsubsection{Chiral properties of the bound state solutions \label{cstranssubsubsec1}}

(a) $D=1+1$ dimensions

In the Pauli matrix representation \eq{matdef} of the Dirac matrices $\gff=\gz\go=\sigma_1$, since it anticommutes with both $\gz$ and $\go$. In the expansion \eq{wf11} of the wave function $\phi_2=\phi_3=0$ when $m=0$, according to \eq{phi23}. Consequently the chiral even states vanish and
\beq\label{chisol11}
\Phi^-(x) = \phi_0(x)\,\mathbb{1}+\phi_1(x)\,\dsi_1 \hspace{2cm}
\gff\Phi^-(x) = \phi_1(x)\,\mathbb{1}+\phi_0(x)\,\dsi_1
\eeq

In the limit $m \to 0$ the local integrability of the norm $\tr\Phi^\dag(x)\Phi(x)$ imposed the discrete spectrum \eq{m0spect}, with no parity degeneracy. The potential singularity at $M-V=0$ was due to the contribution $\propto m^2$ in the bound state equation \eq{m1stord2}. Remarkably, chiral symmetry is broken for states with non-singular wave functions in the limit $m \to 0$. When $m=0$ exactly the norm is regular for any value of the bound state mass $M$. The continuous spectrum then implies also parity doubling.

\vspace{.5cm}
(b) $D=2+1$ dimensions

In odd numbers of dimensions there is no ``$\gff$'' matrix that would commute with all the Dirac matrices. This is seen explicitly in the representation \eq{21dirdef}, where we needed all three Pauli matrices to represent $\gz,\go$ and $\gamma^2$. Consequently chiral invariance is not an issue.

\vspace{.5cm}
(c) $D=3+1$ dimensions

For $m=0$ the chiral partners of the states on the pion trajectory have wave functions\footnote{I add a superscript $\pi$ ($\rho$) to distinguish the radial functions $F_2(r)$ of the $\pi$ ($\rho$) wave functions.}
\beq\label{pidirstr2}
\gff\Phi_\pi(\xv)=\Big[-i\,\mathbb{1} + \frac{2}{M-V}\,\alv\cdot\nv\Big]F_2^\pi(r)Y_{J\lambda}(\theta,\vphi) \hspace{2cm} \com{\gff}{\Phi_\pi}=0
\eeq
Since the pion trajectory has only chiral odd states their parity partners have $\eta_P=\eta_C=(-1)^J$, \ie, they belong to the $\rho$ trajectory in \eq{31traj}. The chiral odd component of the $\rho$ wave function \eq{rhodirstr} is for $m=0$,
\beq\label{rhodirstr2}
\Phi_\rho^- = \Big[-i\alv\cdot\nv+\alv\cdot\xv\,\frac{iV'}{r(M-V)}-\halft(M-V)\,\mathbb{1}\Big]F_2^\rho\, Y_{J\lambda} = \gff\Phi_\pi \ \ \ {\rm for}\ \ \ F_2^\rho(r)=\frac{2i}{M-V}\,F_2^\pi(r) \ \ (J \neq 0)
\eeq
The relation between the wave functions of the parity partners is compatible with \eq{redef}. The case $J=0$ needs to be considered separately because $G_1/F_2^\rho \propto 1/m$ according to \eq{Aconstr}, so that terms $\propto mG_1$ in \eq{rhodirstr} cannot be neglected. 

For $m=0$ the $J=0$ $\rho$ wave function \eq{rhodirstr0} has only a chiral odd component,
\beq\label{rhodirstr3}
\Phi_{\rho}^{J=0,-}(\xv) = -i\alv\cdot\xv\,G_2(r)+\frac{2\big[r^3G_2(r)\big]'}{r^2(M-V)}\,\mathbb{1}
\eeq
Its chiral $0^{-+}$ partner on the pion trajectory has the radial wave function
\beq\label{rhochipion0}
\overline{F_2^\pi} = \frac{2i}{M-V}(rG_2'+3G_2)
\eeq
This gives a singular pion wave function since $\overline{F_2^\pi}(r\to M/V') \sim (M-V)^0$. Thus requiring a locally finite norm implies the breaking of chiral invariance in the $m \to 0$ limit: The $0^{++}$ and $0^{-+}$ states have distinct masses, as seen also by comparing \fig{3+1_pion} with \fig{3+1_rho}.

The chiral odd states on the $a_1$ trajectory \eq{a1dirstr} vanish when $m=0$. The chiral even partners on the $a_1$ and $\rho$ trajectories have wave functions
\beqa\label{a1dirstr2}
\gff\Phi_{a1}(\xv)&=&\Big[\gff\,\gv\cdot \bs{L} - \frac{2}{M-V}\,\gv\cdot\nv\times\bs{L} \Big]F_1\,Y_{J\lambda} \\[2mm]
\Phi_{\rho}^+(\xv)&=& \Big[\gff\,\gv\cdot \bs{L}\,\frac{2i}{M-V}+\gv\cdot\xv \Big]G_1\,Y_{J\lambda}+ \gv\cdot\nv\Big[\frac{4}{M-V}\Big(\frac{rG_1}{M-V}\Big)'\,Y_{J\lambda}\Big]
\eeqa
Comparing the coefficients of $\gff\,\gv\cdot \bs{L}$ gives
\beq\label{a1chirpart}
G_1 = -\frac{i}{2}(M-V)F_1
\eeq
The differential equation \eq{a1rad} ensures that also the other terms agree, $\gff\Phi_{a1}=\Phi_{\rho}^+$. Both wave functions are non-vanishing only for $J \geq 1$.

\subsubsection{Massless bound states \label{csbsubsubsec}}

There is clear evidence that the $u$ and $d$ quarks have small but non-vanishing masses, and that the approximate chiral symmetry is spontaneously broken (CSB) in QCD \cite{Weinberg:1996kr}. An important consequence of CSB is the appearance of a nearly massless Goldstone boson, the pion, with $M_\pi^2\propto m_u,m_d$. The pion cannot be even approximately described using non-relativistic potential models -- the ``hyperfine splitting'' $M_\rho-M_\pi \simeq 630$ MeV is much larger than $M_\pi \simeq 140$ MeV.

The states discussed above have a regular wave function at the two potentially singular points $r=0$ and $r=M/V'$. There are also regular solutions with $M=0$, for which the two points coincide \cite{Dietrich:2012un}. These states have four-momenta $P^\mu=0$ in all reference frames. The $J^{PC}=0^{++}$ state $\sigma$ on the $\rho$ trajectory discussed in section \ref{31rhosubsubsec}(d) is of particular interest. Since $\sigma$ has vacuum quantum numbers and vanishing energy when $M=0$ it can mix with the chiral invariant vacuum to form a ground state that breaks chiral symmetry. 

\begin{figure}[h] \centering
\includegraphics[width=.6\columnwidth]{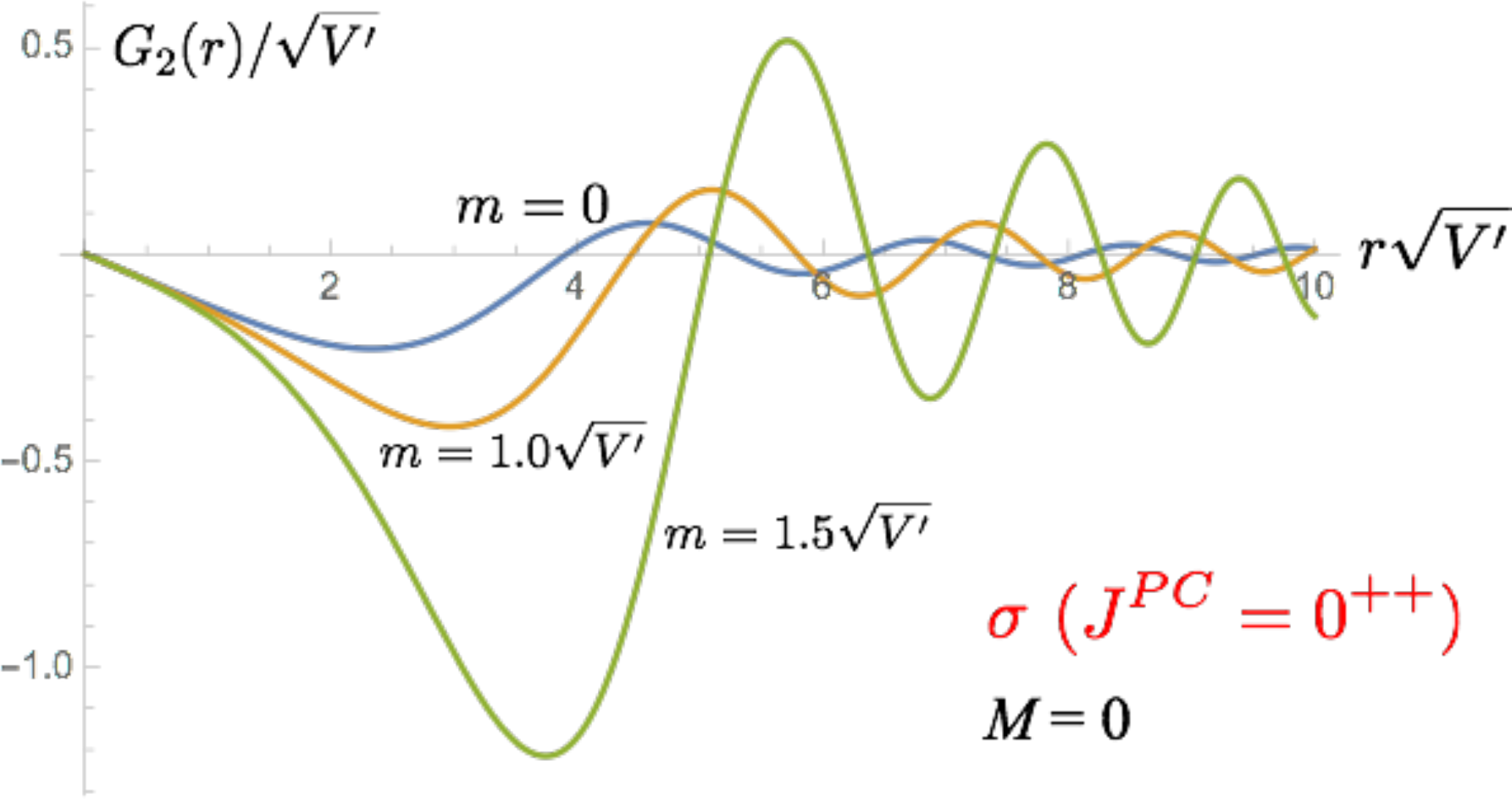}
\caption{The radial function $G_2(r)$ \eq{rhorad00sol} for the $J^{PC}=0^{++}$ $\sigma$ state with $M=0$ on the $\rho$ trajectory, for three values of the quark mass $m$ and $N=1$. \label{3+1_sigma}}
\end{figure}

According to \eq{rhodirstr0} the wave function of the $\sigma$ is for $M=0$, denoting $\hat{\xv}=\xv/r$,
\beq\label{rhochi1}
\Phi_\sigma(\xv)=-\frac{2}{V'}\,\frac{\big[r^3G_2(r)\big]'}{r^3}\,\mathbb{1}-irG_2(r)\,\gz\gv\cdot\hat{\xv}+\frac{2im}{V'}
\,G_2(r)\,\gv\cdot\hat{\xv}
\eeq
The radial wave function satisfies \eq{rhorad0},
\beq\label{rhorad00}
G_2''(r) + \frac{3}{r}\,G_2'(r) + \Big[\quart (V'r)^2-m^2-\frac{3}{r^2}\Big]G_2(r)=0
\eeq
The locally normalizable solution at $r=0$ is
\beq\label{rhorad00sol}
G_2(r) = \frac{N}{r^3}\,\exp\big(-i\quart V' r^2\big)L_{\halft+im^2/2V'}^{-2}\big(i\halft V' r^2\big) \hspace{1cm} (M=0)
\eeq
where $L_n^a$ is a generalized Laguerre function and $N$ an arbitrary normalization constant. \fig{3+1_sigma} shows a plot of $G_2(r)$ for three values of the quark mass $m$. The slope $G_2'(r=0)=-N/8$ is independent of $m$, while the amplitude of the oscillations at large $r$ are $\propto \exp(\pi m^2/2V')/(mr^2)$, increasing exponentially with quark mass. For massless quarks,
\beq\label{G2m0}
G_2(r) = \frac{N}{r}\,J_1(\quart V'r^2) \hspace{1cm} (M=m=0)
\eeq 
where $J_1$ is a Bessel function.

\begin{figure}[h] \centering
\includegraphics[width=.6\columnwidth]{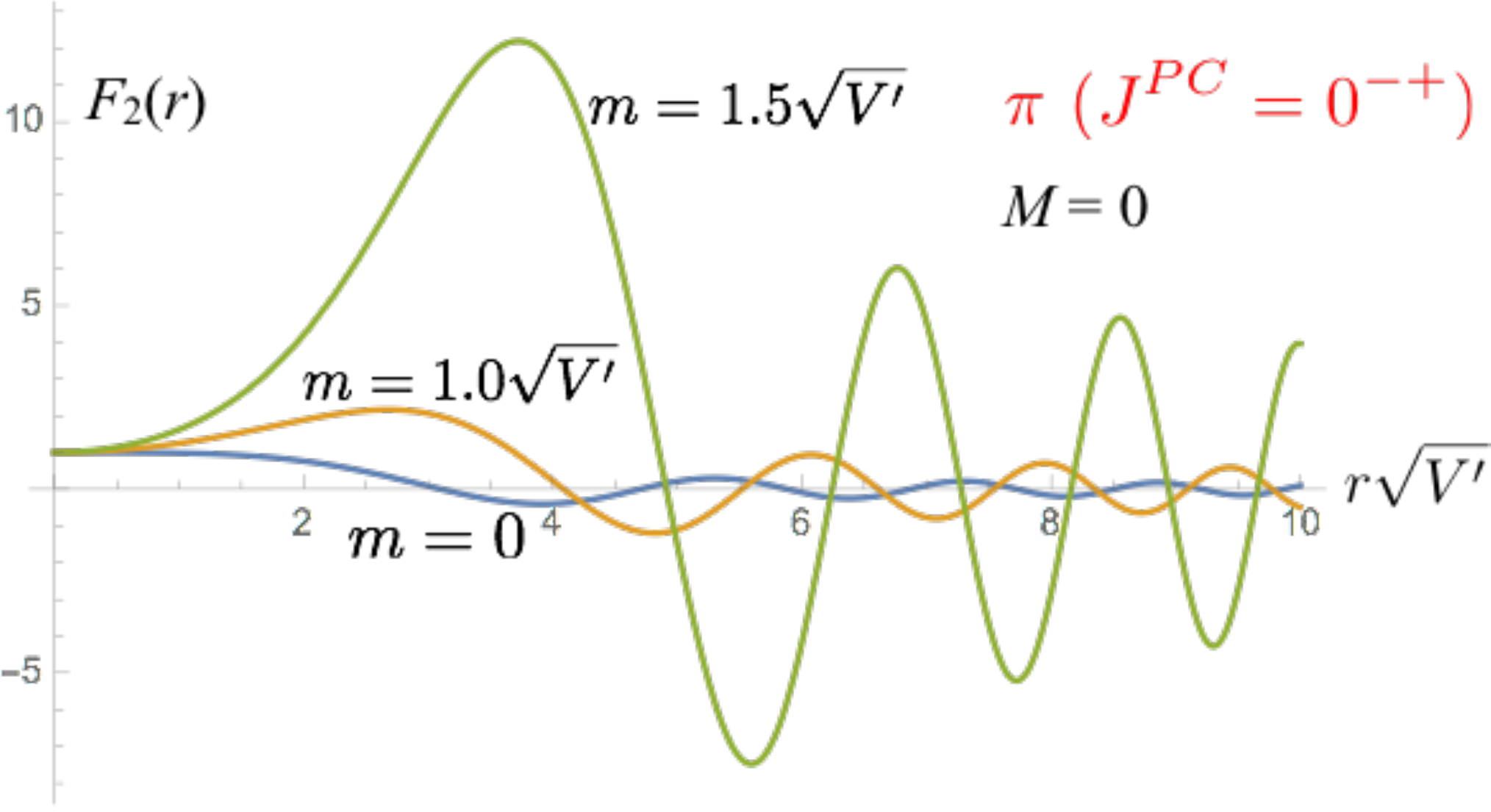}
\caption{The radial function $F_2(r)$ \eq{pisol0} for the $J^{PC}=0^{-+}$ $\pi$ state with $M=0$, for three values of the quark mass $m$ and $N=1$. \label{3+1_gold}}
\end{figure}

The $J^{PC}=0^{-+}$ state with $M=0$ on the pion trajectory has the quantum numbers of the Goldstone boson. Its wave function \eq{pidirstr} is
\beq\label{pichi1}
\Phi_\pi(\xv)= -i\gff\,F_2(r) -\frac{2}{V'r}\,\gff\,\alv\cdot\hat{\xv}\,F_2'(r)+\gz\gff\,\frac{2im}{V'r}\,F_2(r)
\eeq
The radial function satisfies
\beq\label{pirad0}
F_2''(r) + \frac{1}{r}F_2'(r) + \big[\quart (V'r)^2-m^2\big]F_2(r)=0
\eeq
with the regular solution
\beq\label{pisol0}
F_2(r)= N\,e^{-iV'r^2/4} \kum(\halft-i\sfrac{m^2}{2V'},1,i\halft V'r^2) \hspace{1cm} (M=0)
\eeq
where $N$ is an arbitrary normalization constant. 
In the limits of small and large $r$,
\beqa
F_2(r\to 0) &=& N\big[1+\quart (mr)^2+\sfrac{1}{64}(m^4-{V'}^2)\,r^4+\morder{r^6}\big] \label{F20} \\[2mm]
F_2(r\to \infty) &=& N\,\frac{2\sqrt{2}\,e^{\pi m^2/4V'}}{r\sqrt{V'}}{\rm Re}\left\{(\halft V'r^2)^{-im^2/2V'}\frac{e^{i(V'r^2-\pi)/4}}{\Gamma\big[\halft(1-im^2/V')\big]}\right\}\Big[1+\morder{r^{-2}}\Big]  \label{F2inf}
\eeqa
For massless quarks,
\beq
F_2(r) = N\,J_0(\quart V'r^2) \hspace{1cm} (M=m=0)
\eeq
\fig{3+1_gold} shows a plot of $F_2(r)$ for three values of the quark mass $m$.

The existence of $M=0$ solutions is interesting, considering the features of spontaneous chiral symmetry breaking. Further studies of bound state interactions (via quark annihilation/creation such as in \fig{dualdiag}) are required to understand the properties of these solutions.

\section{Frame dependence \label{framesec}}

The QCD action is invariant under Poincar\'e transformations. To address dynamics beyond the mass spectrum, such as form factors or scattering amplitudes, it is necessary to know how bound states transform under boosts, \ie, to know their frame dependence. This is a challenge for spatially extended states because the quantization surface is not invariant under all transformations. An equal time surface is shifted under time translations generated by the Hamiltonian, whereas boosts change the very definition of simultaneity. Consequently part of the Poincar\'e symmetry of bound states is realized dynamically, through interactions that rearrange their Fock expansion. 

In approximations that rely on a power expansion in a fundamental parameter each order of the expansion must have the symmetry of the exact result. The Lorentz invariance of scattering amplitudes at each order in $\alpha$ is a well-known example. Similarly the Born term of an $\hbar$ expansion must be Poincar\'e symmetric. Demonstrating the symmetry provides a non-trivial check that the Born term is correctly evaluated.

The bound states considered here are not interacting with external forces. Thus the generator $\bs\mP$ \eq{momop} of space translations has no interaction term, and its meson eigenstates  with eigenvalue $\Pv$ take the form\footnote{The color algebra is frame independent, so the meson wave function has the (implicit) color structure \eq{mcolor} at any $\Pv$.}
\beq\label{pmeson}
\ket{E,\Pv}=\int d\xv_1 d\xv_2 \, e^{i\Pv\cdot(\xv_1 +\xv_2)/2}\bar\psi(\xv_1)\Phi(\xv_1 -\xv_2)\psi_\beta(\xv_2)\ket{0}
\eeq
In the present approach the generator $\mH$ of time translations has a frame-dependent interaction term, given by the classical gauge field boosted from the rest frame similarly as for Positronium in \eq{Aboost}. The $\Pv$-dependence of the wave function $\Phi(\xv_1 -\xv_2)$ is determined by the requirement that \eq{pmeson} be an eigenstate of the Hamiltonian. Poincar\'e invariance implies that the energy eigenvalue is $E=\sqrt{\Pv^2+M^2}$, where $M$ is the rest mass of the bound state. 

It is useful to gain experience by first considering $D=1+1$ dimensions. The single space dimension allows to solve the bound state equation explicitly for any $P$. A direct solution is more challenging in higher dimensions, where the absence of rotational symmetry for $\Pv \neq 0$ prevents a separation of variables as in section \ref{mes31subsec}.

\subsection{Mesons with $\Pv \neq 0$ in $D=1+1$ dimensions \label{bsboost11subsec}}

In $D=1+1$ the perturbative \order{\alpha_s} potential is linear, as is the confining potential \eq{mpot}. This allows comparing two approaches: (i) the gauge field is treated as an operator with the gauge choice $A^1=0$ (section \ref{popsubsubsec}, \cite{Dietrich:2012iy}) and (ii) the classical $A^0$ potential is boosted analogously to the Positronium case \eq{Aboost} (section \ref{pclsubsubsec}). The gauge choice has a striking effect on the boosted wave function, but both approaches give the same mass spectrum.

\subsubsection{QED$_2$ bound states with $A^1=0$ \label{popsubsubsec}}

The operator version of Gauss' law in QED$_2$
\beq\label{gauss11}
-\partial_x^2 A^0(t,x) =e \psi^\dag(t,x)\psi(t,x)
\eeq
allows to express $A^0$ in terms of the fermion fields. In $A^1=0$ gauge the action then depends on the fermion fields only,
\beq\label{qedact1}
S =\int dtdx\, \psi^\dag\gamma^0\big(i\slashed{\partial}-m\big)\psi +
\frac{e^2}{4}\int dtdxdy\, \psi^\dag\psi(t,x) |x-y| \psi^\dag\psi(t,y)
\eeq
The action is invariant under Poincar\'e transformations generated by
\beqa\label{genop11}
 \mP(t) &=& \int dx\, \psi^\dag(t,x) (-i\partial_x) \psi(t,x)\nn \\[2mm]
\mH(t) &=& \int dx\, \psi^\dag(-i\sigma_1\partial_x+m\sigma_3)\psi
 -\frac{e^2}{4}\int dx dy\,  \psi^\dag\psi(t,x) |x-y| \psi^\dag\psi(t,y)  \\[2mm]
\mK(t) &=& t\, \mP + \int dx\, \psi^\dag\big[ x( i\sigma_1 \partial_x - m\sigma_3) + i\halft \sigma_1 \big] \psi
+ \frac{e^2}{8}\int dx dy\,  \psi^\dag\psi(t,x) (x+y)|x-y| \psi^\dag\psi(t,y) \nn
\eeqa
where the Dirac matrices are represented as in \eq{matdef}. These generators satisfy the Lie algebra
\beq\label{Lie1+1}
\com{\mP}{\mH}=0 \hspace{2cm} \com{\mP}{\mK}=i\mH \hspace{2cm} \com{\mH}{\mK}=i\mP
\eeq
A boost of $A^0$ gives rise to $A^1 \propto A^0$. In order to maintain $A^1=0$ the boost must be combined with a gauge transformation, which is in fact included in the definition of the generator $\mK$ in \eq{genop11}. The gauge parameter is itself an operator \cite{Dietrich:2012iy}, 
\beq\label{gaugepar}
\theta(t,x) = -\frac{e^2}{4}\int dy\, (x-y)|x-y| \psi^\dag\psi(t,y) 
\eeq
Since the $A^0$ potential depends on the separation of the $e^+e^-$ pair, so must the gauge parameter $\theta$.

The bound state \eq{pmeson} is an energy eigenstate, $\mH\ket{E,P} = E\ket{E,P}$, provided the wave function satisfies the bound state equation 
\beq \label{bse3}
i\partial_x\acom{\sigma_1}{\Phi(x)}-\halft P\com{\sigma_1}{\Phi(x)}+m\com{\sigma_3}{\Phi(x)} = \big[E-V(x)\big] \Phi(x)
\eeq
where $V(x)=\halft e^2\,|x| \equiv V'|x|$. The $P$-dependence of $\Phi(x)$ implied by this equation is consistent with that given by a boost generated by $\mK$ \cite{Dietrich:2012iy}. The Lie algebra ensures the correct $P$-dependence of the energy, $E =\sqrt{P^2+M^2}$.

When $\Phi(x)$ is expanded in a Pauli matrix basis as in \eq{wf11} the coefficients $\phi_2(x)$ of $i\sigma_2$ and $\phi_3(x)$ of $\sigma_3$ are not differentiated in \eq{bse3}. Expressing them in terms of $\phi_0$ and $\phi_1$ the $2\times 2$ wave function takes the form
\beq\label{wf11a}
\Phi(x) = \phi_0+\phi_1\sigma_1-\frac{2m\phi_1}{\tau V'}\sigma_1\Pisl^\dag
\hspace{1cm} \Pi=(E-V(x),P) \hspace{1cm} \tau \equiv \Pi^2/V' = \big[(E-V)^2-P^2\big]/V'
\eeq
where the 2-vector $\Pi$ is the kinetic momentum ($V=eA^0,\ eA^1=0$). The two remaining components of the BSE \eq{bse3} give coupled differential equations for $\phi_0$ and $\phi_1$. With a change of variables $x \to \tau(x)$ they become (for $x\geq 0$)
\beq\label{wf11b}
 -4i\partial_\tau\phi_0=\Big[1-\frac{4m^2}{\tau V'}\Big]\phi_1 \hspace{2cm} -4i\partial_\tau\phi_1=\phi_0
\eeq
Since these equations are frame independent (\ie, they do not depend explicitly on  $P$), so are $\phi_0(\tau)$ and $\phi_1(\tau)$. The differential equation \eq{msig2ndord} obtained in the rest frame is indeed compatible with \eq{wf11b}, and the solution is given by \eq{mgensol}. However, the functional dependence of $\tau(x)$ on $x$ does depend on $P$. Thus the wave function components $\phi_0(x)$ and $\phi_1(x)$ are frame dependent when expressed as functions of $x$. In the expression \eq{wf11a} for the $2\times 2$ wave function $\Phi(x)$ the kinetic momentum $\Pi(x)$ is explicitly $P$-dependent.

In section \ref{discretesubsubsec} we saw that requiring the rest frame wave function to be locally normalizable at $\tau=0$ implied $\phi_1(\tau\to 0) \propto \tau$. The same condition ensures that $\Phi(x)$ in \eq{wf11a} is regular at $\tau=0$ in a general frame. 

The continuity condition \eq{mbc} requires that either $\phi_0(x=0)$ or $\phi_1(x=0)$ vanishes, depending on the parity $\eta$. Since $\phi_0(\tau)$ and $\phi_1(\tau)$ are frame independent, so are the values of $\tau$ where they vanish. According to \eq{wf11a} $\tau(x=0)=(E^2-P^2)/V'$, hence we must have $E^2-P^2=M^2$ in all frames. The discrete values of the energy $E$ for which the wave function is regular at $\tau=0$ and continuous at $x=0$ therefore has the correct dependence on the momentum $P$.

\subsubsection{QED$_2$ bound states with a boosted classical gauge field \label{pclsubsubsec}}

The Hamiltonian with a classical gauge field $(A^0,A^1)$ is
\beq\label{hcl11}
\mH= \int dx\,\psi^\dag\big[-i\sigma_1\partial_x+m\sigma_3+g(A^0-\sigma_1 A^1)\big]\psi
\eeq
Operating on a state of the form \eq{pmeson} we get
\beqa\label{hep11}
\mH\ket{E,P} &=& \int dx_1dx_2\,\psi^\dag(x_1)\Big\{\big[i\sigma_1\lder_1+m\sigma_3+gA^0(x_1)-\sigma_1 gA^1(x_1)\big]\sigma_3\,e^{iP(x_1+x_2)/2}\Phi(x_1-x_2) \nn\\[2mm]
&& -\sigma_3\,e^{iP(x_1+x_2)/2}\Phi(x_1-x_2)\big[-i\sigma_1\rder_2+m\sigma_3+gA^0(x_2)-\sigma_1gA^1(x_2)\big]\Big\}\psi(x_2)\ket{0}  \nn\\[2mm]
 &=& E\ket{E,P}
\eeqa
where $\partial_j = \partial/\partial x_j$. The potentials $A^0,\,A^1$ should transform like in \eq{Aboost} under the boost from the rest frame (we are boosting the state rather than the field). In the rest frame the (symmetrized) interaction potential is
\beq\label{Rpots}
gA_R^0(x_{1R})=\halft V(x_{1R}-x_{2R}) = \halft V'|x_{1}-x_{2}|_R=-gA_R^0(x_{2R})
\hspace{2cm} gA_R^1(x_{1R})=gA_R^1(x_{2R})=0
\eeq
Lengths are Lorentz contracted in equal-time wave functions. As in \eq{restcoord},
\beq\label{lcontr}
x_1-x_2 = \frac{(x_1-x_2)_R}{\cosh\xi} \hspace{2cm} \cosh\xi = \frac{E}{M}
\eeq
Consequently the potentials in \eq{hep11} are
\beqa\label{Ppots}
gA^0(x_1) &=& \cosh\xi\,gA^0_R(x_{1R})=\halft V'\cosh^2\xi\,|x_1-x_2| = -gA^0(x_2) \equiv \halft gA^0 \nn\\[2mm]
gA^1(x_1) &=& \sinh\xi\,gA^0_R(x_{1R})=\halft V'\sinh\xi\,\cosh\xi\,|x_1-x_2| = -gA^1(x_2) \equiv \halft gA^1
\eeqa

In terms of these potentials and  $x=x_1-x_2$ the bound state condition implied by \eq{hep11} is, 
\beq\label{bsep11}
i\partial_x\acom{\sigma_1}{\Phi(x)}=(E-gA^0)\Phi(x)+\halft(P-gA^1)\com{\sigma_1}{\Phi(x)}-m\com{\sigma_3}{\Phi(x)}
\eeq
The kinetic energy and momentum are
\beqa\label{kinep11}
\Pi^0 \equiv E-gA^0 &=& E-V'\cosh^2\xi\,|x| = E\Big(1-\frac{V'}{M}\,\cosh\xi\,|x|\Big) \equiv \sqrt{\tau V'}\,\frac{E}{M} \nn\\[2mm]
\Pi^1 \equiv P-gA^1 &=& P-V'\sinh\xi\cosh\xi\,|x| = P\Big(1-\frac{V'}{M}\,\cosh\xi\,|x|\Big) = \sqrt{\tau V'}\,\frac{P}{M}
\eeqa
This variable $\tau$ is related in the same way to the square of the kinetic 2-momentum as its namesake in \eq{wf11a},
\beq\label{kinmom11}
\Pi^2 \equiv (E-gA^0)^2-(P-gA^1)^2 = V'\tau
\eeq
Nevertheless, this function $\tau(x)$ differs from the function $\tau(x)$ defined above in $A^1=0$ gauge. For $P=0$ both reduce to the rest frame function in \eq{msig}.

Expanding $\Phi(x)$ in the basis of Pauli matrices as in \eq{wf11} we find
\beq\label{wf11c}
\Phi(x) = \phi_0+\phi_1\sigma_1-\frac{2m\phi_1}{\tau V'}\sigma_1\Pisl^\dag
\eeq
which is identical in form to the expression in \eq{wf11a}. However, the kinetic 2-momentum $\Pi=(\Pi^0,\Pi^1)$ given in \eq{kinep11} as well as $\tau$ \eq{kinmom11} differ from the corresponding variables in $A^1=0$ gauge. 

The differential equations for $\phi_0$ and $\phi_1$ are identical in form to those in \eq{wf11b}. Consequently the solution for $\phi_1(\tau)$ is again given by \eq{mgensol}, with $b=0$ for a wave function that is regular at $\tau=0$. Since $V'\tau(x=0)=E^2-P^2$ as in $A^1=0$ gauge the energy eigenvalues $E$ have the correct dependence on $P$.

\begin{figure}[h] \centering
\includegraphics[width=.6\columnwidth]{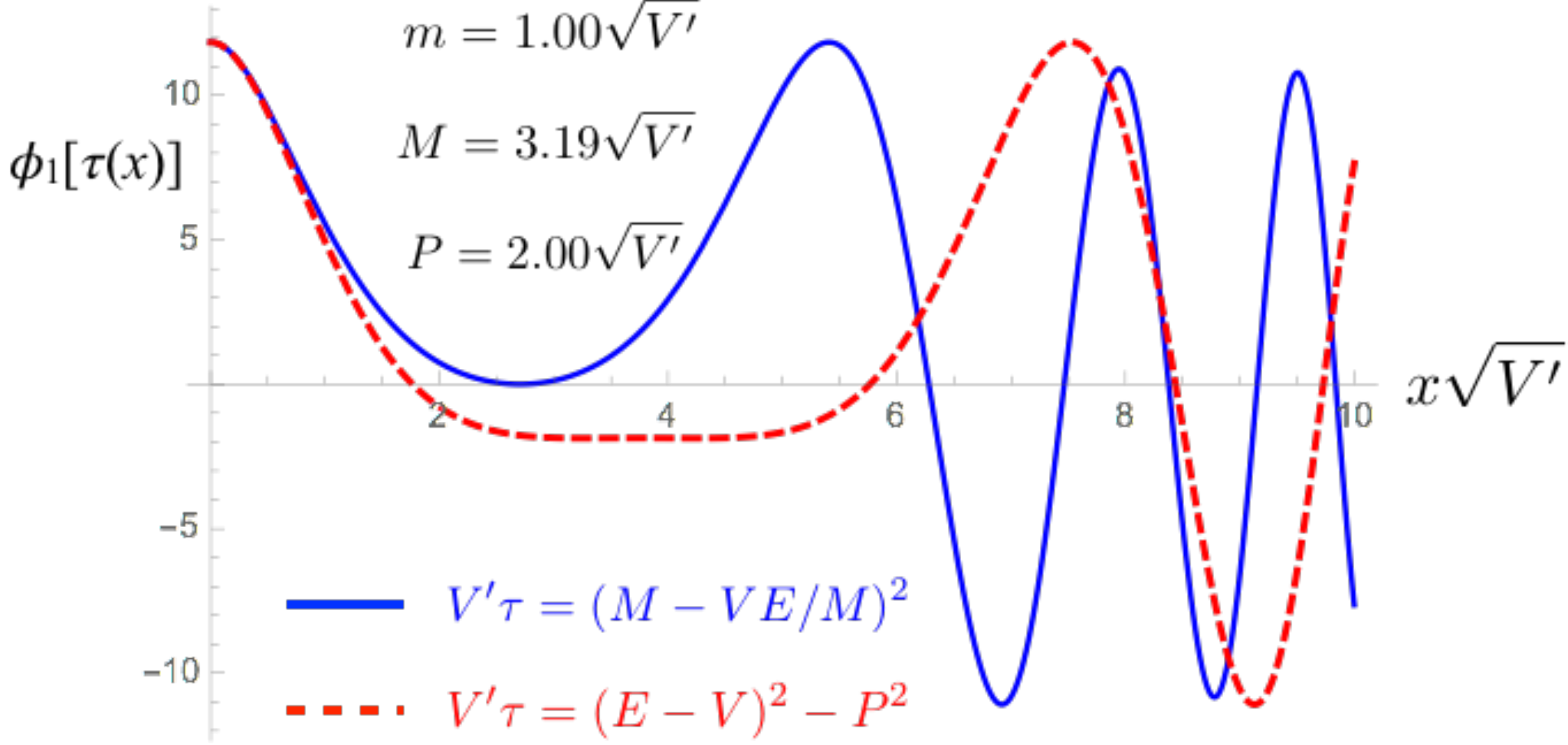}
\caption{The wave function $\phi_1\big[\tau(x)\big]$ \eq{mgensol} $(a=1,\ b=0)$ for the positive parity ground state in $D=1+1$, using the function $\tau(x)$ given by \eq{kinep11} (solid blue curve) and \eq{wf11a} (dashed red curve). \label{P-dep11}}
\end{figure}

\fig{P-dep11} compares the $x$-dependence of the the wave function $\phi_1\big[\tau(x)\big]$ \eq{mgensol} (with $b=0$) calculated using the function $\tau(x)$ given by \eq{kinep11} (solid blue) and \eq{wf11a} (dashed red). The two curves describe the same function, plotted at displaced values of $x$. The difference between the two expressions for $\tau(x)$ increases with $x$, since the gauge transformation to $A^1=0$ is $\propto x$. This illustrates the gauge dependence of the wave function, corresponding to the same energy eigenvalue (here $M=3.19\sqrt{V'}$).

As I noted above, the components $\phi_0(\tau)$ and $\phi_1(\tau)$ of the $2\times 2$ wave function $\Phi$ are frame-independent functions of $\tau$ \eq{kinep11}. However, the kinetic momentum $\Pisl^\dag(P)$ in \eq{wf11c} depends explicitly on $P$. In the rest frame
\beq\label{kinmom11a}
\Pisl^\dag(P=0) = \sqrt{V'\tau}\,\sigma_3
\eeq
Using the definition \eq{kinep11} of $\tau$ it is readily verified that 
\beq\label{kinmom11b}
\Pisl^\dag(P) = (E\sigma_3+P\,i\sigma_2)\frac{\sqrt{V'\tau}}{M} = e^{-\sigma_1\xi/2}\,\Pisl^\dag(P=0)\,e^{\sigma_1\xi/2}
\eeq
Consequently the explicit dependence of the full wave function $\Phi^{(\xi)}$ on $P=M\sinh\xi$ is
\beq\label{wfxi11}
\Phi^{(\xi)}(\tau) = e^{-\sigma_1\xi/2}\,\Phi^{(\xi=0)}(\tau)\,e^{\sigma_1\xi/2}
\eeq
Since the wave functions are related at the same value of $\tau$ their $x$-dependence Lorentz contracts in the standard way when $\tau$ is defined as in \eq{kinep11}.

The transformation \eq{wfxi11} of the wave function in finite boosts emerged from the explicit solution of the bound state equation \eq{bsep11}. In higher dimensions it is less straightforward to solve the bound state equation when $P \neq 0$. It is therefore interesting to ask whether a relation analogous to \eq{wfxi11} is valid in other dimensions as well. In the next section I show that this transformation is correct in $D=2+1$, and holds in $D=3+1$ for some (but not all) states.

\subsection{Mesons with $\Pv \neq 0$ in $D=2+1$ and $D=3+1$ dimensions \label{bsboost31subsec}}

\subsubsection{The bound state equation \label{bsecl}}

In $D=3+1$ I choose the $z$-axis along the bound state momentum,
\beq\label{mom31}
\Pv = (M\cosh\xi,0,0,M\sinh\xi)
\eeq
The boost of the rest frame field $gA^0_R(\xv_{1R})=-gA^0(\xv_{2R})=\halft V'|\xv_1-\xv_2|_R$ leaves $A^1=A^2=0$ and
\beqa\label{Aboost3}
gA^0(\xv_1)&=& -gA^0(\xv_2)=\halft\cosh\xi\,V'|\xv_1-\xv_2|_\xi \equiv \halft gA^0 \nn\\[2mm]
gA^3(\xv_1)&=& -gA^3(\xv_2)=\halft\sinh\xi\,V'|\xv_1-\xv_2|_\xi \equiv \halft gA^3
\eeqa
As in \eq{restcoord} the rest frame separation $\xv_\xi$ is
\beq\label{coordboost3}
\xv_\xi = (x,y,z\cosh\xi)
\eeq

Requiring that the state \eq{pmeson} be an eigenstate of the Hamiltonian 
\beq\label{hcl31}
\mH= \int d\xv\,\psi^\dag(\xv)\Big\{-i\gz\gv\cdot\nv+m\gz+g\big[A^0(\xv)-\gz\gamma^3 A^3(\xv)\big]\Big\}\psi(\xv)
\eeq
gives the bound state equation for the wave function $\Phi^{(\xi)}(\xv_1-\xv_2)$,
\beq\label{bse31}
i\nv\cdot\acomb{\gz\gv}{\Phi^{(\xi)}(\xv)}=(E-gA^0)\Phi^{(\xi)}(\xv) +\halft(P-gA^3)\comb{\gz\gamma^3}{\Phi^{(\xi)}(\xv)}-m\comb{\gz}{\Phi^{(\xi)}(\xv)}
\eeq
where $gA^0$ and $gA^3$ are given by \eq{Aboost3}. The absence of spherical symmetry when $\Pv\neq 0$ means that the radial and angular variables cannot be separated similarly as in the rest frame (section \ref{mes31subsec}). Even in $D=2+1$ where the Dirac matrices can be represented in terms of the $2\times 2$ Pauli matrices \eq{21dirdef} the solution of the coupled partial differential equations \eq{bse31} is not obvious.

\subsubsection{An ansatz for the boost property of meson wave functions \label{ansatzsubsubsec}}

Here I shall study only a simple guess of the boost dependence of the wave function, based on the assumption that the quark separation $\xv$ Lorentz transforms as in classical relativity. The longitudinal direction (denoted $\parallel$, $z$ or $3$) contracts by $\cosh\xi$ whereas the directions transverse to $\Pv$ ($\perp$ or $x,y$) are invariant. It is then convenient to introduce the boost-invariant longitudinal distance $s$
\beq\label{sdef}
s=\frac{E}{M}\,z = \cosh\xi\,z \hspace{2cm} \partial_z = \cosh\xi\, \partial_s \hspace{2cm} \xv_\xi=(\xtr,s)
\eeq
The boost invariance of $|\xv_\xi|$ ensures the boost invariance of the variable $\tau$ defined by
\beq\label{Aboost3b}
E-gA^0 = E\Big(1-\frac{V'}{M}|\xv_\xi|\Big) \equiv \sqrt{V'\tau}\,\cosh\xi \hspace{2cm} 
P-gA^3 = P\Big(1-\frac{V'}{M}|\xv_\xi|\Big) = \sqrt{V'\tau}\,\sinh\xi
\eeq
The bound state equation \eq{bse31} can be expressed in terms of $s,\,\tau$ and $\bs{\alpha} \equiv \gz\gv$, writing $\wfxi(\xtr,z(s))$ as $\wfxi(\xtr,s)$,
\beqa\label{cbs2}
\frac{E}{M}\,i\partial_s\acomb{\apa}{\wfxi(\xtr,s)}=&& \\[2mm]
&& \hspace{-3.5cm}\frac{E}{M}\sqrt{V'\tau}\,\wfxi(\xtr,s) +\frac{P}{2M}\sqrt{V'\tau}\comb{\apa}{\wfxi(\xtr,s)}-m\comb{\gz}{\wfxi(\xtr,s)}- i\ntr\cdot\acomb{\atr}{\wfxi(\xtr,s)}\nn
\eeqa

The simplest {\it Ansatz} for the relation between the boosted and rest frame wave functions is the generalization of \eq{wfxi11},
\beq\label{wftran}
\wfxi(\xtr,s)=\exp(-\halft\apa\xi)\wfr(\xtr,s)\exp(\halft\apa\xi)
\eeq
The derivative of the wave function wrt. $\xi$ at constant $\xv_\xi=(\xtr,s)$ is then given by
\beq\label{xider}
\partial_\xi\wfxi(\xtr,s)= -\halft\comb{\apa}{\wfxi(\xtr,s)}
\eeq
This allows to differentiate the BSE \eq{cbs2} wrt. $\xi$ at constant $\xv_\xi$. Noting that $\acomb{\apa}{\comb{\apa}{\wfxi}}=0$ one gets,
\beq\label{cbs4}
\frac{P}{M}i\partial_s\acomb{\apa}{\wfxi}=\frac{P}{M}\sqrt{V'\tau}\,\wfxi -\frac{P}{4M}\sqrt{V'\tau}\comb{\apa}{\comb{\apa}{\wfxi}}+\halft m\comb{\gz}{\comb{\apa}{\wfxi}}+ \halft i\ntr\cdot\acomb{\atr}{\comb{\apa}{\wfxi}}
\eeq
where $\wfxi=\wfxi(\xtr,s)$.

It is straightforward to check whether the $\xi$-dependence of $\wfxi(\xtr,s)$ given by \eq{wftran} is correct near $\xi=0$. Setting $\xi=0$ in \eq{cbs4} we find a constraint on the rest frame wave function,
\beq\label{delconstr}
\Delta \equiv m\comb{\gz}{\comb{\apa}{\wfr}}+ i\ntr\cdot\acomb{\atr}{\comb{\apa}{\wfr}} =0
\eeq

\vspace{.3cm}

{\it Check of $\Delta=0$ in $D=2+1$ dimensions}

Inserting the  Pauli matrix expansion \eq{wfdef1} of $\wfr$ in $D=2+1$ dimensions it is readily seen that the condition $\Delta=0$ is satisfied because the coefficient of $\sigma_3$ vanishes according to \eq{bsd}. The correctness of the boost dependence \eq{wftran} for finite $\xi$ was verified numerically by solving the radial equations \eq{21coupled} numerically, thus determining $\wfr$. The transformed wave function \eq{wftran} satisfied the bound state equation \eq{cbs2} for general $\xi$.

\vspace{.5cm}

{\it Check of $\Delta=0$ in $D=3+1$ dimensions}

The necessary condition \eq{delconstr} for the Ansatz \eq{wftran} to be correct turns out to be satisfied for all values of the angular momentum $J$, under the following conditions on the quark mass $m$ and $J^z=\lambda$:
\begin{itemize}
\item $\pi$ trajectory: $m=0$ and $\lambda=0$.
\item $a_1$ trajectory: $m=0$.
\item $\rho$ trajectory: $\lambda=0$.
\end{itemize}

In each case a numerical check similar to the one in $D=2+1$ showed that the Ansatz \eq{wftran} is correct for general $\xi$ when the $\xi=0$ condition \eq{delconstr} is satisfied. A more systematic study is needed to identify the correct boost dependence of the wave function for the states with $m,\lambda \neq 0$ in $D=3+1$ dimensions.

\break

\section{What's new -- and what's next? \label{discsec}}

In this concluding section I briefly remark on the novel aspects of the issues discussed in these lectures, and mention some of the (many) opportunities for further studies.

\setcounter{subsubsection}{0}
\subsubsection{QED bound states of lowest order in $\hbar$ \label{discsubsec1}}

The similarity of atomic and quarkonium spectra (\fig{atoms}) was a main motivation for the bound state studies presented above. The similarity suggests that the QCD description of quarkonia is related to the QED description of atoms. 
Could perturbative QCD really provide a first principles approach to hadron physics -- including confinement? Perturbation theory is our main analytic tool for the Standard Model, so this question merits serious thought.

The perturbative expansion for bound states is less straightforward than for scattering processes (section \ref{basics}). Since no single Feynman diagram has a bound state pole we sum a infinite number of ladder diagrams (\fig{sladder4}).  The geometric sum builds the classical QED potential $V(r)=-\alpha/r$, which is missing in the \order{\alpha^0} $in$ and $out$ states of the perturbative $S$-matrix.

Positronium at the Born (Schr\"odinger equation) level is an eigenstate of $H_{QED}$ when the classical $A^0$ (Coulomb) field of the $e^-e^+$ pair is used in the Hamiltonian. Since the formulation is based on the relativistic Hamiltonian also bound states in motion can be addressed (section \ref{scheq}).

This method for deriving the Schr\"odinger equation as a Born approximation of QED bound states appears to be new. Higher order (loop) corrections in $\hbar$ are calculable in the rest frame using the known methods of NRQED \cite{Kinoshita,Baker:2014sua}. In an arbitrary frame the expression \eq{smatrix} for the $S$-matrix should be applicable, using the bound states of lowest order in $\hbar$ as asymptotic states. This remains to be demonstrated.

\subsubsection{Dirac states vs. wave functions \label{discsubsec2}}

The Dirac equation for an electron bound in an external field specifies a wave function with both positive and negative energy components ($\propto\psl\pm m$). The structure of the {\it state} described by this wave function is rarely discussed. The Klein paradox \cite{Klein:1929zz} demonstrates that the Dirac state is composed not only of a single $e^-$, but also has $e^+e^-$ pairs.

In section \ref{diracsec} the Dirac bound states were expanded in the basis of free electron and positron Fock states \cite{blaizot}. The Dirac wave function determines the Bogoliubov transform \eq{cndef1} of the free creation and annihilation operators. The distribution of the $e^+e^-$ pairs in the ground state (vacuum) depends on the complete set of wave functions. 

The energy spectrum specified by the Dirac equation is continuous when the potential $V(r)$ is polynomial in $r$ \cite{plesset,Giachetti:2007vq}. This contrasts with the discrete spectrum of the Schr\"odinger equation and is caused by the presence of positrons in the Dirac states. Potentials that confine electrons repulse positrons. The positrons appear at large $r$, with kinetic energies that match their (negative) potential energy.

It would be interesting to study the distribution of the $e^+e^-$ pairs in, \eg, the Dirac states of the Hydrogen atom.

\subsubsection{Confinement from an \order{\alpha_s^0} classical gluon field \label{discsubsec3}}

The scale $\lqcd \sim 200$ MeV is central to hadron physics but does not appear in the QCD action. At the level of the classical field equations such a scale can arise only via a boundary condition. The QED potential $V(r)=-\alpha/r$ is given by Gauss' law \eq{gausslaw0} with the boundary condition $\lim_{|\xv|\to\infty}A^0(\xv)=0$. The requirement of Poincar\'e invariance appears to select a unique homogeneous solution of \order{\alpha_s^0} (section \ref{confsubsec}). It is characterized by the strength $\propto\la^2$ of the asymptotic color electric field. Only color singlet $q\bar q$ and $qqq$ states are allowed. The $A_a^0(\xv)$ field for each color component of mesons and baryons is given in \eq{mfield} and \eq{bfield}, respectively.

The coherent sum of the gauge fields in a color singlet hadron vanishes for all $\xv$. This non-abelian feature means that the \order{\alpha_s^0} color field of one hadron does not induce interactions with the quarks of another hadron. Two hadrons interact only when their quark constituents coincide, causing $q\bar q$ annihilation/creation as in dual diagrams (\fig{dualdiag}).

As in Positronium there is also a Born level, \order{\alpha_s} gluon exchange contribution. Its inclusion with the \order{\alpha_s^0} confining field remains to be addressed. This may be done in the framework of the expression \eq{smatrix} for the perturbative $S$-matrix, with the \order{\alpha_s^0} solutions as $in$ and $out$ states. A further step would be to consider perturbative loop corrections.

\subsubsection{Hadron spectra \label{discsubsec4}}

The meson eigenstates of $H_{QCD}$ with the classical color field \eq{mfield} are readily found when string breaking is neglected. The light $q\bar q$ states lie on straight Regge and daughter trajectories (\fig{3+1_pion}a). Heavy quarkonia have spectra given by the Schr\"odinger equation with a linear potential.

The states have features of quark-hadron duality (\fig{dual}). This may allow to describe the average features of particle production processes even when string breaking is neglected. The wave functions of single hadrons are sensitive to string breaking effects at large color separations. These aspects require further study.

There are $q\bar q$ bound states with vanishing mass, $M=0$ (section \ref{csbsubsubsec}). They may be relevant for chiral symmetry breaking.

The spectra of mesons with quarks of unequal mass was so far studied only in $D=1+1$ dimensions \cite{Dietrich:2012un}. The solutions of the baryon ($qqq$) bound state equation \eq{bbse} have not been addressed.

\subsubsection{Boost covariance \label{discsubsec5}}

The frame dependence of equal-time Positronium states was first given in \cite{Jarvinen:2004pi}, using the Bethe-Salpeter equation. The present, equivalent formulation is based on boosting the classical gauge field (section \ref{scheq}). For relativistically bound states I am not aware of any other framework where the transformation of equal-time bound state wave functions under boosts would be explicitly known. The fact that some (albeit not all) wave functions transform according to \eq{wftran} in $D=3+1$ dimensions is published here for the first time. It is important to identify the correct transformation law for all states.

\subsubsection{String breaking and hadron loops \label{discsubsec6}}

The \order{\hbar^0} dynamics must be unitary. The non-linear unitarity conditions cannot be satisfied exactly, only order by order in some expansion. In the absence of string-breaking all states have zero width like in dual models \cite{Veneziano:1968yb,Mandelstam:1974fq}. The string breaking effects described by \fig{dualdiag} can be calculated and is of order $1/\sqrt{N_C}$, where $N_C$ is the number of colors. The square of this diagram gives rise to an \order{1/N_C} hadron loop.

This suggests that unitarity may be satisfied analogously to standard perturbation theory. The expansion parameter would be $1/N_C$ and the loops would be made of hadrons rather than quarks and gluons. For QCD $N_C=3$ is a moderate number. Nevertheless, the data on hadron dynamics generally supports an approximation scheme based on narrow resonance widths.

\subsubsection{Scattering amplitudes \label{discsubsec7}}

The Lorentz covariant expressions for hadron states allows the evaluation of scattering amplitudes, including electro-magnetic form factors and parton distributions. So far Deep inelastic scattering was studied only in $D=1+1$ dimensions \cite{Dietrich:2012un}. The quark distributions were found to be enhanced for $x_B \to 0$, suggesting the contribution of sea quarks. This supports the notion that the relativistic wave functions implicitly describe multiparticle Fock states, similarly as in the Dirac case (section \ref{diracsec}).

\acknowledgments
I have benefitted from many discussions, particularly with Jean-Paul Blaizot, Stan Brodsky, Dennis D.~Dietrich, Matti J\"arvinen, J\"orn Knoll and Stephane Peign\'e. During the preparation of this material I enjoyed visits of 1-2 months at NIKHEF (Amsterdam), IPhT (Saclay), CP$^3$ (Odense), GSI (Darmstadt) and ECT* (Trento). I am very grateful for their hospitality, and to the Department of Physics at Helsinki University for my privileges of Professor Emeritus. Travel grants from the Magnus Ehrnrooth Foundation have allowed me to maintain contacts and present my research to colleagues.

\end{document}